\documentclass%
[prb,twocolumn,superscriptaddress,nobibnotes,showkeys,amsmath,amssymb,citeautoscript]%
{revtex4}

\usepackage[dvips]{graphicx}
\usepackage{amsmath}
\usepackage{bm}
\usepackage{times}
\usepackage{txfonts}
\usepackage{mathrsfs} 

\begin{document}
\title{
 Resistance functions for two unequal spheres 
 in linear flow at low Reynolds number 
 with the Navier slip boundary condition
}
\author{Kengo Ichiki}%
\email{kengoichiki@gmail.com}
\affiliation{Department of Mechanical Engineering,
University of Alberta, Edmonton, AB, T6G 2G8, Canada}
\affiliation{National Institute for Nanotechnology,
National Research Council of Canada,
11421 Saskatchewan Drive, Edmonton, AB, T6G 2M9, Canada}

\author{Alexander E. Kobryn}
\affiliation{National Institute for Nanotechnology,
National Research Council of Canada,
11421 Saskatchewan Drive, Edmonton, AB, T6G 2M9, Canada}

\author{Andriy Kovalenko}
\affiliation{National Institute for Nanotechnology,
National Research Council of Canada,
11421 Saskatchewan Drive, Edmonton, AB, T6G 2M9, Canada}
\affiliation{Department of Mechanical Engineering,
University of Alberta, Edmonton, AB, T6G 2G8, Canada}

\date{March 24, 2009}

\begin{abstract}
Resistance functions for two spherical particles 
with the Navier slip boundary condition 
in general linear flows, including 
rigid translation, rigid rotation, and strain, at low Reynolds number 
are derived by the method of reflections as well as 
twin multipole expansions. 
In the solutions, 
particle radii and slip lengths can be chosen independently. 
In the course of calculations, 
single-sphere problem with the slip boundary condition 
is solved by Lamb's general solution 
and the expression of multipole expansions, 
and Fax\'{e}n's laws of force, torque, and stresslet 
for slip particle are also derived. 
The solutions of two-body problem are confirmed 
to recover the existing results in the no-slip limit and 
for the case of equal scaled slip lengths. 
\end{abstract}


\maketitle

\section{Introduction}
According to increasing scientific interests in micro- and nanofluidics 
and nanotechnology in recent years, 
fluid mechanics is applied to such small-scale systems, 
in addition to molecular-level theories, 
where the characteristic Reynolds number is generally small enough 
to take the Stokes approximation governed by 
linear partial differential equations. 
In fluid mechanics, historically, 
both no-slip and slip boundary conditions were proposed 
in nineteenth century \cite{NetoEtal2005} 
when the proper boundary conditions were discussed in the first place. 
Navier \cite{Navier1823} gave the slip boundary condition 
where the slip velocity is proportional to the tangential component 
of the surface force density. 
For gas flows, Maxwell \cite{Maxwell1879} had shown that 
the surface slip is related to the non-continuous nature of the gas and 
the slip length is proportional to the mean-free path.
For liquids, on the other hand, from experiments at that age, 
the no-slip boundary condition was accepted 
and since then had been treated as a fundamental law. 
However, 
by recent extensive studies on the surface slip in micro and nano scales, 
the physics of the liquid-solid slip is 
recognized to be much more complicated than that for gases.
Actually apparent violations of the no-slip boundary condition 
at the liquid-solid interface in nano scale have been reported 
\cite{NetoEtal2005,LaugaEtal2007,Vinogradova1999,LaugaStone2003}.

Although the importance of the surface slip is realized, 
theoretical studies and analytical solutions for 
the slip boundary condition are very limited 
compared with those for the no-slip boundary condition. 
Basset solved the flow of single sphere with slip surface \cite{Basset}, 
Felderhof derived Fax\'{e}n's law and solutions expressed by 
multipole expansions for single sphere \cite{Felderhof1976} 
and two spheres \cite{Felderhof1977}, 
B\l{}awzdziewicz \textit{et al.} showed 
the interaction between the slip spheres and 
lubrication functions for the axisymmetric motion \cite{BlawzdziewiczEtal1999}, 
and Luo and Pozrikidis studied two slip spheres under the shear flow 
\cite{LuoPozrikidis2007a}.
Recently, the present authors extended the Stokesian dynamics method 
(without lubrication) 
for slip particles using multipole expansions and Fax\'{e}n's laws 
and obtained the slip dependencies 
for the drag coefficient and effective viscosity \cite{IchikiEtal2008}.
With no-slip boundary condition, 
the problem of two spherical particles 
is solved by Jeffrey and Onishi \cite{JeffreyOnishi1984} 
and Jeffrey \cite{Jeffrey1992} 
for arbitrary size ratio of the particles
in arbitrary linear flows. 
The extension to the slip particles was done 
by Ying and Peters \cite{YingPeters1989} for the gas-solid system 
and by Keh and Chen \cite{KehChen1997} for the liquid-solid system, 
but they lack the strain flows. 
Keh and Chen \cite{KehChen1997} applied the Navier slip boundary condition 
under a condition that the ratios of 
the slip length and radius for two particles are equal. 
Based on theory by Felderhof \cite{Felderhof1976, Felderhof1977}, 
there is alternative formulation of two-sphere problem which 
covers boundary conditions of surface slip as well as permeability 
\cite{SchmitzFelderhof1982b,SchmitzFelderhof1982c,JonesSchmitz1988,CichockiEtal1988};
they gave mobility functions analytically \cite{SchmitzFelderhof1982c}, 
computationally \cite{JonesSchmitz1988}, 
and numerically \cite{CichockiEtal1988}, 
and resistance functions analytically \cite{SchmitzFelderhof1982b}. 
The analytical expression of resistance function, 
which is the subject of present paper,  
is limited to lower orders. 

In this paper, we will show the exact solution of two spheres 
in the form of resistance functions 
with arbitrary size ratio
under the Navier slip boundary condition with arbitrary slip lengths 
in general linear flows including strain and shear flows.
The present formulation is based on the no-slip case by 
Jeffrey and Onishi \cite{JeffreyOnishi1984} and Jeffrey \cite{Jeffrey1992}, 
but we will show all the necessary equations in order that 
the present paper be self-contained. 
We refer equations in the references as 
Eq. (JO-1) for Jeffrey and Onishi \cite{JeffreyOnishi1984}, 
Eq. (J-1) for Jeffrey \cite{Jeffrey1992}, and 
Eq. (KC-1) for Keh and Chen \cite{KehChen1997}. 

The paper is organized as follows. 
In Sec \ref{sec:formulae}, the definition of resistance functions
and Lamb's general solution are summarized. 
In Sec \ref{sec:single-sphere}, the solution of 
single sphere with slip boundary condition is shown. 
In Sec \ref{sec:two-body-problem}, two-body problem 
is solved by twin multipole expansions 
comparing with the results by method of reflections 
(shown in Appendix \ref{sec:method-of-reflections}). 
Concluding remarks are given in Sec \ref{sec:concluding-remarks}.

\section{Formulas of the Stokes Flow}
\label{sec:formulae}
\subsection{Resistance Functions}
\label{sec:resistance-functions}
At low Reynolds number, the incompressible viscous fluid is governed by 
the Stokes equation 
\begin{eqnarray}
  \bm{0}
  =
  -\bm{\nabla}p
  +\mu\nabla^2\bm{u},
\end{eqnarray}
with the incompressibility condition 
\begin{equation}
  \bm{\nabla}\cdot\bm{u}
  =
  0
  ,
\end{equation}
where $p$ is the pressure, $\bm{u}$ is the velocity, 
and $\mu$ is the shear viscosity of the fluid, 
Let us consider spherical particles in a linear flow $\bm{u}^{\infty}$ given 
at position $\bm{x}$ by 
\begin{equation}
  \bm{u}^{\infty}(\bm{x})
  =
  \bm{U}^{\infty}
  +
  \bm{\Omega}^{\infty}\times\bm{x}
  +
  \bm{E}^{\infty}\cdot\bm{x}
  ,
  \label{eq:u-imposed-by-UOE}
\end{equation}
where the three constants $\bm{U}^{\infty}$, $\bm{\Omega}^{\infty}$, 
and $\bm{E}^{\infty}$ are the rigid translational velocity, 
rigid rotational velocity, and rate of strain 
of the imposed flow, respectively. 
According to the linearity of the Stokes equation, 
dynamics of the particles is completely characterized by 
the resistance equation (or, equivalently, 
the mobility equation, that is, the inverse of the resistance equation). 
For two-body problem, the equation is given [in (J-2)] by 
\begin{equation}
  \left[
    \begin{array}{c}
      \bm{F}^{(1)}\\
      \bm{F}^{(2)}\\
      \bm{T}^{(1)}\\
      \bm{T}^{(2)}\\
      \bm{S}^{(1)}\\
      \bm{S}^{(2)}
    \end{array}
  \right]
  =
  \mu
  \left[
    \begin{array}{cccccc}
      \mathsf{A}_{11} & \mathsf{A}_{12}
      & \widetilde{\mathsf{B}}_{11} & \widetilde{\mathsf{B}}_{12}
      & \widetilde{\mathsf{G}}_{11} & \widetilde{\mathsf{G}}_{12}\\
      \mathsf{A}_{21} & \mathsf{A}_{22}
      & \widetilde{\mathsf{B}}_{21} & \widetilde{\mathsf{B}}_{22}
      & \widetilde{\mathsf{G}}_{21} & \widetilde{\mathsf{G}}_{22}\\
      \mathsf{B}_{11} & \mathsf{B}_{12}
      & \mathsf{C}_{11} & \mathsf{C}_{12}
      & \widetilde{\mathsf{H}}_{11} & \widetilde{\mathsf{H}}_{12}\\
      \mathsf{B}_{21} & \mathsf{B}_{22}
      & \mathsf{C}_{21} & \mathsf{C}_{22}
      & \widetilde{\mathsf{H}}_{21} & \widetilde{\mathsf{H}}_{22}\\
      \mathsf{G}_{11} & \mathsf{G}_{12}
      & \mathsf{H}_{11} & \mathsf{H}_{12}
      & \mathsf{M}_{11} & \mathsf{M}_{12}\\
      \mathsf{G}_{21} & \mathsf{G}_{22}
      & \mathsf{H}_{21} & \mathsf{H}_{22}
      & \mathsf{M}_{21} & \mathsf{M}_{22}
    \end{array}
  \right]
  \cdot
  \left[
    \begin{array}{c}
      \bm{U}^{(1)} - \bm{u}^{\infty}(\bm{x}_{1})\\
      \bm{U}^{(2)} - \bm{u}^{\infty}(\bm{x}_{2})\\
      \bm{\Omega}^{(1)} - \bm{\Omega}^{\infty}\\
      \bm{\Omega}^{(2)} - \bm{\Omega}^{\infty}\\
      \bm{E}^{(1)} - \bm{E}^{\infty}\\
      \bm{E}^{(2)} - \bm{E}^{\infty}
    \end{array}
  \right]
  ,
  \label{eq:resistance-equation}
\end{equation}
where 
$\bm{F}^{(\alpha)}$, $\bm{T}^{(\alpha)}$, and $\bm{S}^{(\alpha)}$ 
are the force, torque, and stresslet of the particle $\alpha$, and 
$\bm{U}^{(\alpha)}$, $\bm{\Omega}^{(\alpha)}$, and $\bm{E}^{(\alpha)}$ 
are the translational and angular velocities and strain 
of the particle $\alpha$, respectively, and 
$\bm{x}_{\alpha}$ denotes the center of particle $\alpha$. 
In the equation, the grand resistance matrix 
is decomposed into $6\times 6$ submatrices. 
Because of the symmetry of the grand resistance matrix, 
the matrices with tilde are obtained from the counterparts as 
$\widetilde{\mathsf{B}}_{\alpha\beta} = \mathsf{B}_{\beta\alpha}^{\dagger}$, 
$\widetilde{\mathsf{G}}_{\alpha\beta} = \mathsf{G}_{\beta\alpha}^{\dagger}$, and 
$\widetilde{\mathsf{H}}_{\alpha\beta} = \mathsf{H}_{\beta\alpha}^{\dagger}$, 
(where $\dagger$ denotes the transpose) and, 
therefore, we need to calculate, at least, the rest. 
Following Jeffrey \textit{et al.} \cite{JeffreyOnishi1984,Jeffrey1992}, 
we scale these submatrices 
[in (JO-1.7a,b,c) and (J-3a,b,c)] as 
\begin{subequations}
\begin{eqnarray}
  \mathsf{A}_{\alpha\beta} &=& 3\pi\left(a_{\alpha} + a_{\beta}\right)
  \widehat{\mathsf{A}}_{\alpha\beta},\\
  \mathsf{B}_{\alpha\beta} &=& \pi\left(a_{\alpha} + a_{\beta}\right)^2
  \widehat{\mathsf{B}}_{\alpha\beta},\\
  \mathsf{C}_{\alpha\beta} &=& \pi\left(a_{\alpha} + a_{\beta}\right)^3
  \widehat{\mathsf{C}}_{\alpha\beta},\\
  \mathsf{G}_{\alpha\beta} &=& \pi\left(a_{\alpha} + a_{\beta}\right)^2
  \widehat{\mathsf{G}}_{\alpha\beta},\\
  \mathsf{H}_{\alpha\beta} &=& \pi\left(a_{\alpha} + a_{\beta}\right)^3
  \widehat{\mathsf{H}}_{\alpha\beta},\\
  \mathsf{M}_{\alpha\beta} &=& \frac{5\pi}{6}\left(a_{\alpha} + a_{\beta}\right)^3
  \widehat{\mathsf{M}}_{\alpha\beta},
\end{eqnarray}
\end{subequations}
where $a_{\alpha}$ is the radius of particle $\alpha$, 
and the matrices with hat are dimensionless. 
For spherical particles, the matrices can be further reduced, 
because the geometry of the problem is completely characterized 
by the single vector $\bm{r} = \bm{x}_{\beta} - \bm{x}_{\alpha}$. 
These submatrices are then given by scalar functions 
[in (JO-16a,b,c) and (J-4a,b,c)] as 
\begin{subequations}
\begin{eqnarray}
  \widehat{A}^{\alpha\beta}_{ij}
  &=&
  X^{A}_{\alpha\beta}
  e_ie_j
  +
  Y^{A}_{\alpha\beta}
  \left(
    \delta_{ij}
    -
    e_ie_j
  \right)
  ,
  \label{eq:A-by-XA-and-YA}
  \\
  \widehat{B}^{\alpha\beta}_{ij}
  &=&
  Y^{B}_{\alpha\beta}
  \epsilon_{ijk}e_k
  ,
  \label{eq:B-by-YB}
  \\
  \widehat{C}^{\alpha\beta}_{ij}
  &=&
  X^{C}_{\alpha\beta}
  e_ie_j
  +
  Y^{C}_{\alpha\beta}
  \left(
    \delta_{ij}
    -
    e_ie_j
  \right)
  ,\\
  \widehat{G}^{\alpha\beta}_{ijk}
  &=&
  X^{G}_{\alpha\beta}
  \left(
    e_ie_j
    -
    \frac{1}{3}
    \delta_{ij}
  \right)
  e_k
  \nonumber\\
  &&
  +
  Y^{G}_{\alpha\beta}
  \left(
    e_i\delta_{jk}
    +
    e_j\delta_{ik}
    -
    2e_ie_je_k
  \right)
  ,\\
  \widehat{H}^{\alpha\beta}_{ijk}
  &=&
  Y^{H}_{\alpha\beta}
  \left(
    e_i\epsilon_{jkl}e_l
    +
    e_j\epsilon_{ikl}e_l
  \right)
  ,\\
  \widehat{M}^{\alpha\beta}_{ijkl}
  &=&
  \frac{3}{2}
  X^{M}_{\alpha\beta}
  \left(
    e_ie_j
    -
    \frac{\delta_{ij}}{3}
  \right)
  \left(
    e_ke_l
    -
    \frac{\delta_{kl}}{3}
  \right)
  \nonumber\\
  &&
  +
  \frac{Y^{M}_{\alpha\beta}}{2}
  \left(
    e_i\delta_{jl}e_k
    +
    e_j\delta_{il}e_k
    +
    e_i\delta_{jk}e_l
    +
    e_j\delta_{ik}e_l
  \right.
  \nonumber\\
  &&
  \quad
  \left.
    -
    4e_ie_je_ke_l
  \right)
  \nonumber\\
  &&
  +
  \frac{Z^{M}_{\alpha\beta}}{2}
  \left(
    \delta_{ik}\delta_{jl}
    +
    \delta_{jk}\delta_{il}
    -
    \delta_{ij}\delta_{kl}
  \right.
  \nonumber\\
  &&
  \quad
  \left.
    +
    e_ie_j\delta_{kl}
    +
    \delta_{ij}e_ke_l
    +
    e_ie_je_ke_l
  \right.
  \nonumber\\
  &&
  \quad
  \left.
    -
    e_i\delta_{jl}e_k
    -
    e_j\delta_{il}e_k
    -
    e_i\delta_{jk}e_l
    -
    e_j\delta_{ik}e_l
  \right)
  ,
\end{eqnarray}
\end{subequations}
where $\bm{e} = \bm{r}/|\bm{r}|$, 
$\delta_{ij}$ is Kronecker's delta, 
and $\epsilon_{ijk}$ is the Levi-Civita tensor. 
The scalar functions $X$, $Y$, and $Z$ above 
are called the resistance functions. 
We have 11 functions for each pair $\alpha\beta$. 
Note that for particles of other shape, such as spheroid 
for which orientation vectors should be included, 
the above factorizations by the single vector $\bm{e}$ cannot be achieved. 
From the symmetry on the exchange of particle indices 
$\alpha$ and $\beta$, we have the relations 
[in (JO-19a) -- (JO-19e) and (J-5a) -- (J-5f)] as 
\begin{subequations}
\begin{eqnarray}
  X^{A}_{\alpha\beta}(s,\lambda) &=& X^{A}_{(3-\alpha)(3-\beta)}(s,\lambda^{-1}),
  \label{eq:XA-particle-exchange}\\
  Y^{A}_{\alpha\beta}(s,\lambda) &=& Y^{A}_{(3-\alpha)(3-\beta)}(s,\lambda^{-1}),
  \label{eq:YA-particle-exchange}\\
  Y^{B}_{\alpha\beta}(s,\lambda) &=& -Y^{B}_{(3-\alpha)(3-\beta)}(s,\lambda^{-1}),
  \label{eq:YB-particle-exchange}\\
  X^{C}_{\alpha\beta}(s,\lambda) &=& X^{C}_{(3-\alpha)(3-\beta)}(s,\lambda^{-1}),
  \label{eq:XC-particle-exchange}\\
  Y^{C}_{\alpha\beta}(s,\lambda) &=& Y^{C}_{(3-\alpha)(3-\beta)}(s,\lambda^{-1}),
  \label{eq:YC-particle-exchange}\\
  X^{G}_{\alpha\beta}(s,\lambda) &=& -X^{G}_{(3-\alpha)(3-\beta)}(s,\lambda^{-1}),
  \label{eq:XG-particle-exchange}\\
  Y^{G}_{\alpha\beta}(s,\lambda) &=& -Y^{G}_{(3-\alpha)(3-\beta)}(s,\lambda^{-1}),
  \label{eq:YG-particle-exchange}\\
  Y^{H}_{\alpha\beta}(s,\lambda) &=& Y^{H}_{(3-\alpha)(3-\beta)}(s,\lambda^{-1}),
  \label{eq:YH-particle-exchange}\\
  X^{M}_{\alpha\beta}(s,\lambda) &=& X^{M}_{(3-\alpha)(3-\beta)}(s,\lambda^{-1}),
  \label{eq:XM-particle-exchange}\\
  Y^{M}_{\alpha\beta}(s,\lambda) &=& Y^{M}_{(3-\alpha)(3-\beta)}(s,\lambda^{-1}),
  \label{eq:YM-particle-exchange}\\
  Z^{M}_{\alpha\beta}(s,\lambda) &=& Z^{M}_{(3-\alpha)(3-\beta)}(s,\lambda^{-1}),
  \label{eq:ZM-particle-exchange}
\end{eqnarray}
\end{subequations}
where 
\begin{equation}
  s
  =
  \frac{2r}{a_1+a_2}
  ,\quad
  \lambda
  =
  \frac{a_2}{a_1}
  ,
  \label{eq:two-body-s-and-lambda}
\end{equation}
and $r = |\bm{r}|$. 
Therefore, once we have obtained 
22 resistance functions for $(\alpha\beta) = (11)$ and $(12)$, 
we can construct the grand resistance matrix completely. 
We will see the calculations in Sec \ref{sec:two-body-problem}.

\subsection{Lamb's General Solution}
In this paper, we utilize Lamb's general solution \cite{Lamb,HappelBrenner} 
to solve the problem. 
Lamb's general solution in the exterior region for
the pressure $p$ and velocity $\bm{u}$ is given by 
\begin{equation}
  p(\bm{r})
  =
  \sum_{n=0}^{\infty}
  p_{-n-1}
  ,
  \label{eq:Lamb-p}
\end{equation}
\begin{eqnarray}
  \bm{v}(\bm{r})
  &=&
  \bm{u}(\bm{r})-\bm{u}^{\infty}(\bm{r})
  \nonumber\\
  &=&
  \sum_{n=0}^{\infty}
  \left\{
    \bm{\nabla}
    \times
    \left(
      \bm{r}
      \chi_{-n-1}
    \right)
    +
    \bm{\nabla}
    \Phi_{-n-1}
  \right\}
  \nonumber\\
  &&
  +
  \frac{1}{\mu}
  \sum_{n=1}^{\infty}
  \left\{
    -
    \frac{n-2}{2n(2n-1)}
    r^2
    \bm{\nabla}
    \frac{p_{-n-1}}{\mu}
  \right.
  \nonumber\\
  &&
  \left.
    +
    \frac{n+1}{n(2n-1)}
    \bm{r}
    \frac{p_{-n-1}}{\mu}
  \right\}
  ,
  \label{eq:Lamb-vel}
\end{eqnarray}
where $\bm{u}^{\infty}$ is the imposed velocity 
and $\bm{v}$ is the disturbance velocity field. 
The solid spherical harmonics $p_{-n-1}$, $\chi_{-n-1}$, and $\Phi_{-n-1}$ 
are expressed [in (JO-2.3)] by
\begin{subequations}
\begin{eqnarray}
  \frac{p_{-n-1}}{\mu}
  &=&
  \sum_{m=0}^{n}
  p_{mn}
  \frac{1}{a}
  \left(
    \frac{a}{r}
  \right)^{n+1}
  Y_{mn}(\theta,\phi)
  ,
  \label{eq:spherical-harmonics-expansion-p}
  \\
  \chi_{-n-1}
  &=&
  \sum_{m=0}^{n}
  q_{mn}
  \left(
    \frac{a}{r}
  \right)^{n+1}
  Y_{mn}(\theta,\phi)
  ,
  \label{eq:spherical-harmonics-expansion-q}
  \\
  \Phi_{-n-1}
  &=&
  \sum_{m=0}^{n}
  v_{mn}
  a
  \left(
    \frac{a}{r}
  \right)^{n+1}
  Y_{mn}(\theta,\phi)
  ,
  \label{eq:spherical-harmonics-expansion-v}
\end{eqnarray}
\end{subequations}
where $Y_{mn}$ is the spherical harmonics defined by 
\begin{equation}
  Y_{mn}(\theta,\phi)
  =
  P_{n}^{m}(\cos\theta)
  e^{{\rm i}m\phi}
  ,
\end{equation}
with the associated Legendre function $P_{n}^{m}$, 
and $p_{mn}$, $q_{mn}$, and $v_{mn}$ are the coefficients to be determined 
from the boundary conditions.

\section{Single Sphere}
\label{sec:single-sphere}
First, let us consider a single sphere with radius $a$ at the origin. 
On the particle surface $|\bm{r}| = a$, 
the conventional no-slip boundary condition is given by 
\begin{equation}
  \bm{u}(\bm{r})
  =
  \bm{U}
  +
  \bm{\Omega}\times\bm{r}
  +
  \bm{E}\cdot\bm{r}
  ,
  \label{eq:no-slip-BC}
\end{equation}
where $\bm{U}$ and $\bm{\Omega}$ are the 
translational and rotational velocities of the particle, respectively. 
Here, we also introduce the strain tensor $\bm{E}$ of 
the particle surface, so that the boundary condition (\ref{eq:no-slip-BC}) 
is applicable to the deformable particle at instance of spherical shape. 
For rigid spherical particle, $\bm{E} = \bm{0}$.

\subsection{The Navier Slip Boundary Condition}
Navier \cite{Navier1823} proposed the slip boundary condition, 
where the slip velocity on the surface is proportional to 
the tangential force density, as 
\begin{equation}
  \bm{u}(\bm{r})
  =
  \bm{U}
  +
  \bm{\Omega}\times\bm{r}
  +
  \bm{E}\cdot\bm{r}
  +
  \frac{\gamma}{\mu}
  \left(
    \bm{I}
    -
    \bm{nn}
  \right)
  \cdot
  \left(
    \bm{\sigma}
    \cdot
    \bm{n}
  \right)
  ,
  \label{eq:slip-BC}
\end{equation}
where $\gamma$ is the slip length, 
$\bm{I}$ is the unit tensor, 
$\bm{n}$ is the surface normal (equal to $\bm{r}/r$ for sphere), 
and $\bm{\sigma}$ is the stress tensor defined by 
\begin{equation}
  \bm{\sigma}
  =
  -p\bm{I}
  +\mu
  \left[
    \bm{\nabla}\bm{u}
    +
    \left(\bm{\nabla}\bm{u}\right)^{\dagger}
  \right]
  .
\end{equation}
Rewriting Eq. (\ref{eq:slip-BC}) by using 
the disturbance field $\bm{v}$ and the imposed flow $\bm{u}^{\infty}$, 
we have 
\begin{equation}
  \bm{v}
  -
  \frac{\gamma}{\mu}
  \bm{t}
  =
  \bm{w}^{\Delta}
  +
  \frac{\gamma}{\mu}
  \bm{t}^{\infty}
  ,
  \label{eq:slip-BC-split}
\end{equation}
where the disturbance part $\bm{t}$ and imposed part $\bm{t}^{\infty}$ 
of the tangential force density are defined by 
\begin{subequations}
\begin{eqnarray}
  \bm{t}
  &=&
  \left(\bm{I}-\bm{nn}\right)\cdot
  \left(\bm{\sigma}^{v}\cdot\bm{n}\right)
  ,
  \label{eq:def-t}
  \\
  \bm{t}^{\infty}
  &=&
  \left(\bm{I}-\bm{nn}\right)\cdot
  \left(\bm{\sigma}^{\infty}\cdot\bm{n}\right)
  ,
\end{eqnarray}
\end{subequations}
and the corresponding stresses are 
\begin{subequations}
\begin{eqnarray}
  \bm{\sigma}^{v}
  &=&
  -p\bm{I}
  +\mu
  \left[
    \bm{\nabla}\bm{v}
    +
    \left(\bm{\nabla}\bm{v}\right)^{\dagger}
  \right]
  ,\\
  \bm{\sigma}^{\infty}
  &=&
  \mu
  \left[
    \bm{\nabla}\bm{u}^{\infty}
    +
    \left(\bm{\nabla}\bm{u}^{\infty}\right)^{\dagger}
  \right]
  ,
\end{eqnarray}
\end{subequations}
$\bm{w}^{\Delta}$ is defined by 
\begin{equation}
  \bm{w}^{\Delta}
  =
  \Delta\bm{U}
  +
  \Delta\bm{\Omega}
  \times
  \bm{r}
  +
  \Delta\bm{E}
  \cdot
  \bm{r}
  ,
  \label{eq:def-wDelta}
\end{equation}
and 
$\Delta\bm{U} = \bm{U}-\bm{U}^{\infty}$,
$\Delta\bm{\Omega} = \bm{\Omega}-\bm{\Omega}^{\infty}$, and 
$\Delta\bm{E} = \bm{E}-\bm{E}^{\infty}$. 
From the imposed flow in Eq. (\ref{eq:u-imposed-by-UOE}), 
$\bm{t}^{\infty}$ becomes 
\begin{equation}
  \bm{t}^{\infty}
  =
  \frac{2\mu}{r}
  \left(\bm{I}-\bm{nn}\right)
  \cdot\bm{E}^{\infty}
  \cdot\bm{r}
  .
  \label{eq:tinf-by-Einf}
\end{equation}
Note that, in the slip boundary condition (\ref{eq:slip-BC-split}), 
the left-hand side is the disturbance quantities 
and the right-hand side is the imposed quantities. 
Also note that, on the imposed part, 
the slip contribution appears only on the flow 
with $\bm{E}^{\infty}\neq\bm{0}$ 
as shown in Eq. (\ref{eq:tinf-by-Einf}). 

In terms of Lamb's general solution for the disturbance field $\bm{v}$ 
in Eq. (\ref{eq:Lamb-vel}), 
the corresponding surface force density $\bm{f}$ 
is given by \cite{Lamb,HappelBrenner} 
\begin{eqnarray}
  \bm{f}
  &=&
  \bm{\sigma}^{v}
  \cdot\bm{n}
  \nonumber\\
  &=&
  \frac{\mu}{r}
  \sum_{n}
  \Biggl\{
    -
    (n+2)
    \bm{\nabla}
    \times
    \left(
      \bm{r}
      \chi_{-n-1}
    \right)
  \nonumber\\
  &&
    -
    2(n+2)
   \bm{\nabla}
    \Phi_{-n-1}
  \nonumber\\
  &&
    +
    \frac{1}{\mu}
    \frac{(n+1)(n-1)}{n(2n-1)}
    r^2
    \bm{\nabla}
    p_{-n-1}
  \nonumber\\
  &&
    -
    \frac{1}{\mu}
    \frac{2n^2+1}{n(2n-1)}
    \bm{r}
    p_{-n-1}
  \Biggr\}
  ,
  \label{eq:lamb-force-density}
\end{eqnarray}
and $\bm{t}$ defined in Eq. (\ref{eq:def-t}) is expressed by 
\begin{eqnarray}
  \bm{t}
  &=&
  \frac{\mu}{r}
  \sum_{n}
  \Biggl\{
    -
    (n+2)
    \bm{\nabla}
    \times
    \left(
      \bm{r}
      \chi_{-n-1}
    \right)
  \nonumber\\
  &&
  -
  2(n+2)
  \left(
    \bm{\nabla}
    -
    \frac{\bm{r}}{r}
    \frac{\partial}{\partial r}
  \right)
  \Phi_{-n-1}
  \nonumber\\
  &&
  \left.
    +
    \frac{1}{\mu}
    \frac{(n+1)(n-1)}{n(2n-1)}
    r^2
    \left(
      \bm{\nabla}
      -
      \frac{\bm{r}}{r}
      \frac{\partial}{\partial r}
    \right)
    p_{-n-1}
  \right\}
  .
  \label{eq:lamb-tangential-force-density}
\end{eqnarray}

\subsubsection{Three Scalar Functions}
In order to achieve the boundary condition for 
Lamb's general solutions, 
Jeffrey and Onishi \cite{JeffreyOnishi1984}
used three scalar functions as in Happel and Brenner \cite{HappelBrenner}, 
\S 3.2. 
Consider a general vector field $\bm{g}$ and its surface vectors $\bm{G}$ 
defined by 
\begin{equation}
  \bm{G}(\theta,\phi)
  =
  \bm{g}
  \Bigr|_{|\bm{r}| = a}
  ,
\end{equation}
so that
\begin{equation}
  \frac{\partial\bm{G}}{\partial r}
  \equiv
  \bm{0}
  .
  \label{eq:dVdr-is-zero}
\end{equation}
We define the following three scalar functions 
\begin{subequations}
\begin{eqnarray}
  G_{\text{rad}}
  &=&
  \frac{\bm{r}}{r}
  \cdot
  \bm{G}
  ,\\
  G_{\text{div}}
  &=&
  -
  r
  \bm{\nabla}
  \cdot
  \bm{G}
  ,\\
  G_{\text{rot}}
  &=&
  \bm{r}
  \cdot
  \bm{\nabla}
  \times
  \bm{G}
  .
\end{eqnarray}
\end{subequations}
Obviously, the first scalar $G_{\text{rad}}$ is the radial component
$G_r = (\bm{r}/r)\cdot\bm{G}$ itself. The other two, 
$G_{\text{div}}$ and $G_{\text{rot}}$, are related to 
the tangential components (\textit{i.e.}, 
$G_{\theta}$ and $G_{\phi}$ in polar coordinates), 
except for the factor $-2G_r$ on the divergence, 
as 
\begin{subequations}
\begin{eqnarray}
  G_{\text{div}}
  &=&
  -
  2
  G_r
  -
  \left(
    \frac{\partial}{\partial \theta}
    +
    \frac{\cos\theta}{\sin\theta}
  \right)
  G_\theta
  -
  \frac{1}{\sin\theta}
  \frac{\partial G_\phi}{\partial \phi}
  ,\\
  G_{\text{rad}}
  &=&
  \mathscr{S}
  \left[
    -
    \frac{1}{\sin\theta}
    \frac{\partial G_\theta}{\partial \phi}
    +
    \left(
      \frac{\partial}{\partial \theta}
      +
      \frac{\cos\theta}{\sin\theta}
    \right)
    G_\phi
  \right]
  ,
  \label{eq:def-Vrad}
\end{eqnarray}
\end{subequations}
where $\mathscr{S}$ is $+1$ in the right-handed coordinates 
and $-1$ in the left-handed coordinates. 
It should be noted that the divergence of the surface vector $\bm{G}$ 
is related to the 3D vector field $\bm{g}$ as
\begin{equation}
  G_{\text{div}}
  =
  -
  r
  \bm{\nabla}
  \cdot
  \bm{g}
  \Bigr|_{|\bm{r}| = a}
  +
  r
  \frac{\partial}{\partial r}
  g_r
  \Bigr|_{|\bm{r}| = a}
  ,
  \label{eq:r-div-V}
\end{equation}
where the substitution of $|\bm{r}| = a$ is applied after the derivatives.

\paragraph{Velocity Field}
As a first example, consider the disturbance velocity $\bm{v}$, 
whose surface vector is defined by $\bm{V}$ as 
\begin{equation}
  \bm{V}(\theta,\phi)
  =
  \bm{v}
  \Bigr|_{|\bm{r}| = a}
  .
\end{equation}
\begin{subequations}
By definition, the first scalar $V_{\text{rad}}$ is given by $\bm{v}$ as 
\begin{equation}
  V_{\text{rad}}
  =
  \frac{\bm{r}}{r}
  \cdot
  \bm{V}
  =
  \frac{\bm{r}}{r}
  \cdot
  \bm{v}
  \Bigr|_{|\bm{r}| = a}
  .
  \label{eq:Vrad-by-v}
\end{equation}
Because $\bm{v}$ satisfies $\bm{\nabla}\cdot\bm{v} = 0$, 
$V_{\text{div}}$ is given by 
\begin{equation}
  V_{\text{div}}
  =
  -
  r
  \bm{\nabla}
  \cdot
  \bm{V}
  =
  r
  \frac{\partial}{\partial r}
  v_r
  \Bigr|_{|\bm{r}| = a}
  ,
  \label{eq:Vdiv-by-v}
\end{equation}
from Eq. (\ref{eq:r-div-V}). 
$V_{\text{rad}}$ is independent of its radial component $V_{r}$
as shown in Eq. (\ref{eq:def-Vrad}), 
so that it is simply written by $\bm{v}$ as 
\begin{equation}
  V_{\text{rot}}
  =
  \bm{r}
  \cdot
  \bm{\nabla}
  \times
  \bm{V}
  =
  \bm{r}\cdot\bm{\nabla}\times\bm{v}
  \Bigr|_{|\bm{r}| = a}
  .
  \label{eq:Vrot-by-v}
\end{equation}
\end{subequations}
From Lamb's general solution for $\bm{v}$ in Eq. (\ref{eq:Lamb-vel}), 
then, the three scalars are obtained as in 
Jeffrey and Onishi \cite{JeffreyOnishi1984}.

\paragraph{Tangential Surface Force}
Next, let us consider $\bm{t}$ which is necessary for 
the slip boundary condition (\ref{eq:slip-BC-split}). 
Its surface vector is defined by 
\begin{equation}
  \bm{T}(\theta,\phi)
  =
  \bm{t}
  \Bigr|_{|\bm{r}| = a}
  .
\end{equation}
\begin{subequations}
The radial component of $\bm{t}$ is zero by definition as 
\begin{equation}
  T_{\text{rad}}
  =
  \frac{\bm{r}}{r}
  \cdot
  \bm{T}
  =
  0
  .
\end{equation}
From Eq. (\ref{eq:r-div-V}), therefore, we have 
\begin{equation}
  T_{\text{div}}
  =
  -
  r
  \bm{\nabla}
  \cdot
  \bm{T}
  =
  -
  r
  \bm{\nabla}
  \cdot
  \bm{t}
  \Bigr|_{|\bm{r}| = a}
  .
\end{equation}
Because the rotation has no radial component 
for an arbitrary vector field, we can use the bare surface force $\bm{f}$ 
for the boundary condition for the tangential force $\bm{t}$ as
\begin{equation}
  T_{\text{rot}}
  =
  \bm{r}
  \cdot
  \bm{\nabla}
  \times
  \bm{T}
  =
  \bm{r}
  \cdot
  \bm{\nabla}
  \times
  \bm{t}
  \Bigr|_{|\bm{r}| = a}
  =
  \bm{r}
  \cdot
  \bm{\nabla}
  \times
  \bm{f}
  \Bigr|_{|\bm{r}| = a}
  .
\end{equation}
\end{subequations}
Using Lamb's general solution in Eq. (\ref{eq:lamb-tangential-force-density}), 
the three scalar components for $\bm{t}$ are given by 
\begin{subequations}
\begin{eqnarray}
  \frac{r_i}{r}
  t_{i}
  &=&
  0
  ,
  \label{eq:Trad-by-t}
  \\
  -
  r\bm{\nabla}\cdot\bm{t}
  &=&
  -
  \mu
  \sum_{n}
  \left[
    \frac{2n(n+1)(n+2)}{r^2}
    \Phi_{-n-1}
  \right.
  \nonumber\\
  &&
  \left.
    -
    \frac{(n+1)^2(n-1)}{2n-1}
    \frac{p_{-n-1}}{\mu}
  \right]
  ,
  \label{eq:Tdiv-by-t}
  \\
  \bm{r}\cdot\bm{\nabla}\times\bm{t}
  &=&
  -
  \frac{\mu}{r}
  \sum_{n}
  (n+2)n(n+1)\chi_{-n-1}
  .
  \label{eq:Trot-by-t}
\end{eqnarray}
\end{subequations}

\paragraph{Disturbance Part}
Three scalars for $\bm{V}$ 
are obtained by Eqs. (\ref{eq:Vrad-by-v}), (\ref{eq:Vdiv-by-v}), 
and (\ref{eq:Vrot-by-v}), 
and the slip contribution $-(\gamma/\mu)\bm{T}$ 
by Eqs. (\ref{eq:Trad-by-t}), (\ref{eq:Tdiv-by-t}), and (\ref{eq:Trot-by-t}). 
Substituting Lamb's solution (\ref{eq:Lamb-vel}) 
with the expansions in Eqs. (\ref{eq:spherical-harmonics-expansion-p}), 
(\ref{eq:spherical-harmonics-expansion-q}), and 
(\ref{eq:spherical-harmonics-expansion-v}) and putting $r=a$, 
the three scalars of the disturbance part, \textit{i.e.} 
the left-hand side, of the slip boundary condition (\ref{eq:slip-BC-split}) 
are given by 
\begin{subequations}
\begin{eqnarray}
  \left(\bm{V}-\frac{\gamma}{\mu}\bm{T}\right)_{\text{rad}}
  &=&
  \sum_{n=0}^{\infty}
  \sum_{m=0}^{n}
  \left[
    -
    (n+1)
    v_{mn}
  \right.
  \nonumber\\
  &&
  \quad\quad
  \left.
    +
    \frac{n+1}{2(2n-1)}
    p_{mn}
  \right]
  Y_{mn}(\theta,\phi)
  ,
  \label{eq:slip-BC-rhs-rad}
  \\
  \left(\bm{V}-\frac{\gamma}{\mu}\bm{T}\right)_{\text{div}}
  &=&
  \sum_{n=0}^{\infty}
  \sum_{m=0}^{n}
  \Biggl[
    (n+1)(n+2)
    \left(
      1
      +
      2n
      \widehat{\gamma}
    \right)
    v_{mn}
  \nonumber\\
  &&
    -
    \frac{n(n+1)}{2(2n-1)}
    \left(
      1
      +
      \frac{2(n+1)(n-1)}{n}
      \widehat{\gamma}
    \right)
    p_{mn}
  \Biggr]
  \nonumber\\
  &&
  Y_{mn}(\theta,\phi)
  ,
  \label{eq:slip-BC-rhs-div}
  \\
  \left(\bm{V}-\frac{\gamma}{\mu}\bm{T}\right)_{\text{rot}}
  &=&
  \sum_{n=0}^{\infty}
  \sum_{m=0}^{n}
  n(n+1)
  \left(
    1
    +
    (n+2)
    \widehat{\gamma}
  \right)
  q_{mn}
  Y_{mn}(\theta,\phi)
  ,\nonumber\\
  \label{eq:slip-BC-rhs-rot}
\end{eqnarray}
\end{subequations}
where the scaled slip length $\widehat{\gamma}$ is defined by 
\begin{equation}
  \widehat{\gamma}
  =
  \frac{\gamma}{a}
  .
\end{equation}

\paragraph{Imposed Part}
Let us look at the three components for the vector $\bm{w}^{\Delta}$ 
in Eq. (\ref{eq:def-wDelta}). 
Note that the divergence is zero as shown by
\begin{equation}
  \partial_i
  w^{\Delta}_i
  =
  \epsilon_{ijk}
  \Delta\Omega_j
  \delta_{ik}
  +
  \Delta E_{ij}
  \delta_{ij}
  =
  0
  ,
\end{equation}
because $E_{kk} = 0$.
Therefore, we need to calculate the divergence component 
through the derivative of the radial velocity (as for $\bm{v}$). 
The three components for $\bm{w}^{\Delta}$ are then given by 
\begin{subequations}
\begin{eqnarray}
  \frac{r_i}{r}
  w^{\Delta}_i
  &=&
  \frac{r_i}{r}
  \Delta U_i
  +
  \frac{r_ir_j}{r}
  \Delta E_{ij}
  ,\\
  r_j
  \partial_j
  \frac{r_i}{r}
  w^{\Delta}_i
  &=&
  \frac{r_ir_j}{r}
  \Delta E_{ij}
  ,\\
  r_i
  \epsilon_{ijk}
  \partial_j
  w^{\Delta}_k
  &=&
  2
  r_i
  \Delta\Omega_{i}
  .
\end{eqnarray}
\end{subequations}
We use the identity 
$
  \epsilon_{ijk}
  \epsilon_{jkl}
  =
  2
  \delta_{il}
$
for the last equation. 
For $\bm{t}^{\infty}$, 
the three components are given as follows. 
The normal component is zero by definition as
\begin{equation}
  \frac{r_i}{r}
  t^{\infty}_i
  =
  0
  ,
\end{equation}
and, therefore, 
the divergence component is obtained 
through Eq. (\ref{eq:r-div-V}) as 
\begin{equation}
  -
  r
  \partial_i
  t^{\infty}_k
  =
  -
  2\mu
  r
  \partial_i
  \left(
    \delta_{ij}
    \frac{r_k}{r}
    -
    \frac{r_ir_jr_k}{r^3}
  \right)
  E^{\infty}_{jk}
  =
  6\mu
  \frac{r_jr_k}{r^2}
  E^{\infty}_{jk}
  ,
\end{equation}
where we use $E^{\infty}_{kk} = 0$. 
The rotation vanishes as 
\begin{equation}
  r_i
  \epsilon_{ijk}
  \partial_j
  t^{\infty}_k
  =
  2\mu
  r_i
  \epsilon_{ijk}
  \partial_j
  \left(
    \delta_{kl}
    \frac{r_m}{r}
    -
    \frac{r_kr_lr_m}{r^3}
  \right)
  E^{\infty}_{lm}
  =
  0
  .
\end{equation}

Define the surface vector of the right-hand side 
of the slip boundary condition (\ref{eq:slip-BC-split}) by 
\begin{equation}
  \bm{W}
  =
  \left.
    \left(
      \bm{w}^{\Delta}
      +
      \frac{\gamma}{\mu}
      \bm{t}^{\infty}
    \right)
  \right|_{|\bm{r}|=a}
  .
\end{equation}
The three scalars for $\bm{W}$ are then given by 
\begin{subequations}
\begin{eqnarray}
  W_{\text{rad}}
  &=&
  e_i
  \Delta U_i
  +
  e_ie_j
  a
  \Delta E_{ij}
  ,
  \label{eq:BC-W-rad}
  \\
  W_{\text{div}}
  &=&
  e_ie_j
  a
  \Delta E_{ij}
  +
  6
  \widehat{\gamma}
  e_ie_j
  a
  E^{\infty}_{ij}
  ,
  \label{eq:BC-W-div}
  \\
  W_{\text{rot}}
  &=&
  2
  e_i
  a
  \Delta\Omega_{i}
  .
  \label{eq:BC-W-rot}
\end{eqnarray}
\end{subequations}

\subsubsection{Recurrence Relations}
Let us introduce the spherical harmonics expansion 
for the three components of the imposed part by
\begin{subequations}
\begin{eqnarray}
  W_{\text{rad}}
  &=&
  \sum_{n=0}^{\infty}
  \sum_{m=0}^{n}
  \chi_{mn}
  Y_{mn}(\theta,\phi)
  ,\\
  W_{\text{div}}
  &=&
  \sum_{n=0}^{\infty}
  \sum_{m=0}^{n}
  \psi_{mn}
  Y_{mn}(\theta,\phi)
  ,\\
  W_{\text{rot}}
  &=&
  \sum_{n=0}^{\infty}
  \sum_{m=0}^{n}
  \omega_{mn}
  Y_{mn}(\theta,\phi)
  .
\end{eqnarray}
\end{subequations}
From Eqs. (\ref{eq:BC-W-rad}), 
(\ref{eq:BC-W-div}), and (\ref{eq:BC-W-rot}), 
the coefficients $\chi_{mn}$, $\psi_{mn}$, and $\omega_{mn}$ 
are given by the parameters 
$\Delta\bm{U}$, $\Delta\bm{\Omega}$, $\Delta\bm{E}$, and $\bm{E}^{\infty}$. 
Therefore, by the boundary condition (\ref{eq:slip-BC-split}) 
at the surface $|\bm{r}| = a$ 
with the scalars of the disturbance fields in Eqs. 
(\ref{eq:slip-BC-rhs-rad}), (\ref{eq:slip-BC-rhs-div}), 
and (\ref{eq:slip-BC-rhs-rot}), 
the coefficients $(p_{mn}, q_{mn}, v_{mn})$ 
are given by the boundary condition 
$(\chi_{mn}, \psi_{mn}, \omega_{mn})$ as 
\begin{subequations}
\begin{eqnarray}
  p_{mn}
  &=&
  \frac{2n-1}{n+1}
  \Gamma_{0,2n+1}
  \psi_{mn}
  \nonumber\\
  &&
  +
  \frac{(n+2)(2n-1)}{n+1}
  \Gamma_{2n,2n+1}
  \chi_{mn}
  ,
  \label{eq:1b-slip-pmn}
  \\
  v_{mn}
  &=&
  \frac{1}{2(n+1)}
  \Gamma_{0,2n+1}
  \psi_{mn}
  \nonumber\\
  &&
  +
  \frac{n}{2(n+1)}
  \Gamma_{2(n+1)(n-1)/n,2n+1}
  \chi_{mn}
  ,
  \label{eq:1b-slip-vmn}
  \\
  q_{mn}
  &=&
  \frac{1}{n(n+1)}
  \Gamma_{0,n+2}
  \omega_{mn}
  ,
  \label{eq:1b-slip-qmn}
\end{eqnarray}
\end{subequations}
where 
\begin{equation}
  \Gamma_{m,n}
  =
  \frac{1+m\widehat{\gamma}}{1+n\widehat{\gamma}}
  .
\end{equation}
Note that in the no-slip ($\widehat{\gamma}=0$) 
and perfect-slip ($\widehat{\gamma}=\infty$) limits, 
$\Gamma_{m,n}$ reduces to 
\begin{equation}
  \Gamma_{m,n}
  =
  \begin{cases}
    1 & \text{ for $\widehat{\gamma}=0$},\\
    m/n & \text{ for $\widehat{\gamma}=\infty$}.
  \end{cases}
\end{equation}

\subsection{Single Body Solutions}
In the following, we solve single-body problem 
with the slip boundary condition through Eqs. 
(\ref{eq:1b-slip-pmn}), (\ref{eq:1b-slip-vmn}), and (\ref{eq:1b-slip-qmn}).

\subsubsection{Translating Sphere}
Consider translating sphere with the velocity 
$\bm{U} = (0,0,U)$, which is given by 
\begin{equation}
  \chi_{m,n}
  =
  U
  \delta_{0m}
  \delta_{1n}
  .
  \label{eq:1b-bc-XA}
\end{equation}
Substituting the condition (\ref{eq:1b-bc-XA}) 
into the recurrence relations 
(\ref{eq:1b-slip-pmn}), (\ref{eq:1b-slip-vmn}), 
and (\ref{eq:1b-slip-qmn}), 
we have the solution 
\begin{subequations}
\begin{eqnarray}
  p_{mn}
  &=&
  \frac{3}{2}
  U
  \Gamma_{2,3}
  \delta_{m0}
  \delta_{n1}
  ,
  \label{eq:1b-XA-p}
  \\
  v_{mn}
  &=&
  \frac{1}{4}
  U
  \Gamma_{0,3}
  \delta_{m0}
  \delta_{n1}
  ,
  \label{eq:1b-XA-v}
  \\
  q_{mn}
  &=&
  0
  .
  \label{eq:1b-XA-q}
\end{eqnarray}
\end{subequations}
The force acting on the particle is given by the coefficients 
of Lamb's general solution [in (JO-2.10)] as
\begin{equation}
  \bm{F}
  =
  4\pi\mu a
  \left[
    p_{01}
    \hat{\bm{z}}
    -
    p_{11}
    \left(
      \hat{\bm{x}}
      +
      {\rm i}
      \hat{\bm{y}}
    \right)
  \right]
  ,
\end{equation}
where $\hat{\bm{x}}$, $\hat{\bm{y}}$, and $\hat{\bm{z}}$ 
are the unit vectors in $x$, $y$, and $z$ directions, respectively. 
Therefore, the force on the sphere translating with the velocity $U$
in $z$ direction is
\begin{equation}
  \bm{F}
  =
  6\pi\mu a
  \Gamma_{2,3}
  U
  \hat{\bm{z}}
  .
  \label{eq:F-by-U-1b-slip}
\end{equation}
This is identical to the result by Basset \cite{Basset}. 
(See also Lamb \cite{Lamb} Art. 337, 3$^{\circ}$ 
and Felderhof \cite{Felderhof1976}.)
Substituting the coefficients (\ref{eq:1b-XA-p}), (\ref{eq:1b-XA-v}), and 
(\ref{eq:1b-XA-q}) into Lamb's general solution in Eq. (\ref{eq:Lamb-vel}) 
and rewriting the parameter $U$ 
by the strength of the force $F$ through Eq. (\ref{eq:F-by-U-1b-slip}), 
the disturbance field is given by 
\begin{equation}
  \bm{v}
  =
  \frac{1}{8\pi\mu}
  \left(
    1
    +
    \Gamma_{0,2}
    \frac{a^2}{6}
    \nabla^2
  \right)
  \bm{J}
  \cdot
  \bm{F}
  ,
  \label{eq:dist-vel-F-1b-slip}
\end{equation}
where $\bm{J}$ is the Oseen-Burgers tensor 
\begin{equation}
  J_{ij}(\bm{r})
  =
  \frac{1}{r}
  \left(
    \delta_{ij}
    +
    \frac{r_ir_j}{r^2}
  \right)
  .
\end{equation}

\subsubsection{Rotating Sphere}
For the problem of rotating sphere, 
$W_{\text{rot}}$ in Eq. (\ref{eq:BC-W-rot}) is the only non-zero component.
Consider a sphere with the angular velocity 
$\bm{\Omega} = (0, 0, \Omega)$,
which reduces to 
\begin{equation}
  \omega_{m,n}
  =
  2
  a
  \Omega
  \delta_{0m}
  \delta_{1n}
  .
  \label{eq:1b-bc-XC}
\end{equation}
Substituting the condition (\ref{eq:1b-bc-XA})
into the recurrence relations 
(\ref{eq:1b-slip-pmn}), (\ref{eq:1b-slip-vmn}), 
and (\ref{eq:1b-slip-qmn}), 
we have the solution 
\begin{subequations}
\begin{eqnarray}
  p_{mn}
  &=&
  0
  ,
  \label{eq:1b-XC-p}
  \\
  v_{mn}
  &=&
  0
  ,
  \label{eq:1b-XC-v}
  \\
  q_{mn}
  &=&
  a\Omega
  \Gamma_{0,3}
  \delta_{m0}
  \delta_{n1}
  .
  \label{eq:1b-XC-q}
\end{eqnarray}
\end{subequations}
The torque acting on the particle is given by the coefficients 
of Lamb's general solution [in (JO-2.11)] as
\begin{equation}
  \bm{T}
  =
  8\pi\mu a^2
  \left[
    q_{01}
    \hat{\bm{z}}
    -
    q_{11}
    \left(
      \hat{\bm{x}}
      +
      {\rm i}
      \hat{\bm{y}}
    \right)
  \right]
  .
\end{equation}
Therefore, the torque on the sphere rotating with the angular velocity
$\Omega$ in $z$ direction is
\begin{equation}
  \bm{T}
  =
  8\pi\mu a^3
  \Gamma_{0,3}
  \Omega
  \hat{\bm{z}}
  .
  \label{eq:T-by-O-1b-slip}
\end{equation}
This is consistent with the result by Felderhof \cite{Felderhof1976} 
and Padmavathi {\it et al.} \cite{PadmavathiEtal1993} 
Note that the torque $\bm{T}$ would vanish for the sphere 
with the perfect-slip surface (for $\widehat{\gamma}=\infty$).
Substituting coefficients (\ref{eq:1b-XC-p}), (\ref{eq:1b-XC-v}), and 
(\ref{eq:1b-XC-q}) into Lamb's general solution in Eq. (\ref{eq:Lamb-vel}) 
and using Eq. (\ref{eq:T-by-O-1b-slip}), 
the disturbance field is given by 
\begin{equation}
  \bm{v}
  =
  \frac{1}{8\pi\mu}
  \bm{R}
  \cdot
  \bm{T}
  ,
  \label{eq:dist-vel-T-1b-slip}
\end{equation}
where
\begin{equation}
  R_{ij}(\bm{r})
  =
  \epsilon_{ijk}
  \frac{r_k}{r^3}
  .
\end{equation}

\subsubsection{Sphere in Strain Flow}
For the problem of sphere in strain flow, 
we have two non-zero components on $\bm{W}$. 
Here we assume the rigid sphere, so that $\bm{E} = \bm{0}$ and 
from Eqs. (\ref{eq:BC-W-rad}), (\ref{eq:BC-W-div}), and (\ref{eq:BC-W-rot}), 
\begin{subequations}
\begin{eqnarray}
  W_{\text{rad}}
  &=&
  -
  e_ie_j
  a
  E^{\infty}_{ij}
  ,\\
  W_{\text{div}}
  &=&
  -
  e_ie_j
  a
  E^{\infty}_{ij}
  \left(
    1
    -
    6
    \widehat{\gamma}
  \right)
  ,\\
  W_{\text{rot}}
  &=&
  0
  .
\end{eqnarray}
\end{subequations}
Let us consider the strain given by 
\begin{equation}
  -
  E^{\infty}_{ij}
  =
  E
  \left(
    \hat{z}_{i}\hat{z}_{j}
    -
    \frac{1}{3}
    \delta_{ij}
  \right)
  .
\end{equation}
This is achieved by 
\begin{subequations}
\begin{eqnarray}
  \chi_{m,n}
  &=&
  \frac{2}{3}
  a
  E
  \delta_{0m}
  \delta_{2n}
  ,
  \label{eq:init-cond-E-X-chi}
  \\
  \psi_{m,n}
  &=&
  \frac{2}{3}
  a
  E
  \left(
    1
    -
    6
    \widehat{\gamma}
  \right)
  \delta_{0m}
  \delta_{2n}
  .
  \label{eq:init-cond-E-X-psi}
\end{eqnarray}
\end{subequations}
Substituting the boundary conditions 
(\ref{eq:init-cond-E-X-chi}) and (\ref{eq:init-cond-E-X-psi}) 
into the recurrence relations 
(\ref{eq:1b-slip-pmn}), (\ref{eq:1b-slip-vmn}), 
and (\ref{eq:1b-slip-qmn}), 
we have the solution 
\begin{subequations}
\begin{eqnarray}
  p_{mn}
  &=&
  \frac{10}{3}
  aE
  \Gamma_{2,5}
  \delta_{m0}
  \delta_{n2}
  ,
  \label{eq:1b-XM-p}
  \\
  v_{mn}
  &=&
  \frac{1}{3}
  aE
  \Gamma_{0,5}
  \delta_{m0}
  \delta_{n2}
  ,
  \label{eq:1b-XM-v}
  \\
  q_{mn}
  &=&
  0
  .
  \label{eq:1b-XM-q}
\end{eqnarray}
\end{subequations}
The stresslet acting on the particle is given by the coefficients 
of Lamb's general solution [in (J-6)] as
\begin{eqnarray}
  \bm{S}
  &=&
  2\pi\mu a^2
  \left\{
    p_{02}
    \left(
      \hat{\bm{z}}
      \hat{\bm{z}}
      -
      \frac{1}{3}
      \bm{I}
    \right)
  \right.
  \nonumber\\
  &&
  \left.
    -
    p_{12}
    \left[
      \hat{\bm{x}}
      \hat{\bm{z}}
      +
      \hat{\bm{z}}
      \hat{\bm{x}}
      +
      {\rm i}
      \left(
        \hat{\bm{y}}
        \hat{\bm{z}}
        +
        \hat{\bm{z}}
        \hat{\bm{y}}
      \right)
    \right]
  \right.
  \nonumber\\
  &&
  \left.
    +
    2
    p_{22}
    \left[
      \hat{\bm{x}}
      \hat{\bm{x}}
      -
      \hat{\bm{y}}
      \hat{\bm{y}}
      +
      {\rm i}
      \left(
        \hat{\bm{x}}
        \hat{\bm{y}}
        +
        \hat{\bm{y}}
        \hat{\bm{x}}
      \right)
    \right]
  \right\}
  .
\end{eqnarray}
Therefore, the stresslet on the sphere in the strain flow with 
the parameter $E$ is
\begin{equation}
  \bm{S}
  =
  \frac{20}{3}
  \pi\mu a^3
  \Gamma_{2,5}
  \bm{E}
  ,
  \label{eq:S-by-E-1b-slip}
\end{equation}
which is identical to the result by Felderhof \cite{Felderhof1976}. 
This yields to the effective viscosity $\mu^*$ of the suspension 
in the dilute limit up to $O(\phi)$ as 
\begin{equation}
  \frac{\mu^*}{\mu}
  =
  1
  +
  \frac{5}{2}
  \Gamma_{2,5}
  \phi
  ,
\end{equation}
where $\phi$ is the volume fraction. 
This is identical to the expression (9-5.11) in 
Happel and Brenner \cite{HappelBrenner}.
The effective viscosity of slip particles has two extremes as 
\begin{equation}
  \frac{\mu^*}{\mu}
  =
  \begin{cases}
    1
    +
    \frac{5}{2}
    \phi
    & \text{for no-slip particles},\\
    1
    +
    \phi
    & \text{for perfect-slip particles}.
  \end{cases}
\end{equation}
The latter agrees with the result for spherical gas bubbles. 
Substituting the coefficients (\ref{eq:1b-XM-p}), (\ref{eq:1b-XM-v}), and 
(\ref{eq:1b-XM-q}) into Lamb's general solution in Eq. (\ref{eq:Lamb-vel}) 
and using Eq. (\ref{eq:S-by-E-1b-slip}), 
the disturbance field is given by 
\begin{equation}
  \bm{v}
  =
  -
  \frac{1}{8\pi\mu}
  \left(
    1
    +
    \Gamma_{0,2}
    \frac{a^2\nabla^2}{10}
  \right)
  \bm{K}
  :
  \bm{S}
  ,
  \label{eq:dist-vel-S-1b-slip}
\end{equation}
where 
\begin{equation}
  K_{ijk}(\bm{r})
  =
  -3
  \frac{r_ir_jr_k}{r^5}
  .
\end{equation}

\section{Two-Body Problem}
\label{sec:two-body-problem}
Now, we study two-body problem. 
We will determine 22 resistance functions mentioned 
in Sec. \ref{sec:resistance-functions}. 
Following Jeffrey \textit{et al.} \cite{JeffreyOnishi1984,Jeffrey1992}, 
we write these functions in terms of the coefficients $f_{m}$ 
and determine the coefficients. 
Here we summarize the definitions of the coefficients: 
$X^{A}_{\alpha\beta}$ are given [in (JO-3.13) and (JO-3.14)] by 
\begin{subequations}
\begin{eqnarray}
  X^{A}_{11}(s,\lambda)
  &=&
  \sum_{m=0,\text{even}}^{\infty}\frac{f^{XA}_{m}}{\left[(1+\lambda)s\right]^m},\\
  X^{A}_{12}(s,\lambda)
  &=&
  \frac{-2}{1+\lambda}
  \sum_{m=1,\text{odd}}^{\infty}\frac{f^{XA}_{m}}{\left[(1+\lambda)s\right]^m},
  \label{eq:XA12-by-f}
\end{eqnarray}
\end{subequations}
$Y^{A}_{\alpha\beta}$ [in (JO-4.13) and (JO-4.14)] by 
\begin{subequations}
\begin{eqnarray}
  Y^{A}_{11}(s,\lambda)
  &=&
  \sum_{m=0,\text{even}}^{\infty}\frac{f^{YA}_{m}}{\left[(1+\lambda)s\right]^m},\\
  Y^{A}_{12}(s,\lambda)
  &=&
  \frac{-2}{1+\lambda}
  \sum_{m=1,\text{odd}}^{\infty}\frac{f^{YA}_{m}}{\left[(1+\lambda)s\right]^m},
  \label{eq:YA12-by-f}
\end{eqnarray}
\end{subequations}
$Y^{B}_{\alpha\beta}$ [in (JO-5.3) and (JO-5.4)] by 
\begin{subequations}
\begin{eqnarray}
  Y^{B}_{11}(s,\lambda)
  &=&
  \sum_{m=1,\text{odd}}^{\infty}\frac{f^{YB}_{m}}{\left[(1+\lambda)s\right]^m},
  \label{eq:YB11-by-f}
  \\
  Y^{B}_{12}(s,\lambda)
  &=&
  \frac{-4}{(1+\lambda)^2}
  \sum_{m=0,\text{even}}^{\infty}\frac{f^{YB}_{m}}{\left[(1+\lambda)s\right]^m},
  \label{eq:YB12-by-f}
\end{eqnarray}
\end{subequations}
$X^{C}_{\alpha\beta}$ [in (JO-6.7) and (JO-6.8)] by
\begin{subequations}
\begin{eqnarray}
  X^{C}_{11}(s,\lambda)
  &=&
  \sum_{m=0,\text{even}}^{\infty}\frac{f^{XC}_{m}}{\left[(1+\lambda)s\right]^m},\\
  X^{C}_{12}(s,\lambda)
  &=&
  \frac{-8}{(1+\lambda)^3}
  \sum_{m=1,\text{odd}}^{\infty}\frac{f^{XC}_{m}}{\left[(1+\lambda)s\right]^m},
  \label{eq:XC12-by-f}
\end{eqnarray}
\end{subequations}
$Y^{C}_{\alpha\beta}$ [in (JO-7.7) and (JO-7.8)] by
\begin{subequations}
\begin{eqnarray}
  Y^{C}_{11}(s,\lambda)
  &=&
  \sum_{m=0,\text{even}}^{\infty}\frac{f^{YC}_{m}}{\left[(1+\lambda)s\right]^m},\\
  Y^{C}_{12}(s,\lambda)
  &=&
  \frac{8}{(1+\lambda)^3}
  \sum_{m=1,\text{odd}}^{\infty}\frac{f^{YC}_{m}}{\left[(1+\lambda)s\right]^m},
  \label{eq:YC12-by-f}
\end{eqnarray}
\end{subequations}
$X^{G}_{\alpha\beta}$ [in (J-18a,b)] by
\begin{subequations}
\begin{eqnarray}
  X^{G}_{11}(s,\lambda)
  &=&
  \sum_{m=1,\text{odd}}^{\infty}\frac{f^{XG}_{m}}{\left[(1+\lambda)s\right]^m},\\
  X^{G}_{12}(s,\lambda)
  &=&
  \frac{-4}{(1+\lambda)^2}
  \sum_{m=2,\text{even}}^{\infty}\frac{f^{XG}_{m}}{\left[(1+\lambda)s\right]^m},
  \label{eq:XG12-by-f}
\end{eqnarray}
\end{subequations}
$Y^{G}_{\alpha\beta}$ [in (J-26a,b)] by
\begin{subequations}
\begin{eqnarray}
  Y^{G}_{11}(s,\lambda)
  &=&
  \sum_{m=1,\text{odd}}^{\infty}\frac{f^{YG}_{m}}{\left[(1+\lambda)s\right]^m},\\
  Y^{G}_{12}(s,\lambda)
  &=&
  \frac{-4}{(1+\lambda)^2}
  \sum_{m=0,\text{even}}^{\infty}\frac{f^{YG}_{m}}{\left[(1+\lambda)s\right]^m},
  \label{eq:YG12-by-f}
\end{eqnarray}
\end{subequations}
$Y^{H}_{\alpha\beta}$ [in (J-34a,b)] by
\begin{subequations}
\begin{eqnarray}
  Y^{H}_{11}(s,\lambda)
  &=&
  \sum_{m=0,\text{even}}^{\infty}\frac{f^{YH}_{m}}{\left[(1+\lambda)s\right]^m},\\
  Y^{H}_{12}(s,\lambda)
  &=&
  \frac{8}{(1+\lambda)^3}
  \sum_{m=1,\text{odd}}^{\infty}\frac{f^{YH}_{m}}{\left[(1+\lambda)s\right]^m},
  \label{eq:YH12-by-f}
\end{eqnarray}
\end{subequations}
$X^{M}_{\alpha\beta}$ [in (J-47a,b)] by
\begin{subequations}
\begin{eqnarray}
  X^{M}_{11}(s,\lambda)
  &=&
  \sum_{m=0,\text{even}}^{\infty}\frac{f^{XM}_{m}}{\left[(1+\lambda)s\right]^m},\\
  X^{M}_{12}(s,\lambda)
  &=&
  \frac{8}{(1+\lambda)^3}
  \sum_{m=1,\text{odd}}^{\infty}\frac{f^{XM}_{m}}{\left[(1+\lambda)s\right]^m},
  \label{eq:XM12-by-f}
\end{eqnarray}
\end{subequations}
$Y^{M}_{\alpha\beta}$ [in (J-63a,b)] by
\begin{subequations}
\begin{eqnarray}
  Y^{M}_{11}(s,\lambda)
  &=&
  \sum_{m=0,\text{even}}^{\infty}\frac{f^{YM}_{m}}{\left[(1+\lambda)s\right]^m},\\
  Y^{M}_{12}(s,\lambda)
  &=&
  \frac{8}{(1+\lambda)^3}
  \sum_{m=1,\text{odd}}^{\infty}\frac{f^{YM}_{m}}{\left[(1+\lambda)s\right]^m},
  \label{eq:YM12-by-f}
\end{eqnarray}
\end{subequations}
$Z^{M}_{\alpha\beta}$ [in (J-78a,b)] by
\begin{subequations}
\begin{eqnarray}
  Z^{M}_{11}(s,\lambda)
  &=&
  \sum_{m=0,\text{even}}^{\infty}\frac{f^{ZM}_{m}}{\left[(1+\lambda)s\right]^m},\\
  Z^{M}_{12}(s,\lambda) &=&
  \frac{-8}{(1+\lambda)^3}
  \sum_{m=1,\text{odd}}^{\infty}\frac{f^{ZM}_{m}}{\left[(1+\lambda)s\right]^m}.
  \label{eq:ZM12-by-f}
\end{eqnarray}
\end{subequations}

\subsection{Twin Multipole Expansions}
\label{sec:twin-multipole-expansions}
Let us consider two particles $\alpha = 1$ and $2$, 
whose centers, radii, and slip lengths are given by $\bm{x}_{\alpha}$, 
$a_{\alpha}$, and $\gamma_{\alpha}$, respectively. 
The scaled slip length for particle $\alpha$ is defined by 
\begin{equation}
  \widehat{\gamma}_{\alpha} = \frac{\gamma_{\alpha}}{a_{\alpha}}.
\end{equation}

First, we outline the derivation of equations 
among coefficients $(p_{mn}, q_{mn}, v_{mn})$ and 
$(\psi_{mn}, \chi_{mn}, \omega_{mn})$ for the slip spheres. 
Then, we solve the recurrence relations for each problem 
and obtain all the resistance functions.

\subsubsection{Outline}
In Sec. \ref{sec:single-sphere}, the problem of single slip sphere 
has been solved by Lamb's general solution (\ref{eq:Lamb-vel}) 
through three scalars of the surface vector on 
the both sides of the slip boundary condition (\ref{eq:slip-BC-split}). 
Jeffrey \textit{et al.} \cite{JeffreyOnishi1984,Jeffrey1992} 
solved two-sphere problem with no-slip boundary condition, 
\textit{i.e.} $\gamma = 0$ in Eq. (\ref{eq:slip-BC-split}). 
To complete the boundary condition for two slip spheres, 
we need to obtain the tangential force density 
caused by particle $(3-\alpha)$ on the surface of particle $\alpha$. 
Let us denote it by ${\bm{t}'}^{(\alpha)}$, that is, 
\begin{equation}
  {\bm{t}'}^{(\alpha)}
  =
  \left(
    \bm{I}-\bm{n}^{(\alpha)}\bm{n}^{(\alpha)}
  \right)
  \cdot
  \left(
    \bm{\sigma}^{(3-\alpha)}
    \cdot
    \bm{n}^{(\alpha)}
  \right)
  ,
\end{equation}
where $\bm{n}^{(\alpha)}$ is the surface normal of particle $\alpha$ 
($\bm{r}^{(\alpha)}/r_{\alpha}$ for a sphere), and 
$\bm{\sigma}^{(3-\alpha)}$ is the disturbance part of the stress 
caused by particle $(3-\alpha)$ given by 
\begin{equation}
  \bm{\sigma}^{(3-\alpha)}
  =
  -p^{(3-\alpha)}\bm{I}
  +\mu\left[
    \bm{\nabla}\bm{v}^{(3-\alpha)}
    +
    \left(\bm{\nabla}\bm{v}^{(3-\alpha)}\right)^{\dagger}
  \right]
  .
\end{equation}
Here, $p^{(3-\alpha)}$ and $\bm{v}^{(3-\alpha)}$ are expressed 
in terms of Lamb's general solution 
for the polar coordinates of particle $(3-\alpha)$ given by 
Eqs. (\ref{eq:Lamb-p}) and (\ref{eq:Lamb-vel}), 
respectively. 
Because $\bm{\sigma}^{(3-\alpha)}\cdot\bm{n}^{(\alpha)}\neq\bm{f}^{(3-\alpha)}$, 
we cannot use the surface force density $\bm{f}$ 
in Eq. (\ref{eq:lamb-force-density}). 

Following similar calculations by Jeffrey and Onishi \cite{JeffreyOnishi1984} 
for the disturbance velocity, 
we can write $\bm{\sigma}^{(3-\alpha)}$ 
by the spherical harmonics with respect to the particle $\alpha$ 
in terms of the transformation [in (JO-2.1)] 
\begin{equation}
  \left(
    \frac{a_\alpha}{r_{\alpha}}
  \right)^{n+1}
  Y_{mn}
  \left(
    \theta_\alpha,
    \phi
  \right)
  =
  \left(
    \frac{a_\alpha}{r}
  \right)^{n+1}
  \sum_{s=m}^\infty
  \left(
    \begin{array}{c}
      n+s\\
      s+m
    \end{array}
  \right)
  \left(
    \frac{r_{3-\alpha}}{r}
  \right)^s
  Y_{ms}
  \left(
    \theta_{3-\alpha},
    \phi
  \right)
  ,
  \label{eq:shift-harmonics}
\end{equation}
and the following relations [in (JO-2.7)] 
\begin{subequations}
\begin{equation}
  \bm{r}_\alpha
  =
  \hat{\bm{r}}_{3-\alpha}
  \left(
    r_{3-\alpha}
    -
    r\cos\theta_{3-\alpha}
  \right)
  +
  \hat{\bm{\theta}}_{3-\alpha}
  r\sin\theta_{3-\alpha}
\end{equation}
\begin{equation}
  r_\alpha^2
  =
  r_{3-\alpha}^2
  +
  r^2
  -
  2
  r_{3-\alpha}
  r
  \cos\theta_{3-\alpha}
  .
\end{equation}
\end{subequations}
After substituting the expansions for 
the solid spherical harmonics 
$p^{(3-\alpha)}_{-n-1}$, $\chi^{(3-\alpha)}_{-n-1}$, and $\Phi^{(3-\alpha)}_{-n-1}$ 
in Eqs. (\ref{eq:spherical-harmonics-expansion-p}), 
(\ref{eq:spherical-harmonics-expansion-q}), and 
(\ref{eq:spherical-harmonics-expansion-v}), 
the three scalars of the surface vector of ${\bm{t}'}^{(\alpha)}$ 
are obtained in the form of the expansion 
with spherical harmonics $Y_{mn}(\theta_{\alpha},\phi)$. 
Combining the results of ${\bm{t}'}^{(\alpha)}$ 
and those of the disturbance velocity 
on particle $\alpha$ caused by particle $(3-\alpha)$ 
given by Jeffrey and Onishi \cite{JeffreyOnishi1984} 
with the single-sphere problem in Eqs. (\ref{eq:1b-slip-pmn}), 
(\ref{eq:1b-slip-vmn}), and (\ref{eq:1b-slip-qmn}), 
we have three equations for the coefficients, 
corresponding to Eqs. (JO-2.9a), (JO-2.9b), and (JO-2.9c) 
for the no-slip case, as 
\begin{subequations}
\begin{eqnarray}
  &&
  \psi^{(\alpha)}_{mn}
  -
  (n-1)
  \left(
    1
    -
    2(n+1)
    \widehat{\gamma}_{\alpha}
  \right)
  \chi^{(\alpha)}_{mn}
  \nonumber\\
  &=&
  (n+1)
  (2n+1)
  \left(
    1
    +
    2
    \widehat{\gamma}_{\alpha}
  \right)
  v^{(\alpha)}_{mn}
  -
  \frac{n+1}{2}
  p^{(\alpha)}_{mn}
  \nonumber\\
  &&
  +
  \sum_{s=m}^{\infty}
  \left(
    \begin{array}{c}
      n+s \\ n+m
    \end{array}
  \right)
  t_{\alpha}^{n-1}
  t_{3-\alpha}^{s}
  \frac{n}{2n+3}
  \left(
    1
    -
    (2n+1)
    \widehat{\gamma}_{\alpha}
  \right)
  p^{(3-\alpha)}_{ms}
  t_{\alpha}^2
  ,
  \nonumber\\
  \label{eq:psi-chi-slip}
\end{eqnarray}
\begin{eqnarray}
  &&
  \psi^{(\alpha)}_{mn}
  +
  (n+2)
  \left(
    1
    +
    2n
    \widehat{\gamma}_{\alpha}
  \right)
  \chi^{(\alpha)}_{mn}
  \nonumber\\
  &=&
  \frac{n+1}{2n-1}
  \left(
    1
    +
    (2n+1)
    \widehat{\gamma}_{\alpha}
  \right)
  p^{(\alpha)}_{mn}
  +
  \sum_{s=m}^{\infty}
  \left(
    \begin{array}{c}
      n+s \\ n+m
    \end{array}
  \right)
  t_{\alpha}^{n-1}
  t_{3-\alpha}^{s}
  \nonumber\\
  &&
  \times
  \left[
    {\rm i}
    (-1)^{\alpha}
    m
    (2n+1)
    \left(
      1
      +
      2
      \widehat{\gamma}_{\alpha}
    \right)
    q^{(3-\alpha)}_{ms}
    t_{3-\alpha}
  \right.
  \nonumber\\
  &&
  \left.
    +
    n
    (2n+1)
    \left(
      1
      +
      2
      \widehat{\gamma}_{\alpha}
    \right)
    v^{(3-\alpha)}_{ms}
    t_{3-\alpha}^2
  \right.
  \nonumber\\
  &&
  \left.
    +
    \frac{2n+1}{2n-1}
    \left(
      1
      +
      2
      \widehat{\gamma}_{\alpha}
    \right)
  \right.
  \nonumber\\
  &&
  \left.
    \times
    \frac{ns(n+s-2ns-2)-m^2(2ns-4s-4n+2)}{2s(2s-1)(n+s)}
    p^{(3-\alpha)}_{ms}
  \right.
  \nonumber\\
  &&
  \left.
    +
    \frac{n}{2}
    p^{(3-\alpha)}_{ms}
    t_{\alpha}^2
  \right]
  ,
  \label{eq:psi-chi-slip-2}
\end{eqnarray}
\begin{eqnarray}
  \omega^{(\alpha)}_{mn}
  &=&
  n(n+1)
  \left(
    1
    +
    (n+2)
    \widehat{\gamma}_{\alpha}
  \right)
  q^{(\alpha)}_{mn}
  \nonumber\\
  &&
  +\sum_{s=m}^{\infty}
  \left(
    \begin{array}{c}
      n+s \\ n+m
    \end{array}
  \right)
  t_{\alpha}^{n}
  t_{3-\alpha}^{s}
  \left(
    1
    -
    (n-1)
    \widehat{\gamma}_{\alpha}
  \right)
  \nonumber\\
  &&
  \quad
  \times
  \left[
    -ns
    q^{(3-\alpha)}_{ms}
    t_{3-\alpha}
    +
    {\rm i}
    (-1)^\alpha
    \frac{m}{s}
    p^{(3-\alpha)}_{ms}
  \right]
  ,
  \label{eq:omega-slip}
\end{eqnarray}
\end{subequations}
where 
\begin{equation}
  t_{\alpha}
  =
  \frac{a_{\alpha}}{r}
  .
\end{equation}
In writing these three equations, 
we can take any independent linear combinations in principle. 
For the single-sphere problem, 
we may write three equations for $p_{mn}$, $q_{mn}$, and $q_{mn}$ 
as in Eqs. (\ref{eq:1b-slip-pmn}), 
(\ref{eq:1b-slip-vmn}), and (\ref{eq:1b-slip-qmn}), 
or those for the coefficients of the boundary condition 
$\chi_{mn}$, $\psi_{mn}$, and $\omega_{mn}$, instead. 
Jeffrey and Onishi \cite{JeffreyOnishi1984} take equations for 
$\psi^{(\alpha)}_{mn}-(n-1)\chi^{(\alpha)}_{mn}$, 
$\psi^{(\alpha)}_{mn}+(n+2)\chi^{(\alpha)}_{mn}$, and 
$\omega^{(\alpha)}_{mn}$ for no-slip particles. 
Here we extend the equations for slip particles so that 
the structures of the equations for no-slip case would hold, that is, 
the interaction terms (with the summation of $s$) 
contain only $p^{(3-\alpha)}_{mn}$ in Eq. (\ref{eq:psi-chi-slip}), 
and the term of $v^{(\alpha)}_{mn}$ is eliminated 
in Eq. (\ref{eq:psi-chi-slip-2}). 
Equation (\ref{eq:omega-slip}) for $\omega^{(\alpha)}_{mn}$ 
is just the same choice to the no-slip case. 

Note that 
Keh and Chen \cite{KehChen1997} take a different form for the first equation, 
that is, $
  \psi^{(\alpha)}_{mn}
  -
  \left(
    (n-1)
    +
    (2n^2+1)
    \widehat{\gamma}_{\alpha}
  \right)
  \chi^{(\alpha)}_{mn}
$ in Eq. (KC-20a). 
Although they are mathematically equivalent, 
Eq. (\ref{eq:psi-chi-slip}) is simpler and 
we will use it later in this paper. 
Also note that 
there are typos in Keh and Chen \cite{KehChen1997} at Eqs. (KC-20a,b,c) 
where $\widehat{\beta}^{-1}_{(3-\alpha)}$ 
($\widehat{\gamma}_{(3-\alpha)}$ in present notations) 
should be replaced by $\widehat{\beta}^{-1}_{(\alpha)}$. 
If we look at the slip boundary condition from which these 
three equations are derived, it is obvious that 
only the slip length of particle $\alpha$ would appear there. 
It should be noted that the results such as coefficients $f_{k}$ 
in Keh and Chen \cite{KehChen1997} 
are correct, because they took a simplification that 
the scaled slip lengths for two particles are the same 
($\widehat{\gamma}_{1} = \widehat{\gamma}_{2}$ in present notations).

\subsubsection{Recurrence Relations}
For resistance functions, the boundary conditions are given 
completely by $\chi_{mn}$, $\psi_{mn}$, and $\omega_{mn}$, which 
are independent of the distance between the particle $r$ and 
therefore $t_{\alpha}$ and $t_{3-\alpha}$. This means that 
the coefficients $P_{npq}$, $V_{npq}$, and $Q_{npq}$ of 
the ($p,q$)-term in the expansion by $t_{\alpha}^pt_{3-\alpha}^q$ 
(see, for example, Eqs. (\ref{eq:Pnpq-for-X}) and (\ref{eq:Vnpq-for-X}) 
in the following)
are solved by the recurrence relations 
for $p\ge 0$ and $q\ge 0$ with the initial condition 
at $p=0$ and $q=0$.
Therefore, we split the above three equations into two parts, the 
initial conditions and the recurrence relations. 
The initial conditions are
\begin{subequations}
\begin{eqnarray}
  p^{(\alpha)}_{mn}
  &=&
  \frac{2n-1}{n+1}
  \Gamma^{(\alpha)}_{0,2n+1}
  \psi^{(\alpha)}_{mn}
  \nonumber\\
  &&
  +
  \frac{(n+2)(2n-1)}{n+1}
  \Gamma^{(\alpha)}_{2n,2n+1}
  \chi^{(\alpha)}_{mn}
  \label{eq:pmn-slip-init}
  ,\\
  2(2n+1)
  v^{(\alpha)}_{mn}
  &=&
  \frac{2}{n+1}
  \Gamma^{(\alpha)}_{0,2}
  \psi^{(\alpha)}_{mn}
  -
  \frac{2(n-1)}{(n+1)}
  \Gamma^{(\alpha)}_{-2(n+1),2}
  \chi^{(\alpha)}_{mn}
  \nonumber\\
  &&
  +
  \Gamma^{(\alpha)}_{0,2}
  p^{(\alpha)}_{mn}
  \label{eq:vmn-slip-init}
  ,\\
  q^{(\alpha)}_{mn}
  &=&
  \frac{1}{n(n+1)}
  \Gamma^{(\alpha)}_{0,n+2}
  \omega^{(\alpha)}_{mn}
  \label{eq:qmn-slip-init}
  .
\end{eqnarray}
\end{subequations}
The recurrence relations are
\begin{subequations}
\begin{eqnarray}
  p^{(\alpha)}_{mn}
  &=&
  \sum_{s=m}^{\infty}
  \left(
    \begin{array}{c}
      n+s \\ n+m
    \end{array}
  \right)
  \nonumber\\
  &&
  \times
  \left[
    -
    {\rm i}
    (-1)^{\alpha}
    m
    \frac{(2n+1)(2n-1)}{n+1}
    \Gamma^{(\alpha)}_{2,2n+1}
    q^{(3-\alpha)}_{ms}
    t_{\alpha}^{n-1}
    t_{3-\alpha}^{s+1}
  \right.
  \nonumber\\
  &&
  \left.
    -
    \frac{n(2n+1)(2n-1)}{n+1}
    \Gamma^{(\alpha)}_{2,2n+1}
    v^{(3-\alpha)}_{ms}
    t_{\alpha}^{n-1}
    t_{3-\alpha}^{s+2}
  \right.
  \nonumber\\
  &&
  \left.
    -
    \frac{2n+1}{n+1}
    \frac{ns(n+s-2ns-2)-m^2(2ns-4s-4n+2)}{2s(2s-1)(n+s)}
  \right.
  \nonumber\\
  &&
  \quad
  \left.
    \times
    \Gamma^{(\alpha)}_{2,2n+1}
    p^{(3-\alpha)}_{ms}
    t_{\alpha}^{n-1}
    t_{3-\alpha}^{s}
  \right.
  \nonumber\\
  &&
  \left.
    -
    \frac{n(2n-1)}{2(n+1)}
    \Gamma^{(\alpha)}_{0,2n+1}
    p^{(3-\alpha)}_{ms}
    t_{\alpha}^{n+1}
    t_{3-\alpha}^{s}
  \right]
  \label{eq:pmn-slip-rec}
  ,
\end{eqnarray}
\begin{eqnarray}
  &&
  2(2n+1)
  v^{(\alpha)}_{mn}
  =
  \Gamma^{(\alpha)}_{0,2}
  p^{(\alpha)}_{mn}
  \label{eq:vmn-slip-rec}
  \\
  &&
  \quad
  -
  \sum_{s=m}^{\infty}
  \left(
    \begin{array}{c}
      n+s \\ n+m
    \end{array}
  \right)
  \frac{2n}{(n+1)(2n+3)}
  \Gamma^{(\alpha)}_{-(2n+1),2}
  p^{(3-\alpha)}_{ms}
  t_{\alpha}^{n+1}
  t_{3-\alpha}^{s}
  ,
  \nonumber
\end{eqnarray}
\begin{eqnarray}
  q^{(\alpha)}_{mn}
  &=&
  \sum_{s=m}^{\infty}
  \left(
    \begin{array}{c}
      n+s \\ n+m
    \end{array}
  \right)
  \left[
    \frac{s}{(n+1)}
    \Gamma^{(\alpha)}_{-(n-1),n+2}
    q^{(3-\alpha)}_{ms}
    t_{\alpha}^{n}
    t_{3-\alpha}^{s+1}
  \right.
  \nonumber\\
  &&
  \left.
    -
    {\rm i}
    (-1)^\alpha
    \frac{m}{ns(n+1)}
    \Gamma^{(\alpha)}_{-(n-1),n+2}
    p^{(3-\alpha)}_{ms}
    t_{\alpha}^{n}
    t_{3-\alpha}^{s}
  \right]
  .
  \label{eq:qmn-slip-rec}
\end{eqnarray}
\end{subequations}
It should be noted that the initial conditions are independent of $m$, 
while the recurrence relations are not. Therefore, the initial conditions 
are the same for $X$ ($m=0$), $Y$ ($m=1$), and $Z$ ($m=2$) functions 
for each problems (translating, rotating, or in the strain flow).

We also note that 
the recurrence relations have $\alpha$-dependent quantity $\Gamma^{(\alpha)}$, 
so that we need to solve the coefficients 
$P_{npq}$, $V_{npq}$, and $Q_{npq}$ for $\alpha$ 
as well as $(3-\alpha)$, 
while, for the no-slip case, the coefficients for $\alpha$ 
and $(3-\alpha)$ are identical.

The results shown in the following are obtained by the program 
implemented on an open source computer algebra system called 
``Maxima'' \cite{MAXIMA}. 
The program is relatively slow due to its symbolic calculation 
and the coefficients are obtained up to $k=20$, at least. 
We also implement a code in C with floating-point variables 
where the parameters $a_{\alpha}$ and $\gamma_{\alpha}$ must be 
given by numbers for the calculation. 
With this code, we can obtain the coefficients around $k=100$.

\subsection{$X$ Functions ($m=0$)}
For the case of $m=0$, 
$q^{(\alpha)}$ and $q^{(3-\alpha)}$ are decoupled from the others.

\subsubsection{$X^{A}$ Function}
The boundary condition for the $X^{A}$ problem is given by
\begin{equation}
  \chi^{(\alpha)}_{mn}
  =
  U
  \delta_{m0}
  \delta_{n1}
  ,\quad
  \psi^{(\alpha)}_{mn}
  =
  0
  ,\quad
  \omega^{(\alpha)}_{mn}
  =
  0
  .
\end{equation}
To obtain the coefficients for each order of the power of $r$, 
we expand the coefficients [in (JO-3.4) and (JO-3.5)] as
\begin{subequations}
\begin{eqnarray}
  p^{(\alpha)}_{0n}
  &=&
  \frac{3}{2}
  U
  \sum_{p=0}^{\infty}
  \sum_{q=0}^{\infty}
  P^{(\alpha)}_{npq}
  t_{\alpha}^{p}
  t_{3-\alpha}^{q}
  ,
  \label{eq:Pnpq-for-X}
  \\
  v^{(\alpha)}_{0n}
  &=&
  \frac{3}{2}
  U
  \sum_{p=0}^{\infty}
  \sum_{q=0}^{\infty}
  \frac{V^{(\alpha)}_{npq}}{2(2n+1)}
  t_{\alpha}^{p}
  t_{3-\alpha}^{q}
  .
  \label{eq:Vnpq-for-X}
\end{eqnarray}
\end{subequations}
Substituting the expansions, 
we have the initial conditions for $p=0$ and $q=0$ 
from Eqs. (\ref{eq:pmn-slip-init}) and (\ref{eq:vmn-slip-init}) by 
\begin{equation}
  P^{(\alpha)}_{n00}
  =
  \delta_{n1}
  \Gamma^{(\alpha)}_{2,3}
  ,\quad
  V^{(\alpha)}_{n00}
  =
  \delta_{n1}
  \Gamma^{(\alpha)}_{0,3}
  ,
\end{equation}
and the recurrence relations for $p\ge 0$ and $q\ge 0$ 
from Eqs. (\ref{eq:pmn-slip-rec}) and (\ref{eq:vmn-slip-rec}) by 
\begin{subequations}
\begin{eqnarray}
  P^{(\alpha)}_{npq}
  &=&
  \sum_{s=0}^{\infty}
  \left(
    \begin{array}{c}
      n+s \\ n
    \end{array}
  \right)
  \nonumber\\
  &&
  \times
  \left[
    -
    \frac{n(2n-1)(2n+1)}{2(n+1)(2s+1)}
    \Gamma^{(\alpha)}_{2,2n+1}
    V^{(3-\alpha)}_{s(q-s-2)(p-n+1)}
  \right.
  \nonumber\\
  &&
  \left.
    -
    \frac{n(2n+1)(n+s-2ns-2)}{2(n+1)(n+s)(2s-1)}
    \Gamma^{(\alpha)}_{2,2n+1}
    P^{(3-\alpha)}_{s(q-s)(p-n+1)}
  \right.
  \nonumber\\
  &&
  \left.
    -
    \frac{n(2n-1)}{2(n+1)}
    \Gamma^{(\alpha)}_{0,2n+1}
    P^{(3-\alpha)}_{s(q-s)(p-n-1)}
  \right]
  .
\end{eqnarray}
\begin{eqnarray}
  &&
  V^{(\alpha)}_{npq}
  =
  \Gamma^{(\alpha)}_{0,2}
  P^{(\alpha)}_{npq}
  \label{eq:Vnpq-for-XA}
  \\
  &&
  -
  \sum_{s=0}^{\infty}
  \left(
    \begin{array}{c}
      n+s \\ n
    \end{array}
  \right)
  \frac{2n}{(n+1)(2n+3)}
  \Gamma^{(\alpha)}_{-(2n+1),2}
  P^{(3-\alpha)}_{s(q-s)(p-n-1)}
  .
  \nonumber
\end{eqnarray}
\end{subequations}
The initial conditions correspond to Eqs. (KC-26a,b) 
and the recurrence relations to Eqs. (KC-27a,b). 
Note that Eq. (\ref{eq:Vnpq-for-XA}) for $V^{(\alpha)}_{npq}$ 
is simpler than the corresponding equation in Keh and Chan (KC-27b), 
because we use the simpler recurrence relation in Eq. (\ref{eq:psi-chi-slip}).

The coefficient $f^{XA\alpha}_{k}$ is defined [in (JO-3.15)] as 
\begin{equation}
  f^{XA\alpha}_{k}
  =
  2^k
  \sum_{q=0}^{k}
  P^{(\alpha)}_{1(k-q)q}
  \lambda^{q}
  .
\end{equation}
Here we see a slight difference from the no-slip case. 
This is because of the $\alpha$ dependence of $P^{(\alpha)}_{npq}$, 
so that $f^{XA\alpha}_{k}$ also depends on $\alpha$. 
The explicit forms up to $k=7$ are 
\begin{subequations}
\begin{eqnarray}
  f^{XA1}_{0} &=& \left(
     \Gamma^{(1)}_{2,3} 
  \right),\\
  f^{XA1}_{1} &=& \lambda \left(
     3 \Gamma^{(1)}_{2,3} \Gamma^{(2)}_{2,3} 
  \right),\\
  f^{XA1}_{2} &=& \lambda \left(
     9 (\Gamma^{(1)}_{2,3})^{2} \Gamma^{(2)}_{2,3} 
  \right),\\
  f^{XA1}_{3} &=& \lambda \left(
     - 4 \Gamma^{(1)}_{0,3} \Gamma^{(2)}_{2,3} 
  \right)\nonumber\\
  &+& \lambda^{2} \left(
     27 (\Gamma^{(1)}_{2,3})^{2} (\Gamma^{(2)}_{2,3})^{2} 
  \right)\nonumber\\
  &+& \lambda^{3} \left(
     - 4 \Gamma^{(2)}_{0,3} \Gamma^{(1)}_{2,3} 
  \right),\\
  f^{XA1}_{4} &=& \lambda \left(
     - 24 \Gamma^{(1)}_{0,3} \Gamma^{(1)}_{2,3} \Gamma^{(2)}_{2,3} 
  \right)\nonumber\\
  &+& \lambda^{2} \left(
     81 (\Gamma^{(1)}_{2,3})^{3} (\Gamma^{(2)}_{2,3})^{2} 
  \right)\nonumber\\
  &+& \lambda^{3} \left(
     12 (\Gamma^{(1)}_{2,3})^{2} ( 5 \Gamma^{(2)}_{2,5} 
     - 2 \Gamma^{(2)}_{0,3} ) 
  \right),\\
  f^{XA1}_{5} &=& \lambda^{2} \left(
     36 \Gamma^{(1)}_{2,3} (\Gamma^{(2)}_{2,3})^{2} ( 5 \Gamma^{(1)}_{2,5} 
     - 3 \Gamma^{(1)}_{0,3} ) 
  \right)\nonumber\\
  &+& \lambda^{3} \left(
     243 (\Gamma^{(1)}_{2,3})^{3} (\Gamma^{(2)}_{2,3})^{3} 
  \right)\nonumber\\
  &+& \lambda^{4} \left(
     36 (\Gamma^{(1)}_{2,3})^{2} \Gamma^{(2)}_{2,3} ( 5 \Gamma^{(2)}_{2,5} 
     - 3 \Gamma^{(2)}_{0,3} ) 
  \right),\\
  f^{XA1}_{6} &=& \lambda \left(
     16 (\Gamma^{(1)}_{0,3})^{2} \Gamma^{(2)}_{2,3} 
  \right)\nonumber\\
  &+& \lambda^{2} \left(
     108 (\Gamma^{(1)}_{2,3})^{2} (\Gamma^{(2)}_{2,3})^{2} ( 5 \Gamma^{(1)}_{2,5} 
     - 4 \Gamma^{(1)}_{0,3} ) 
  \right)\nonumber\\
  &+& \lambda^{3} \left(
     - \Gamma^{(1)}_{2,3} ( 480 \Gamma^{(1)}_{0,3} \Gamma^{(2)}_{2,5} 
     - 729 (\Gamma^{(1)}_{2,3})^{3} (\Gamma^{(2)}_{2,3})^{3} 
   \right.\nonumber\\
   &&\left.
     - 32 \Gamma^{(1)}_{0,3} \Gamma^{(2)}_{0,3} ) 
  \right)\nonumber\\
  &+& \lambda^{4} \left(
     216 (\Gamma^{(1)}_{2,3})^{3} \Gamma^{(2)}_{2,3} ( 5 \Gamma^{(2)}_{2,5} 
     - 2 \Gamma^{(2)}_{0,3} ) 
  \right)\nonumber\\
  &+& \lambda^{5} \left(
     16 (\Gamma^{(1)}_{2,3})^{2} ( 126 \Gamma^{(2)}_{2,7} 
     - 90 \Gamma^{(2)}_{0,5} 
     + 5 \Gamma^{(2)}_{0,2} \Gamma^{(2)}_{0,3} 
   \right.\nonumber\\
   &&\left.
     + 4 \Gamma^{(2)}_{-3,2} ) / 5 
  \right),\\
  f^{XA1}_{7} &=& \lambda^{2} \left(
     48 (\Gamma^{(2)}_{2,3})^{2} ( 126 \Gamma^{(1)}_{2,3} \Gamma^{(1)}_{2,7} 
     - 70 \Gamma^{(1)}_{0,3} \Gamma^{(1)}_{2,5} 
   \right.\nonumber\\
   &&\left.
     - 45 \Gamma^{(1)}_{0,5} \Gamma^{(1)}_{2,3} 
     + 15 (\Gamma^{(1)}_{0,3})^{2} 
     + 4 \Gamma^{(1)}_{-3,3} ) / 5 
  \right)\nonumber\\
  &+& \lambda^{3} \left(
     1620 (\Gamma^{(1)}_{2,3})^{2} (\Gamma^{(2)}_{2,3})^{3} ( 2 \Gamma^{(1)}_{2,5} 
     - \Gamma^{(1)}_{0,3} ) 
  \right)\nonumber\\
  &+& \lambda^{4} \left(
     3 \Gamma^{(1)}_{2,3} \Gamma^{(2)}_{2,3} ( 800 \Gamma^{(1)}_{2,5} \Gamma^{(2)}_{2,5} 
     - 560 \Gamma^{(1)}_{0,3} \Gamma^{(2)}_{2,5} 
   \right.\nonumber\\
   &&\left.
     - 560 \Gamma^{(2)}_{0,3} \Gamma^{(1)}_{2,5} 
     + 729 (\Gamma^{(1)}_{2,3})^{3} (\Gamma^{(2)}_{2,3})^{3} 
     + 96 \Gamma^{(1)}_{0,3} \Gamma^{(2)}_{0,3} ) 
  \right)\nonumber\\
  &+& \lambda^{5} \left(
     1620 (\Gamma^{(1)}_{2,3})^{3} (\Gamma^{(2)}_{2,3})^{2} ( 2 \Gamma^{(2)}_{2,5} 
     - \Gamma^{(2)}_{0,3} ) 
  \right)\nonumber\\
  &+& \lambda^{6} \left(
     48 (\Gamma^{(1)}_{2,3})^{2} ( 126 \Gamma^{(2)}_{2,3} \Gamma^{(2)}_{2,7} 
     - 70 \Gamma^{(2)}_{0,3} \Gamma^{(2)}_{2,5} 
     - 45 \Gamma^{(2)}_{0,5} \Gamma^{(2)}_{2,3} 
   \right.\nonumber\\
   &&\left.
     + 15 (\Gamma^{(2)}_{0,3})^{2} 
     + 4 \Gamma^{(2)}_{-3,3} ) / 5 
  \right).
\end{eqnarray}
\end{subequations}
The results are identical to those obtained by method of reflections 
in Eqs. (\ref{eq:mofr-fXA-0}), (\ref{eq:mofr-fXA-1}), and (\ref{eq:mofr-fXA-3}) 
for the terms containing one or two $\Gamma$'s, 
because only the first reflection from the particles $1$ to $2$ is taken 
and the higher reflections are missing in the present calculation 
of the method of reflections. 
Therefore, $f^{XA1}_{2}$ and $\lambda^2$ term in $f^{XA}_{3}$ 
do not appear.

Also the results reduce to those by Jeffrey and Onishi \cite{JeffreyOnishi1984} 
in the no-slip limit $\widehat{\gamma} = 0$, 
and those by Keh and Chen \cite{KehChen1997} 
in the case of $\widehat{\gamma}_{1} = \widehat{\gamma}_{2}$. 
Therefore, they also reduce to 
those by Hetsroni and Haber \cite{HetsroniHaber1978} 
in the perfect slip limit $\widehat{\gamma} = \infty$.

\subsubsection{$X^{G}$ Function}
With the same recurrence relations and the initial condition for $X^{A}$, 
that is, for the translating particles, 
the function $X^{G}$ is obtained from the coefficient $P_{2pq}$ for 
the stresslet instead of $P_{1pq}$ for the force.

In this case, the coefficient $f^{XG\alpha}_{k}$ is defined as
\begin{equation}
  f^{XG\alpha}_{k}
  =
  \left(
    \frac{3}{4}
  \right)
  2^k
  \sum_{q=0}^{k}
  P^{(\alpha)}_{2(k-q)q}
  \lambda^{q}
  .
\end{equation}
The explicit forms up to $k=7$ are 
\begin{subequations}
\begin{eqnarray}
  f^{XG1}_{0} &=& 0,\\
  f^{XG1}_{1} &=& 0,\\
  f^{XG1}_{2} &=& \lambda \left(
     15 \Gamma^{(2)}_{2,3} \Gamma^{(1)}_{2,5} 
  \right),\\
  f^{XG1}_{3} &=& \lambda \left(
     45 \Gamma^{(1)}_{2,3} \Gamma^{(2)}_{2,3} \Gamma^{(1)}_{2,5} 
  \right),\\
  f^{XG1}_{4} &=& \lambda \left(
     - 36 \Gamma^{(1)}_{0,5} \Gamma^{(2)}_{2,3} 
  \right)\nonumber\\
  &+& \lambda^{2} \left(
     135 \Gamma^{(1)}_{2,3} (\Gamma^{(2)}_{2,3})^{2} \Gamma^{(1)}_{2,5} 
  \right)\nonumber\\
  &+& \lambda^{3} \left(
     - 60 \Gamma^{(2)}_{0,3} \Gamma^{(1)}_{2,5} 
  \right),\\
  f^{XG1}_{5} &=& \lambda \left(
     - 12 \Gamma^{(2)}_{2,3} ( 5 \Gamma^{(1)}_{0,3} \Gamma^{(1)}_{2,5} 
     + 9 \Gamma^{(1)}_{0,5} \Gamma^{(1)}_{2,3} ) 
  \right)\nonumber\\
  &+& \lambda^{2} \left(
     405 (\Gamma^{(1)}_{2,3})^{2} (\Gamma^{(2)}_{2,3})^{2} \Gamma^{(1)}_{2,5} 
  \right)\nonumber\\
  &+& \lambda^{3} \left(
     120 \Gamma^{(1)}_{2,3} \Gamma^{(1)}_{2,5} ( 5 \Gamma^{(2)}_{2,5} 
     - 2 \Gamma^{(2)}_{0,3} ) 
  \right),\\
  f^{XG1}_{6} &=& \lambda^{2} \left(
     36 (\Gamma^{(2)}_{2,3})^{2} ( 25 (\Gamma^{(1)}_{2,5})^{2} 
     - 10 \Gamma^{(1)}_{0,3} \Gamma^{(1)}_{2,5} 
     - 9 \Gamma^{(1)}_{0,5} \Gamma^{(1)}_{2,3} ) 
  \right)\nonumber\\
  &+& \lambda^{3} \left(
     1215 (\Gamma^{(1)}_{2,3})^{2} (\Gamma^{(2)}_{2,3})^{3} \Gamma^{(1)}_{2,5} 
  \right)\nonumber\\
  &+& \lambda^{4} \left(
     900 \Gamma^{(1)}_{2,3} \Gamma^{(2)}_{2,3} \Gamma^{(1)}_{2,5} ( 2 \Gamma^{(2)}_{2,5} 
     - \Gamma^{(2)}_{0,3} ) 
  \right),\\
  f^{XG1}_{7} &=& \lambda \left(
     144 \Gamma^{(1)}_{0,3} \Gamma^{(1)}_{0,5} \Gamma^{(2)}_{2,3} 
  \right)\nonumber\\
  &+& \lambda^{2} \left(
     108 \Gamma^{(1)}_{2,3} (\Gamma^{(2)}_{2,3})^{2} ( 25 (\Gamma^{(1)}_{2,5})^{2} 
     - 15 \Gamma^{(1)}_{0,3} \Gamma^{(1)}_{2,5} 
     - 9 \Gamma^{(1)}_{0,5} \Gamma^{(1)}_{2,3} ) 
  \right)\nonumber\\
  &+& \lambda^{3} \left(
     - 3 ( 800 \Gamma^{(1)}_{0,3} \Gamma^{(1)}_{2,5} \Gamma^{(2)}_{2,5} 
     + 960 \Gamma^{(1)}_{0,5} \Gamma^{(1)}_{2,3} \Gamma^{(2)}_{2,5} 
   \right.\nonumber\\
   &&\left.
     - 1215 (\Gamma^{(1)}_{2,3})^{3} (\Gamma^{(2)}_{2,3})^{3} \Gamma^{(1)}_{2,5} 
  \right.\nonumber\\
  &&\left.
     - 80 \Gamma^{(1)}_{0,3} \Gamma^{(2)}_{0,3} \Gamma^{(1)}_{2,5} 
     - 48 \Gamma^{(2)}_{0,3} \Gamma^{(1)}_{0,5} \Gamma^{(1)}_{2,3} ) 
  \right)\nonumber\\
  &+& \lambda^{4} \left(
     1620 (\Gamma^{(1)}_{2,3})^{2} \Gamma^{(2)}_{2,3} \Gamma^{(1)}_{2,5} ( 5 \Gamma^{(2)}_{2,5} 
     - 2 \Gamma^{(2)}_{0,3} ) 
  \right)\nonumber\\
  &+& \lambda^{5} \left(
     48 \Gamma^{(1)}_{2,3} \Gamma^{(1)}_{2,5} ( 126 \Gamma^{(2)}_{2,7} 
     - 90 \Gamma^{(2)}_{0,5} 
     + 5 \Gamma^{(2)}_{0,2} \Gamma^{(2)}_{0,3} 
   \right.\nonumber\\
   &&\left.
     + 4 \Gamma^{(2)}_{-3,2} ) 
  \right).
\end{eqnarray}
\end{subequations}
The results are identical to those obtained by method of reflections 
in Eqs. (\ref{eq:mofr-fXG-0}), (\ref{eq:mofr-fXG-2}), and (\ref{eq:mofr-fXG-4}) 
for the terms containing one or two $\Gamma$'s, similarly to $X^{A}$. 
The results reduce to those by Jeffrey \cite{Jeffrey1992} 
in the no-slip limit $\widehat{\gamma} = 0$.

\subsubsection{$X^{C}$ Function}
The function $X^{C}$ gives the torque for the rotating particles 
in the axisymmetric case ($m=0$). 
The boundary condition is given by 
\begin{equation}
  \chi^{(\alpha)}_{mn}
  =
  0
  ,\quad
  \psi^{(\alpha)}_{mn}
  =
  0
  ,\quad
  \omega^{(\alpha)}_{mn}
  =
  2U
  \delta_{m0}
  \delta_{n1}
  .
\end{equation}
Note that $Q_{npq}$ is decoupled from $P_{npq}$ and $V_{npq}$ for $m=0$. 
Using the expansion [in (JO-6.4)] 
\begin{equation}
  q^{(\alpha)}_{0n}
  =
  U
  \sum_{p=0}^{\infty}
  \sum_{q=0}^{\infty}
  Q^{(\alpha)}_{npq}
  t_{\alpha}^{p}
  t_{3-\alpha}^{q}
  ,
\end{equation}
we have the initial condition for $p=0$ and $q=0$ 
from Eq. (\ref{eq:qmn-slip-init}) by
\begin{equation}
  Q_{n00}
  =
  \delta_{n1}
  \Gamma^{(\alpha)}_{0,3}
  ,
\end{equation}
and the recurrence relation for $p\ge 0$ and $q\ge 0$ 
from Eq. (\ref{eq:qmn-slip-rec}) by
\begin{equation}
  Q^{(\alpha)}_{npq}
  =
  \sum_{s=0}^{\infty}
  \left(
    \begin{array}{c}
      n+s \\ n
    \end{array}
  \right)
  \frac{s}{(n+1)}
  \Gamma^{(\alpha)}_{-(n-1),n+2}
  Q^{(3-\alpha)}_{s(q-s-1)(p-n)}
  .
\end{equation}

The coefficient $f^{XC\alpha}_{k}$ is defined as
\begin{equation}
  f^{XC\alpha}_{k}
  =
  2^k
  \sum_{q=0}^{k}
  Q^{(\alpha)}_{1(k-q)q}
  \lambda^{q+j}
  ,
\end{equation}
where $j=0$ for even $k$ and $j=1$ for odd $k$. 
Because many terms of $f^{XC1}_{k}$ in lower orders are zero, 
we show the explicit forms up to $k=11$ as 
\begin{subequations}
\begin{eqnarray}
  f^{XC1}_{0} &=& \left(
     \Gamma^{(1)}_{0,3} 
  \right),\\
  f^{XC1}_{1} &=& 0,\\
  f^{XC1}_{2} &=& 0,\\
  f^{XC1}_{3} &=& \lambda^{3} \left(
     8 \Gamma^{(1)}_{0,3} \Gamma^{(2)}_{0,3} 
  \right),\\
  f^{XC1}_{4} &=& 0,\\
  f^{XC1}_{5} &=& 0,\\
  f^{XC1}_{6} &=& \lambda^{3} \left(
     64 (\Gamma^{(1)}_{0,3})^{2} \Gamma^{(2)}_{0,3} 
  \right),\\
  f^{XC1}_{7} &=& 0,\\
  f^{XC1}_{8} &=& \lambda^{5} \left(
     768 \Gamma^{(2)}_{-1,4} (\Gamma^{(1)}_{0,3})^{2} 
  \right),\\
  f^{XC1}_{9} &=& \lambda^{6} \left(
     512 (\Gamma^{(1)}_{0,3})^{2} (\Gamma^{(2)}_{0,3})^{2} 
  \right),\\
  f^{XC1}_{10} &=& \lambda^{7} \left(
     6144 \Gamma^{(2)}_{-2,5} (\Gamma^{(1)}_{0,3})^{2} 
  \right),\\
  f^{XC1}_{11} &=& \lambda^{6} \left(
     6144 \Gamma^{(1)}_{-1,4} \Gamma^{(1)}_{0,3} (\Gamma^{(2)}_{0,3})^{2} 
  \right)\nonumber\\
  &+& \lambda^{8} \left(
     6144 \Gamma^{(2)}_{-1,4} (\Gamma^{(1)}_{0,3})^{2} \Gamma^{(2)}_{0,3} 
  \right).
\end{eqnarray}
\end{subequations}
The results are identical to those obtained by method of reflections 
in Eqs. (\ref{eq:mofr-fXC-0}), (\ref{eq:mofr-fXC-1}), and (\ref{eq:mofr-fXC-3}) 
for the terms containing one or two $\Gamma$'s. 
The results reduce to those by Jeffrey and Onishi \cite{JeffreyOnishi1984} 
in the no-slip limit $\widehat{\gamma} = 0$ 
and those by Keh and Chen \cite{KehChen1997} 
in the case of $\widehat{\gamma}_{1} = \widehat{\gamma}_{2}$.

\subsubsection{$X^{M}$ Function}
The function $X^{M}$ gives the stresslet under a strain flow
in the axisymmetric case ($m=0$). Therefore, it is derived by 
the coefficient $P_{2pq}$ for the stresslet from the same recurrence 
relations for $X^{A}$ with a different initial condition. 
The boundary condition is given by 
\begin{subequations}
\begin{eqnarray}
  \chi^{(\alpha)}_{mn}
  &=&
  \frac{2}{3}
  a_{\alpha}
  E_{\alpha}
  \delta_{0m}
  \delta_{2n}
  ,\\
  \psi^{(\alpha)}_{mn}
  &=&
  \frac{2}{3}
  a_{\alpha}
  E_{\alpha}
  (1-6\widehat{\gamma})
  \delta_{0m}
  \delta_{2n}
  ,\\
  \omega^{(\alpha)}_{mn}
  &=&
  0
  ,
\end{eqnarray}
\end{subequations}
which corresponds to Eq. (J-41) with the correction due to the slip. 
Using the expansion [in (J-42) and (J-43)]
\begin{subequations}
\begin{eqnarray}
  p^{(\alpha)}_{0n}
  &=&
  \frac{10}{3}
  a_{\alpha}
  E_{\alpha}
  \sum_{p=0}^{\infty}
  \sum_{q=0}^{\infty}
  P^{(\alpha)}_{npq}
  t_{\alpha}^{p}
  t_{3-\alpha}^{q}
  ,\\
  v^{(\alpha)}_{0n}
  &=&
  \frac{10}{3}
  a_{\alpha}
  E_{\alpha}
  \sum_{p=0}^{\infty}
  \sum_{q=0}^{\infty}
  \frac{V^{(\alpha)}_{npq}}{2(2n+1)}
  t_{\alpha}^{p}
  t_{3-\alpha}^{q}
  ,
\end{eqnarray}
\end{subequations}
the initial conditions for $P_{npq}$ and $V_{npq}$ are given 
from Eqs. (\ref{eq:pmn-slip-init}) and (\ref{eq:vmn-slip-init}) by 
\begin{equation}
  P^{(\alpha)}_{n00}
  =
  \delta_{2n}
  \Gamma^{(\alpha)}_{2,5}
  ,\quad
  V^{(\alpha)}_{n00}
  =
  \delta_{2n}
  \Gamma^{(\alpha)}_{0,5}
  .
\end{equation}

The coefficient $f^{XM\alpha}_{k}$ is defined as
\begin{equation}
  f^{XM\alpha}_{k}
  =
  2^k
  \sum_{q=0}^{k}
  P^{(\alpha)}_{2(k-q)q}
  \lambda^{q+j}
  ,
\end{equation}
where $j=0$ for even $k$ and $j=1$ for odd $k$.
The explicit forms up to $k=7$ are 
\begin{subequations}
\begin{eqnarray}
  f^{XM1}_{0} &=& \left(
     \Gamma^{(1)}_{2,5} 
  \right),\\
  f^{XM1}_{1} &=& 0,\\
  f^{XM1}_{2} &=& 0,\\
  f^{XM1}_{3} &=& \lambda^{3} \left(
     40 \Gamma^{(1)}_{2,5} \Gamma^{(2)}_{2,5} 
  \right),\\
  f^{XM1}_{4} &=& \lambda \left(
     60 \Gamma^{(2)}_{2,3} (\Gamma^{(1)}_{2,5})^{2} 
  \right),\\
  f^{XM1}_{5} &=& \lambda^{3} \left(
     - 192 \Gamma^{(1)}_{0,5} \Gamma^{(2)}_{2,5} 
  \right)\nonumber\\
  &+& \lambda^{4} \left(
     180 \Gamma^{(1)}_{2,3} \Gamma^{(2)}_{2,3} \Gamma^{(1)}_{2,5} \Gamma^{(2)}_{2,5} 
  \right)\nonumber\\
  &+& \lambda^{5} \left(
     - 192 \Gamma^{(2)}_{0,5} \Gamma^{(1)}_{2,5} 
  \right),\\
  f^{XM1}_{6} &=& \lambda \left(
     - 288 \Gamma^{(1)}_{0,5} \Gamma^{(2)}_{2,3} \Gamma^{(1)}_{2,5} 
  \right)\nonumber\\
  &+& \lambda^{2} \left(
     540 \Gamma^{(1)}_{2,3} (\Gamma^{(2)}_{2,3})^{2} (\Gamma^{(1)}_{2,5})^{2} 
  \right)\nonumber\\
  &+& \lambda^{3} \left(
     160 (\Gamma^{(1)}_{2,5})^{2} ( 10 \Gamma^{(2)}_{2,5} 
     - 3 \Gamma^{(2)}_{0,3} ) 
  \right),\\
  f^{XM1}_{7} &=& \lambda^{4} \left(
     48 \Gamma^{(2)}_{2,3} ( 50 (\Gamma^{(1)}_{2,5})^{2} 
     - 20 \Gamma^{(1)}_{0,3} \Gamma^{(1)}_{2,5} 
     - 9 \Gamma^{(1)}_{0,5} \Gamma^{(1)}_{2,3} ) \Gamma^{(2)}_{2,5} 
  \right)\nonumber\\
  &+& \lambda^{5} \left(
     1620 (\Gamma^{(1)}_{2,3})^{2} (\Gamma^{(2)}_{2,3})^{2} \Gamma^{(1)}_{2,5} \Gamma^{(2)}_{2,5} 
  \right)\nonumber\\
  &+& \lambda^{6} \left(
     48 \Gamma^{(1)}_{2,3} \Gamma^{(1)}_{2,5} ( 50 (\Gamma^{(2)}_{2,5})^{2} 
     - 20 \Gamma^{(2)}_{0,3} \Gamma^{(2)}_{2,5} 
   \right.\nonumber\\
   &&\left.
     - 9 \Gamma^{(2)}_{0,5} \Gamma^{(2)}_{2,3} ) 
  \right).
\end{eqnarray}
\end{subequations}
The results are identical to those obtained by method of reflections 
in Eqs. (\ref{eq:mofr-fXM-1}), (\ref{eq:mofr-fXM-3}), and (\ref{eq:mofr-fXM-5}) 
for the terms containing one or two $\Gamma$'s. 
The results reduce to those by Jeffrey \cite{Jeffrey1992} 
in the no-slip limit $\widehat{\gamma} = 0$.

\subsection{$Y$ Functions ($m=1$)}

\subsubsection{$Y^{A}$ Functions}
The boundary condition for the $Y^{A}$ problem is given by
\begin{equation}
  \chi^{(\alpha)}_{mn}
  =
  (-1)^\alpha
  U
  \delta_{m1}
  \delta_{n1}
  ,\quad
  \psi^{(\alpha)}_{mn}
  =
  0
  ,\quad
  \omega^{(\alpha)}_{mn}
  =
  0
  .
\end{equation}
(Note that the equation by Jeffrey and Onishi \cite{JeffreyOnishi1984}, 
in p. 271, lost the factor $U$ for $\chi^{(\alpha)}_{mn}$.)
Again, we expand the coefficients by $t_{\alpha}^p$ and $t_{3-\alpha}^q$ as
\begin{subequations}
\begin{eqnarray}
  p^{(\alpha)}_{1n}
  &=&
  (-1)^\alpha
  \frac{3}{2}
  U
  \sum_{p=0}^{\infty}
  \sum_{q=0}^{\infty}
  P^{(\alpha)}_{npq}
  t_{\alpha}^{p}
  t_{3-\alpha}^{q}
  ,\\
  v^{(\alpha)}_{1n}
  &=&
  (-1)^\alpha
  \frac{3}{2}
  U
  \sum_{p=0}^{\infty}
  \sum_{q=0}^{\infty}
  \frac{V^{(\alpha)}_{npq}}{2(2n+1)}
  t_{\alpha}^{p}
  t_{3-\alpha}^{q}
  ,\\
  q^{(\alpha)}_{1n}
  &=&
  -
  {\rm i}
  U
  \sum_{p=0}^{\infty}
  \sum_{q=0}^{\infty}
  Q^{(\alpha)}_{npq}
  t_{\alpha}^{p}
  t_{3-\alpha}^{q}
  .
\end{eqnarray}
\end{subequations}
Also note that 
the minus sign in the right-hand side of (JO-4.5) is missing. 
Substituting these expansions into 
Eqs. (\ref{eq:pmn-slip-init}), (\ref{eq:vmn-slip-init}), 
and (\ref{eq:qmn-slip-init}), the initial conditions are given by 
\begin{equation}
  P^{(\alpha)}_{n00}
  =
  \delta_{n1}
  \Gamma^{(\alpha)}_{2,3}
  ,\quad
  V^{(\alpha)}_{n00}
  =
  \delta_{n1}
  \Gamma^{(\alpha)}_{0,3}
  ,\quad
  Q^{(\alpha)}_{n00}
  =
  0
  ,
\end{equation}
which correspond to (KC-37a,b,c). 
From Eqs. (\ref{eq:pmn-slip-rec}), (\ref{eq:vmn-slip-rec}), 
and (\ref{eq:qmn-slip-rec}), the recurrence relations are given by 
\begin{subequations}
\begin{eqnarray}
  P^{(\alpha)}_{npq}
  &=&
  \sum_{s=1}^{\infty}
  \left(
    \begin{array}{c}
      n+s \\ n+1
    \end{array}
  \right)
  \nonumber\\
  &&
  \times
  \left[
    -
    \frac{2}{3}
    \frac{(2n+1)(2n-1)}{n+1}
    \Gamma^{(\alpha)}_{2,2n+1}
    Q^{(3-\alpha)}_{s(q-s-1)(p-n+1)}
  \right.
  \nonumber\\
  &&
  \left.
    +
    \frac{n(2n+1)(2n-1)}{2(n+1)(2s+1)}
    \Gamma^{(\alpha)}_{2,2n+1}
    V^{(3-\alpha)}_{s(q-s-2)(p-n+1)}
  \right.
  \nonumber\\
  &&
  \left.
    +
    \frac{2n+1}{n+1}
    \frac{ns(n+s-2ns-2)-(2ns-4s-4n+2)}{2s(2s-1)(n+s)}
  \right.
  \nonumber\\
  &&
  \quad
  \left.
    \Gamma^{(\alpha)}_{2,2n+1}
    P^{(3-\alpha)}_{s(q-s)(p-n+1)}
  \right.
  \nonumber\\
  &&
  \left.
    +
    \frac{n(2n-1)}{2(n+1)}
    \Gamma^{(\alpha)}_{0,2n+1}
    P^{(3-\alpha)}_{s(q-s)(p-n-1)}
  \right]
  ,
  \label{eq:rec-Pnpq-Y}
\end{eqnarray}
\begin{eqnarray}
  &&
  V^{(\alpha)}_{npq}
  =
  \Gamma^{(\alpha)}_{0,2}
  P^{(\alpha)}_{npq}
  \label{eq:rec-Vnpq-Y}
  \\
  &&
  +
  \sum_{s=1}^{\infty}
  \left(
    \begin{array}{c}
      n+s \\ n+1
    \end{array}
  \right)
  \frac{2n}{(n+1)(2n+3)}
  \Gamma^{(\alpha)}_{-(2n+1),2}
  P^{(3-\alpha)}_{s(q-s)(p-n-1)}
  ,
  \nonumber
\end{eqnarray}
\begin{eqnarray}
  Q^{(\alpha)}_{npq}
  &=&
  \sum_{s=1}^{\infty}
  \left(
    \begin{array}{c}
      n+s \\ n+1
    \end{array}
  \right)
  \nonumber\\
  &&
  \times
  \left[
    \frac{s}{(n+1)}
    \Gamma^{(\alpha)}_{-(n-1),n+2}
    Q^{(3-\alpha)}_{s(q-s-1)(p-n)}
  \right.
  \nonumber\\
  &&
  \left.
    -
    \frac{3}{2}
    \frac{1}{ns(n+1)}
    \Gamma^{(\alpha)}_{-(n-1),n+2}
    P_{s(q-s)(p-n)}
  \right]
  .
  \label{eq:rec-Qnpq-Y}
\end{eqnarray}
\end{subequations}
Note that Eqs. (\ref{eq:rec-Pnpq-Y}) and (\ref{eq:rec-Qnpq-Y}) 
correspond to (KC-38a) and (KC-38b), while 
Eq. (\ref{eq:rec-Vnpq-Y}) is simpler than Eq. (KC-38c). 
The coefficient $f^{YA\alpha}_{k}$ is defined as
\begin{equation}
  f^{YA\alpha}_{k}
  =
  2^k
  \sum_{q=0}
  P^{(\alpha)}_{1(k-q)q}
  \lambda^{q}
  .
\end{equation}
The explicit forms up to $k=7$ are 
\begin{subequations}
\begin{eqnarray}
  f^{YA1}_{0} &=& \left(
     \Gamma^{(1)}_{2,3} 
  \right),\\
  f^{YA1}_{1} &=& \lambda \left(
     3 \Gamma^{(1)}_{2,3} \Gamma^{(2)}_{2,3} / 2 
  \right),\\
  f^{YA1}_{2} &=& \lambda \left(
     9 (\Gamma^{(1)}_{2,3})^{2} \Gamma^{(2)}_{2,3} / 4 
  \right),\\
  f^{YA1}_{3} &=& \lambda \left(
     2 \Gamma^{(1)}_{0,3} \Gamma^{(2)}_{2,3} 
  \right)\nonumber\\
  &+& \lambda^{2} \left(
     27 (\Gamma^{(1)}_{2,3})^{2} (\Gamma^{(2)}_{2,3})^{2} / 8 
  \right)\nonumber\\
  &+& \lambda^{3} \left(
     2 \Gamma^{(2)}_{0,3} \Gamma^{(1)}_{2,3} 
  \right),\\
  f^{YA1}_{4} &=& \lambda \left(
     6 \Gamma^{(1)}_{0,3} \Gamma^{(1)}_{2,3} \Gamma^{(2)}_{2,3} 
  \right)\nonumber\\
  &+& \lambda^{2} \left(
     81 (\Gamma^{(1)}_{2,3})^{3} (\Gamma^{(2)}_{2,3})^{2} / 16 
  \right)\nonumber\\
  &+& \lambda^{3} \left(
     18 \Gamma^{(2)}_{0,3} (\Gamma^{(1)}_{2,3})^{2} 
  \right),\\
  f^{YA1}_{5} &=& \lambda^{2} \left(
     63 \Gamma^{(1)}_{0,3} \Gamma^{(1)}_{2,3} (\Gamma^{(2)}_{2,3})^{2} / 2 
  \right)\nonumber\\
  &+& \lambda^{3} \left(
     243 (\Gamma^{(1)}_{2,3})^{3} (\Gamma^{(2)}_{2,3})^{3} / 32 
  \right)\nonumber\\
  &+& \lambda^{4} \left(
     63 \Gamma^{(2)}_{0,3} (\Gamma^{(1)}_{2,3})^{2} \Gamma^{(2)}_{2,3} / 2 
  \right),\\
  f^{YA1}_{6} &=& \lambda \left(
     4 (\Gamma^{(1)}_{0,3})^{2} \Gamma^{(2)}_{2,3} 
  \right)\nonumber\\
  &+& \lambda^{2} \left(
     54 \Gamma^{(1)}_{0,3} (\Gamma^{(1)}_{2,3})^{2} (\Gamma^{(2)}_{2,3})^{2} 
  \right)\nonumber\\
  &+& \lambda^{3} \left(
     \Gamma^{(1)}_{2,3} ( 729 (\Gamma^{(1)}_{2,3})^{3} (\Gamma^{(2)}_{2,3})^{3} 
     + 512 \Gamma^{(1)}_{0,3} \Gamma^{(2)}_{0,3} ) / 64 
  \right)\nonumber\\
  &+& \lambda^{4} \left(
     81 \Gamma^{(2)}_{0,3} (\Gamma^{(1)}_{2,3})^{3} \Gamma^{(2)}_{2,3} 
  \right)\nonumber\\
  &+& \lambda^{5} \left(
     4 (\Gamma^{(1)}_{2,3})^{2} ( 21 \Gamma^{(2)}_{2,7} 
     + 5 \Gamma^{(2)}_{0,2} \Gamma^{(2)}_{0,3} 
     + 60 \Gamma^{(2)}_{-1,4} 
   \right.\nonumber\\
   &&\left.
     + 4 \Gamma^{(2)}_{-3,2} ) / 5 
  \right),\\
  f^{YA1}_{7} &=& \lambda^{2} \left(
     6 (\Gamma^{(2)}_{2,3})^{2} ( 21 \Gamma^{(1)}_{2,3} \Gamma^{(1)}_{2,7} 
     + 60 \Gamma^{(1)}_{-1,4} \Gamma^{(1)}_{2,3} 
     + 35 (\Gamma^{(1)}_{0,3})^{2} 
   \right.\nonumber\\
   &&\left.
     + 4 \Gamma^{(1)}_{-3,3} ) / 5 
  \right)\nonumber\\
  &+& \lambda^{3} \left(
     1053 \Gamma^{(1)}_{0,3} (\Gamma^{(1)}_{2,3})^{2} (\Gamma^{(2)}_{2,3})^{3} / 8 
  \right)\nonumber\\
  &+& \lambda^{4} \left(
     3 \Gamma^{(1)}_{2,3} \Gamma^{(2)}_{2,3} ( 729 (\Gamma^{(1)}_{2,3})^{3} (\Gamma^{(2)}_{2,3})^{3} 
     + 5632 \Gamma^{(1)}_{0,3} \Gamma^{(2)}_{0,3} ) / 128 
  \right)\nonumber\\
  &+& \lambda^{5} \left(
     1053 \Gamma^{(2)}_{0,3} (\Gamma^{(1)}_{2,3})^{3} (\Gamma^{(2)}_{2,3})^{2} / 8 
  \right)\nonumber\\
  &+& \lambda^{6} \left(
     6 (\Gamma^{(1)}_{2,3})^{2} ( 21 \Gamma^{(2)}_{2,3} \Gamma^{(2)}_{2,7} 
     + 60 \Gamma^{(2)}_{-1,4} \Gamma^{(2)}_{2,3} 
     + 35 (\Gamma^{(2)}_{0,3})^{2} 
   \right.\nonumber\\
   &&\left.
     + 4 \Gamma^{(2)}_{-3,3} ) / 5 
  \right).
\end{eqnarray}
\end{subequations}
The results are identical to those obtained by method of reflections 
in Eqs. (\ref{eq:mofr-fYA-0}), (\ref{eq:mofr-fYA-1}), and (\ref{eq:mofr-fYA-3}) 
for the terms containing one or two $\Gamma$'s. 
The results reduce to those by Jeffrey and Onishi \cite{JeffreyOnishi1984} 
in the no-slip limit $\widehat{\gamma} = 0$ 
and those by Keh and Chen \cite{KehChen1997} 
in the case of $\widehat{\gamma}_{1} = \widehat{\gamma}_{2}$.

\subsubsection{$Y^{B}$ Functions}
The problem for $Y^{B}$ is exactly the same as for $Y^{A}$.
The difference is that the force is calculated in $Y^{A}$ while
the torque in $Y^{B}$. Correspondingly, 
The coefficient $f^{YB\alpha}_{k}$ is defined as
\begin{equation}
  f^{YB\alpha}_{k}
  =
  2\ 
  2^k
  \sum_{q=0}
  Q^{(\alpha)}_{1(k-q)q}
  \lambda^{q}
  ,
\end{equation}
for $Q_{1pq}$ obtained by the recurrence relations 
for $Y^{A}$.
The explicit forms up to $k=7$ are 
\begin{subequations}
\begin{eqnarray}
  f^{YB1}_{0} &=& 0,\\
  f^{YB1}_{1} &=& 0,\\
  f^{YB1}_{2} &=& \lambda \left(
     - 6 \Gamma^{(1)}_{0,3} \Gamma^{(2)}_{2,3} 
  \right),\\
  f^{YB1}_{3} &=& \lambda \left(
     - 9 \Gamma^{(1)}_{0,3} \Gamma^{(1)}_{2,3} \Gamma^{(2)}_{2,3} 
  \right),\\
  f^{YB1}_{4} &=& \lambda^{2} \left(
     - 27 \Gamma^{(1)}_{0,3} \Gamma^{(1)}_{2,3} (\Gamma^{(2)}_{2,3})^{2} / 2 
  \right),\\
  f^{YB1}_{5} &=& \lambda \left(
     - 12 (\Gamma^{(1)}_{0,3})^{2} \Gamma^{(2)}_{2,3} 
  \right)\nonumber\\
  &+& \lambda^{2} \left(
     - 81 \Gamma^{(1)}_{0,3} (\Gamma^{(1)}_{2,3})^{2} (\Gamma^{(2)}_{2,3})^{2} / 4 
  \right)\nonumber\\
  &+& \lambda^{3} \left(
     - 36 \Gamma^{(1)}_{0,3} \Gamma^{(2)}_{0,3} \Gamma^{(1)}_{2,3} 
  \right),\\
  f^{YB1}_{6} &=& \lambda^{2} \left(
     - 108 (\Gamma^{(1)}_{0,3})^{2} (\Gamma^{(2)}_{2,3})^{2} 
  \right)\nonumber\\
  &+& \lambda^{3} \left(
     - 243 \Gamma^{(1)}_{0,3} (\Gamma^{(1)}_{2,3})^{2} (\Gamma^{(2)}_{2,3})^{3} / 8 
  \right)\nonumber\\
  &+& \lambda^{4} \left(
     - 72 \Gamma^{(1)}_{0,3} \Gamma^{(2)}_{0,3} \Gamma^{(1)}_{2,3} \Gamma^{(2)}_{2,3} 
  \right),\\
  f^{YB1}_{7} &=& \lambda^{2} \left(
     - 189 (\Gamma^{(1)}_{0,3})^{2} \Gamma^{(1)}_{2,3} (\Gamma^{(2)}_{2,3})^{2} 
  \right)\nonumber\\
  &+& \lambda^{3} \left(
     - 3 \Gamma^{(1)}_{0,3} ( 2560 \Gamma^{(1)}_{0,3} \Gamma^{(2)}_{2,5} 
     + 243 (\Gamma^{(1)}_{2,3})^{3} (\Gamma^{(2)}_{2,3})^{3} ) / 16 
  \right)\nonumber\\
  &+& \lambda^{4} \left(
     - 243 \Gamma^{(1)}_{0,3} \Gamma^{(2)}_{0,3} (\Gamma^{(1)}_{2,3})^{2} \Gamma^{(2)}_{2,3} 
  \right)\nonumber\\
  &+& \lambda^{5} \left(
     48 \Gamma^{(1)}_{0,3} \Gamma^{(1)}_{2,3} ( 7 \Gamma^{(2)}_{2,7} 
     - 6 \Gamma^{(2)}_{0,5} 
     - 4 \Gamma^{(2)}_{-1,4} ) 
  \right).
\end{eqnarray}
\end{subequations}
The results are identical to those obtained by method of reflections 
in Eqs. (\ref{eq:mofr-fYB-0}), (\ref{eq:mofr-fYB-1}), and (\ref{eq:mofr-fYB-2}) 
for the terms containing one or two $\Gamma$'s. 
The results reduce to those by Jeffrey and Onishi \cite{JeffreyOnishi1984} 
in the no-slip limit $\widehat{\gamma} = 0$ 
and those by Keh and Chen \cite{KehChen1997} 
in the case of $\widehat{\gamma}_{1} = \widehat{\gamma}_{2}$.

\subsubsection{$Y^{G}$ Function}
With the same recurrence relations and the initial condition for $Y^{A}$, 
that is, for the translating particles, 
the function $Y^{G}$ is obtained from the coefficient $P_{2pq}$ for 
the stresslet instead of $P_{1pq}$ for the force.

In this case, the coefficient $f^{YG\alpha}_{k}$ is defined as
\begin{equation}
  f^{YG\alpha}_{k}
  =
  \left(
    \frac{3}{4}
  \right)
  2^k
  \sum_{q=0}^{k}
  P^{(\alpha)}_{2(k-q)q}
  \lambda^{q}
  .
\end{equation}
The explicit forms up to $k=7$ are 
\begin{subequations}
\begin{eqnarray}
  f^{YG1}_{0} &=& 0,\\
  f^{YG1}_{1} &=& 0,\\
  f^{YG1}_{2} &=& 0,\\
  f^{YG1}_{3} &=& 0,\\
  f^{YG1}_{4} &=& \lambda \left(
     12 \Gamma^{(1)}_{0,5} \Gamma^{(2)}_{2,3} 
  \right)\nonumber\\
  &+& \lambda^{3} \left(
     20 \Gamma^{(2)}_{0,3} \Gamma^{(1)}_{2,5} 
  \right),\\
  f^{YG1}_{5} &=& \lambda \left(
     18 \Gamma^{(1)}_{0,5} \Gamma^{(1)}_{2,3} \Gamma^{(2)}_{2,3} 
  \right)\nonumber\\
  &+& \lambda^{3} \left(
     90 \Gamma^{(2)}_{0,3} \Gamma^{(1)}_{2,3} \Gamma^{(1)}_{2,5} 
  \right),\\
  f^{YG1}_{6} &=& \lambda^{2} \left(
     27 \Gamma^{(1)}_{0,5} \Gamma^{(1)}_{2,3} (\Gamma^{(2)}_{2,3})^{2} 
  \right)\nonumber\\
  &+& \lambda^{4} \left(
     135 \Gamma^{(2)}_{0,3} \Gamma^{(1)}_{2,3} \Gamma^{(2)}_{2,3} \Gamma^{(1)}_{2,5} 
  \right),\\
  f^{YG1}_{7} &=& \lambda \left(
     24 \Gamma^{(1)}_{0,3} \Gamma^{(1)}_{0,5} \Gamma^{(2)}_{2,3} 
  \right)\nonumber\\
  &+& \lambda^{2} \left(
     81 \Gamma^{(1)}_{0,5} (\Gamma^{(1)}_{2,3})^{2} (\Gamma^{(2)}_{2,3})^{2} / 2 
  \right)\nonumber\\
  &+& \lambda^{3} \left(
     - 8 ( 50 \Gamma^{(1)}_{0,3} \Gamma^{(1)}_{2,5} \Gamma^{(2)}_{2,5} 
     - 5 \Gamma^{(1)}_{0,3} \Gamma^{(2)}_{0,3} \Gamma^{(1)}_{2,5} 
     - 3 \Gamma^{(2)}_{0,3} \Gamma^{(1)}_{0,5} \Gamma^{(1)}_{2,3} ) 
  \right)\nonumber\\
  &+& \lambda^{4} \left(
     405 \Gamma^{(2)}_{0,3} (\Gamma^{(1)}_{2,3})^{2} \Gamma^{(2)}_{2,3} \Gamma^{(1)}_{2,5} / 2 
  \right)\nonumber\\
  &+& \lambda^{5} \left(
     8 \Gamma^{(1)}_{2,3} \Gamma^{(1)}_{2,5} ( 56 \Gamma^{(2)}_{2,7} 
     - 30 \Gamma^{(2)}_{0,5} 
     + 5 \Gamma^{(2)}_{0,2} \Gamma^{(2)}_{0,3} 
     + 40 \Gamma^{(2)}_{-1,4} 
   \right.\nonumber\\
   &&\left.
     + 4 \Gamma^{(2)}_{-3,2} ) 
  \right).
\end{eqnarray}
\end{subequations}
The results are identical to those obtained by method of reflections 
in Eqs. (\ref{eq:mofr-fYG-0}), (\ref{eq:mofr-fYG-2}), and (\ref{eq:mofr-fYG-4}) 
for the terms containing one or two $\Gamma$'s. 
The results reduce to those by Jeffrey \cite{Jeffrey1992} 
in the no-slip limit $\widehat{\gamma} = 0$.

\subsubsection{$Y^{C}$ Function}
The function $Y^{C}$ gives the torque for the rotating particles 
with $m=1$. Therefore, it is derived by 
the coefficient $Q_{1pq}$ for the torque from the same recurrence 
relations as for $Y^{A}$, but with different initial condition 
\begin{equation}
  P^{(\alpha)}_{n00}
  =
  0
  ,\quad
  V^{(\alpha)}_{n00}
  =
  0
  ,\quad
  Q^{(\alpha)}_{n00}
  =
  \delta_{1n}
  \Gamma^{(\alpha)}_{0,3}
  .
\end{equation}

In this case, the coefficient $f^{YC\alpha}_{k}$ is defined as
\begin{equation}
  f^{YC\alpha}_{k}
  =
  2^k
  \sum_{q=0}^{k}
  Q^{(\alpha)}_{1(k-q)q}
  \lambda^{q+j}
  ,
\end{equation}
where $j=0$ for even $k$ and $j=1$ for odd $k$.
The explicit forms up to $k=7$ are 
\begin{subequations}
\begin{eqnarray}
  f^{YC1}_{0} &=& \left(
     \Gamma^{(1)}_{0,3} 
  \right),\\
  f^{YC1}_{1} &=& 0,\\
  f^{YC1}_{2} &=& 0,\\
  f^{YC1}_{3} &=& \lambda^{3} \left(
     4 \Gamma^{(1)}_{0,3} \Gamma^{(2)}_{0,3} 
  \right),\\
  f^{YC1}_{4} &=& \lambda \left(
     12 (\Gamma^{(1)}_{0,3})^{2} \Gamma^{(2)}_{2,3} 
  \right),\\
  f^{YC1}_{5} &=& \lambda^{4} \left(
     18 \Gamma^{(1)}_{0,3} \Gamma^{(2)}_{0,3} \Gamma^{(1)}_{2,3} \Gamma^{(2)}_{2,3} 
  \right),\\
  f^{YC1}_{6} &=& \lambda^{2} \left(
     27 (\Gamma^{(1)}_{0,3})^{2} \Gamma^{(1)}_{2,3} (\Gamma^{(2)}_{2,3})^{2} 
  \right)\nonumber\\
  &+& \lambda^{3} \left(
     16 (\Gamma^{(1)}_{0,3})^{2} ( 15 \Gamma^{(2)}_{2,5} 
     + \Gamma^{(2)}_{0,3} ) 
  \right),\\
  f^{YC1}_{7} &=& \lambda^{4} \left(
     72 (\Gamma^{(1)}_{0,3})^{2} \Gamma^{(2)}_{0,3} \Gamma^{(2)}_{2,3} 
  \right)\nonumber\\
  &+& \lambda^{5} \left(
     81 \Gamma^{(1)}_{0,3} \Gamma^{(2)}_{0,3} (\Gamma^{(1)}_{2,3})^{2} (\Gamma^{(2)}_{2,3})^{2} / 2 
  \right)\nonumber\\
  &+& \lambda^{6} \left(
     72 \Gamma^{(1)}_{0,3} (\Gamma^{(2)}_{0,3})^{2} \Gamma^{(1)}_{2,3} 
  \right).
\end{eqnarray}
\end{subequations}
The results are identical to those obtained by method of reflections 
in Eqs. (\ref{eq:mofr-fYC-1}) and (\ref{eq:mofr-fYC-3}) 
for the terms containing one or two $\Gamma$'s. 
The results reduce to those by Jeffrey and Onishi \cite{JeffreyOnishi1984} 
in the no-slip limit $\widehat{\gamma} = 0$ 
and those by Keh and Chen \cite{KehChen1997} 
in the case of $\widehat{\gamma}_{1} = \widehat{\gamma}_{2}$.

\subsubsection{$Y^{H}$ Function}
With the same recurrence relations and the initial condition for $Y^{C}$, 
that is, for the rotating particles, 
the function $Y^{H}$ is obtained from the coefficient $P_{2pq}$ for 
the stresslet instead of $Q_{1pq}$ for the torque.

In this case, the coefficient $f^{YH\alpha}_{k}$ is defined as
\begin{equation}
  f^{YH\alpha}_{k}
  =
  -
  \left(
    \frac{3}{8}
  \right)
  2^k
  \sum_{q=0}^{k}
  P^{(\alpha)}_{2(k-q)q}
  \lambda^{q+j}
  ,
\end{equation}
where $j=0$ for even $k$ and $j=1$ for odd $k$.
The explicit forms up to $k=7$ are 
\begin{subequations}
\begin{eqnarray}
  f^{YH1}_{0} &=& 0,\\
  f^{YH1}_{1} &=& 0,\\
  f^{YH1}_{2} &=& 0,\\
  f^{YH1}_{3} &=& \lambda^{3} \left(
     10 \Gamma^{(2)}_{0,3} \Gamma^{(1)}_{2,5} 
  \right),\\
  f^{YH1}_{4} &=& 0,\\
  f^{YH1}_{5} &=& 0,\\
  f^{YH1}_{6} &=& \lambda \left(
     24 \Gamma^{(1)}_{0,3} \Gamma^{(1)}_{0,5} \Gamma^{(2)}_{2,3} 
  \right)\nonumber\\
  &+& \lambda^{3} \left(
     - 40 \Gamma^{(1)}_{0,3} \Gamma^{(1)}_{2,5} ( 5 \Gamma^{(2)}_{2,5} 
     - 2 \Gamma^{(2)}_{0,3} ) 
  \right),\\
  f^{YH1}_{7} &=& \lambda^{4} \left(
     36 \Gamma^{(2)}_{0,3} \Gamma^{(1)}_{0,5} \Gamma^{(1)}_{2,3} \Gamma^{(2)}_{2,3} 
  \right)\nonumber\\
  &+& \lambda^{6} \left(
     180 (\Gamma^{(2)}_{0,3})^{2} \Gamma^{(1)}_{2,3} \Gamma^{(1)}_{2,5} 
  \right).
\end{eqnarray}
\end{subequations}
The results are identical to those obtained by method of reflections 
in Eqs. (\ref{eq:mofr-fYH-1}) and (\ref{eq:mofr-fYH-3}) 
for the terms containing one or two $\Gamma$'s. 
The results reduce to those by Jeffrey \cite{Jeffrey1992} 
in the no-slip limit $\widehat{\gamma} = 0$.

\subsubsection{$Y^{M}$ Function}
The function $Y^{M}$ gives the stresslet under a strain flow
for $m=1$. Therefore, it is derived by 
the coefficient $P_{2pq}$ for the stresslet from the same recurrence 
relations as for $Y^{A}$, but with different initial condition. 
The boundary conditions are
\begin{subequations}
\begin{eqnarray}
  \chi^{(\alpha)}_{mn}
  &=&
  \frac{2}{3}
  (-1)^\alpha
  a_{\alpha}
  E_{\alpha}
  \delta_{1m}
  \delta_{2n}
  ,\\
  \psi^{(\alpha)}_{mn}
  &=&
  \frac{2}{3}
  (-1)^\alpha
  a_{\alpha}
  E_{\alpha}
  (1-6\widehat{\gamma})
  \delta_{1m}
  \delta_{2n}
  ,\\
  \omega^{(\alpha)}_{mn}
  &=&
  0
  ,
\end{eqnarray}
\end{subequations}
which correspond to Eq. (J-54) with the correction due to the slip. 
The expansions used here are [in (J-55), (J-56), and (J-57)] 
\begin{subequations}
\begin{eqnarray}
  p^{(\alpha)}_{1n}
  &=&
  (-1)^\alpha
  \frac{10}{3}
  a_{\alpha}
  E_{\alpha}
  \sum_{p=0}^{\infty}
  \sum_{q=0}^{\infty}
  P_{npq}
  t_{\alpha}^{p}
  t_{3-\alpha}^{q}
  ,\\
  v^{(\alpha)}_{1n}
  &=&
  (-1)^\alpha
  \frac{10}{3}
  a_{\alpha}
  E_{\alpha}
  \sum_{p=0}^{\infty}
  \sum_{q=0}^{\infty}
  \frac{V_{npq}}{2(2n+1)}
  t_{\alpha}^{p}
  t_{3-\alpha}^{q}
  ,\\
  q^{(\alpha)}_{1n}
  &=&
  -
  {\rm i}
  \frac{10}{3}
  a_{\alpha}
  E_{\alpha}
  \sum_{p=0}^{\infty}
  \sum_{q=0}^{\infty}
  Q_{npq}
  t_{\alpha}^{p}
  t_{3-\alpha}^{q}
  .
\end{eqnarray}
\end{subequations}
The initial conditions are given 
from Eqs. (\ref{eq:pmn-slip-init}), (\ref{eq:vmn-slip-init}), 
and (\ref{eq:qmn-slip-init}) by 
\begin{equation}
  P^{(\alpha)}_{n00}
  =
  \delta_{n2}
  \Gamma^{(\alpha)}_{2,5}
  ,\quad
  V^{(\alpha)}_{n00}
  =
  \delta_{n2}
  \Gamma^{(\alpha)}_{0,5}
  ,\quad
  Q^{(\alpha)}_{n00}
  =
  0
  .
\end{equation}

In this case, the coefficient $f^{YM\alpha}_{k}$ is defined as
\begin{equation}
  f^{YM\alpha}_{k}
  =
  2^k
  \sum_{q=0}^{k}
  P^{(\alpha)}_{2(k-q)q}
  \lambda^{q+j}
  ,
\end{equation}
where $j=0$ for even $k$ and $j=1$ for odd $k$.
The explicit forms up to $k=7$ are 
\begin{subequations}
\begin{eqnarray}
  f^{YM1}_{0} &=& \left(
     \Gamma^{(1)}_{2,5} 
  \right),\\
  f^{YM1}_{1} &=& 0,\\
  f^{YM1}_{2} &=& 0,\\
  f^{YM1}_{3} &=& \lambda^{3} \left(
     - 20 \Gamma^{(1)}_{2,5} \Gamma^{(2)}_{2,5} 
  \right),\\
  f^{YM1}_{4} &=& 0,\\
  f^{YM1}_{5} &=& \lambda^{3} \left(
     128 \Gamma^{(1)}_{0,5} \Gamma^{(2)}_{2,5} 
  \right)\nonumber\\
  &+& \lambda^{5} \left(
     128 \Gamma^{(2)}_{0,5} \Gamma^{(1)}_{2,5} 
  \right),\\
  f^{YM1}_{6} &=& \lambda^{3} \left(
     80 (\Gamma^{(1)}_{2,5})^{2} ( 5 \Gamma^{(2)}_{2,5} 
     + 3 \Gamma^{(2)}_{0,3} ) 
  \right),\\
  f^{YM1}_{7} &=& 0.
\end{eqnarray}
\end{subequations}
The results are identical to those obtained by method of reflections 
in Eqs. (\ref{eq:mofr-fYM-1}), (\ref{eq:mofr-fYM-3}), and (\ref{eq:mofr-fYM-5}) 
for the terms containing one or two $\Gamma$'s. 
The results reduce to those by Jeffrey \cite{Jeffrey1992} 
in the no-slip limit $\widehat{\gamma} = 0$.

\subsection{$Z$ Functions ($m=2$)}
The boundary conditions are given by 
\begin{subequations}
\begin{eqnarray}
  \chi^{(\alpha)}_{mn}
  &=&
  \frac{1}{3}
  (-1)^{3-\alpha}
  a_{\alpha}
  E_{\alpha}
  \delta_{2m}
  \delta_{2n}
  ,\\
  \psi^{(\alpha)}_{mn}
  &=&
  \frac{1}{3}
  (-1)^{3-\alpha}
  a_{\alpha}
  E_{\alpha}
  (1-6\widehat{\gamma})
  \delta_{2m}
  \delta_{2n}
  ,\\
  \omega^{(\alpha)}_{mn}
  &=&
  0
  ,
\end{eqnarray}
\end{subequations}
which correspond to Eq. (J-69) with the correction due to the slip. 
The expansions used here are [in (J-70), (J-71), and (J-72)]
\begin{subequations}
\begin{eqnarray}
  p^{(\alpha)}_{2n}
  &=&
  (-1)^{3-\alpha}
  \frac{5}{3}
  a_{\alpha}
  E_{\alpha}
  \sum_{p=0}^{\infty}
  \sum_{q=0}^{\infty}
  P_{npq}
  t_{\alpha}^{p}
  t_{3-\alpha}^{q}
  ,\\
  v^{(\alpha)}_{2n}
  &=&
  (-1)^{3-\alpha}
  \frac{5}{3}
  a_{\alpha}
  E_{\alpha}
  \sum_{p=0}^{\infty}
  \sum_{q=0}^{\infty}
  \frac{V_{npq}}{2(2n+1)}
  t_{\alpha}^{p}
  t_{3-\alpha}^{q}
  ,\\
  q^{(\alpha)}_{2n}
  &=&
  {\rm i}
  \frac{5}{3}
  a_{\alpha}
  E_{\alpha}
  \sum_{p=0}^{\infty}
  \sum_{q=0}^{\infty}
  Q_{npq}
  t_{\alpha}^{p}
  t_{3-\alpha}^{q}
  .
\end{eqnarray}
\end{subequations}
From Eqs. (\ref{eq:pmn-slip-rec}), (\ref{eq:vmn-slip-rec}), 
and (\ref{eq:qmn-slip-rec}) for $m=2$ and the expansions above, 
the recurrence relations are given by 
\begin{subequations}
\begin{eqnarray}
  P^{(\alpha)}_{npq}
  &=&
  \sum_{s=2}^{\infty}
  \left(
    \begin{array}{c}
      n+s \\ n+2
    \end{array}
  \right)
  \nonumber\\
  &&
  \times
  \left[
    -
    \frac{2(2n+1)(2n-1)}{n+1}
    \Gamma^{(\alpha)}_{2,2n+1}
    Q^{(3-\alpha)}_{s(q-s-1)(p-n+1)}
  \right.
  \nonumber\\
  &&
  \left.
    +
    \frac{n(2n+1)(2n-1)}{2(n+1)(2s+1)}
    \Gamma^{(\alpha)}_{2,2n+1}
    V^{(3-\alpha)}_{s(q-s-2)(p-n+1)}
  \right.
  \nonumber\\
  &&
  \left.
    +
    \frac{2n+1}{n+1}
    \frac{ns(n+s-2ns-2)-2^2(2ns-4s-4n+2)}{2s(2s-1)(n+s)}
  \right.
  \nonumber\\
  &&
  \quad
  \left.
    \times
    \Gamma^{(\alpha)}_{2,2n+1}
    P^{(3-\alpha)}_{s(q-s)(p-n+1)}
  \right.
  \nonumber\\
  &&
  \left.
    +
    \frac{n(2n-1)}{2(n+1)}
    \Gamma^{(\alpha)}_{0,2n+1}
    P^{(3-\alpha)}_{s(p-s)(p-n-1)}
  \right]
  ,\\
  V^{(\alpha)}_{npq}
  &=&
  \Gamma^{(\alpha)}_{0,2}
  P^{(\alpha)}_{npq}
  +
  \sum_{s=2}^{\infty}
  \left(
    \begin{array}{c}
      n+s \\ n+2
    \end{array}
  \right)
  \nonumber\\
  &&
  \times
  \frac{2n}{(n+1)(2n+3)}
  \Gamma^{(\alpha)}_{-(2n+1),2}
  P^{(3-\alpha)}_{s(q-s)(p-n-1)}
  ,\\
  Q^{(\alpha)}_{npq}
  &=&
  \sum_{s=2}^{\infty}
  \left(
    \begin{array}{c}
      n+s \\ n+2
    \end{array}
  \right)
  \nonumber\\
  &&
  \times
  \left[
    \frac{s}{(n+1)}
    \Gamma^{(\alpha)}_{-(n-1),n+2}
    Q^{(3-\alpha)}_{s(q-s-1)(p-n)}
  \right.
  \nonumber\\
  &&
  \left.
    -
    \frac{2}{ns(n+1)}
    \Gamma^{(\alpha)}_{-(n-1),n+2}
    P^{(3-\alpha)}_{s(q-s)(p-n)}
  \right]
  .
\end{eqnarray}
\end{subequations}
The  initial conditions are obtained 
from Eqs. (\ref{eq:pmn-slip-init}), (\ref{eq:vmn-slip-init}), 
and (\ref{eq:qmn-slip-init}) as 
\begin{equation}
  P^{(\alpha)}_{n00}
  =
  \delta_{2n}
  \Gamma^{(\alpha)}_{2,5}
  ,\quad
  V^{(\alpha)}_{n00}
  =
  \delta_{2n}
  \Gamma^{(\alpha)}_{0,5}
  ,\quad
  Q^{(\alpha)}_{n00}
  =
  0
  .
\end{equation}

In this case, the coefficient $f^{ZM\alpha}_{k}$ is defined as
\begin{equation}
  f^{ZM\alpha}_{k}
  =
  2^k
  \sum_{q=0}^{k}
  P^{(\alpha)}_{2(k-q)q}
  \lambda^{q+j}
  ,
\end{equation}
where $j=0$ for even $k$ and $j=1$ for odd $k$.
The explicit forms up to $k=11$ are 
\begin{subequations}
\begin{eqnarray}
  f^{ZM1}_{0} &=& \left(
     \Gamma^{(1)}_{2,5} 
  \right),\\
  f^{ZM1}_{1} &=& 0,\\
  f^{ZM1}_{2} &=& 0,\\
  f^{ZM1}_{3} &=& 0,\\
  f^{ZM1}_{4} &=& 0,\\
  f^{ZM1}_{5} &=& \lambda^{3} \left(
     32 \Gamma^{(1)}_{0,5} \Gamma^{(2)}_{2,5} 
  \right)\nonumber\\
  &+& \lambda^{5} \left(
     32 \Gamma^{(2)}_{0,5} \Gamma^{(1)}_{2,5} 
  \right),\\
  f^{ZM1}_{6} &=& 0,\\
  f^{ZM1}_{7} &=& 0,\\
  f^{ZM1}_{8} &=& \lambda^{5} \left(
     160 (\Gamma^{(1)}_{2,5})^{2} ( 7 \Gamma^{(2)}_{2,7} 
     + 8 \Gamma^{(2)}_{-1,4} ) / 3 
  \right),\\
  f^{ZM1}_{9} &=& 0,\\
  f^{ZM1}_{10} &=& \lambda^{3} \left(
     1024 (\Gamma^{(1)}_{0,5})^{2} \Gamma^{(2)}_{2,5} 
  \right)\nonumber\\
  &+& \lambda^{5} \left(
     - 256 \Gamma^{(1)}_{0,5} \Gamma^{(1)}_{2,5} ( 35 \Gamma^{(2)}_{2,7} 
     - 8 \Gamma^{(2)}_{0,5} ) 
  \right)\nonumber\\
  &+& \lambda^{7} \left(
     128 (\Gamma^{(1)}_{2,5})^{2} ( 1620 \Gamma^{(2)}_{2,9} 
     - 525 \Gamma^{(2)}_{0,2} \Gamma^{(2)}_{2,7} 
     - 525 \Gamma^{(2)}_{0,7} 
  \right.\nonumber\\
  &&\left.
     + 168 \Gamma^{(2)}_{0,2} \Gamma^{(2)}_{0,5} 
     + 700 \Gamma^{(2)}_{-2,5} 
     + 32 \Gamma^{(2)}_{-5,2} ) / 21 
  \right),\\
  f^{ZM1}_{11} &=& 0.
\end{eqnarray}
\end{subequations}
The results are identical to those obtained by method of reflections 
in Eqs. (\ref{eq:mofr-fZM-1}), (\ref{eq:mofr-fZM-3}), and (\ref{eq:mofr-fZM-5}) 
for the terms containing one or two $\Gamma$'s. 
The results reduce to those by Jeffrey \cite{Jeffrey1992} 
in the no-slip limit $\widehat{\gamma} = 0$.

\section{Concluding Remarks}
\label{sec:concluding-remarks}
We have extended the calculations of resistance functions 
of two spheres with arbitrary size by the method of twin multipole expansions 
in general linear flows by Jeffrey and Onishi \cite{JeffreyOnishi1984} 
and Jeffrey \cite{Jeffrey1992} 
to the slip particles with the Navier slip boundary condition 
with arbitrary slip lengths. 
This extension complements the previous results of slip particles 
obtained by Keh and Chen \cite{KehChen1997} 
for the same scaled slip lengths without strain flow. 
In limiting cases, the present calculations recover the existing results, 
that is, 
those by Jeffrey \textit{et al.} \cite{JeffreyOnishi1984,Jeffrey1992} 
in the no-slip limit, 
and those by Keh and Chen \cite{KehChen1997} 
in the case of equal scaled slip lengths. 
We have also derived the resistance functions 
by the method of reflections and demonstrated its consistency 
with the twin multipole expansions. 

The present solutions of two-sphere problem cover 
much wider range than the previous solutions. 
Because the particle radii and slip lengths can be chosen independently, 
the solutions are not only applicable to the problem of two bubbles 
(demonstrated in Keh and Chen \cite{KehChen1997}) 
but also to that of solid particle and gas bubble, for example, 
with arbitrary sizes. 
In addition to these fundamental aspects in fluid dynamics, 
the solutions of slip particles is quite important 
for applications to micro- and nanofluidics, 
where the no-slip boundary condition may break 
\cite{NetoEtal2005,LaugaEtal2007,Vinogradova1999,LaugaStone2003}. 
Furthermore, the importance of the exact solution should be emphasized, 
because of the fact that 
the slip boundary condition is solved under relatively limited cases 
compared to the no-slip case. 

Using the multipole expansions and Fax\'{e}n's laws 
derived in the present paper, 
recently the Stokesian dynamics method \cite{BradyBossis1988} 
is extended from the no-slip particles 
to the slip particles \cite{IchikiEtal2008}.  
Because the lubrication corrections are missing in the formulation, 
the applicability is limited to relatively dilute configurations. 
The present work is a first step to improve the Stokesian dynamics method 
for slip particles at the level of the no-slip particles. 
To complete the program, 
we have to obtain the asymptotic forms of resistance functions 
by lubrication theory. To the authors' knowledge, 
just a few functions \cite{BlawzdziewiczEtal1999} 
are obtained for slip particles by now. 
On the other hand, the present exact solution 
expressed by $1/r$ expansion is the complete set 
for the motion of rigid (slip) particles, 
that is, it contains all 11 scalar functions 
for each pair of particles $\alpha\beta$, 
so that it is quite helpful to complete the lubrication theory 
for slip particles and to develop the Stokesian dynamics method 
with lubrication effect for arbitrary slip particles. 

The computer programs used in the paper and the results of coefficients 
for higher orders (up to $k=20$) 
are available on the open source project 
``RYUON-twobody''\cite{RYUON-twobody}.

\begin{acknowledgments}
This work was supported by the National Research Council (NRC) of Canada. 
One of the authors (KI) thanks Professor David Jeffrey 
for his kind support and fruitful discussions. 
\end{acknowledgments}

\appendix
\section{Method of Reflections}
\label{sec:method-of-reflections}
Here we summarize the results of lower coefficients 
obtained by the method of reflections functions.

\subsection{Fax\'{e}n's Laws}
From Eqs. (\ref{eq:dist-vel-F-1b-slip}), (\ref{eq:dist-vel-T-1b-slip}), 
and (\ref{eq:dist-vel-S-1b-slip}) in the previous section, 
the disturbance velocity field at position $\bm{x}$ 
caused by a single sphere $\alpha$ at $\bm{x}_{\alpha}$ 
with slip length $\gamma_{\alpha}$ is given by
\begin{eqnarray}
  \bm{v}(\bm{x})
  &=&
  \frac{1}{8\pi\mu}
  \left[
    \left(
      1
      +
      \Gamma^{(\alpha)}_{0,2}
      \frac{a_{\alpha}^2}{6}
      \nabla^2
    \right)
    \bm{J}(\bm{x}-\bm{x}_{\alpha})
    \cdot
    \bm{F}^{(\alpha)}
  \right.
  \nonumber\\
  &&
  \quad
  \left.
    +
    \bm{R}(\bm{x}-\bm{x}_{\alpha})
    \cdot
    \bm{T}^{(\alpha)}
  \right.
  \nonumber\\
  &&
  \quad
  \left.
    -
    \left(
      1
      +
      \Gamma^{(\alpha)}_{0,2}
      \frac{a_{\alpha}^2\nabla^2}{10}
    \right)
    \bm{K}(\bm{x}-\bm{x}_{\alpha})
    :\bm{S}^{(\alpha)}
  \right]
  ,
  \label{eq:dist-vel-FTS-1b-slip}
\end{eqnarray}
where 
\begin{equation}
  \Gamma^{(\alpha)}_{m,n}
  =
  \frac{1+m\widehat{\gamma}_{\alpha}}{1+n\widehat{\gamma}_{\alpha}}
  ,
\end{equation}
and the force $\bm{F}^{(\alpha)}$, torque $\bm{T}^{(\alpha)}$, 
and stresslet $\bm{S}^{(\alpha)}$ on the sphere are given by
\begin{subequations}
\begin{eqnarray}
  \bm{F}^{(\alpha)}
  &=&
  6\pi\mu a_{\alpha}
  \Gamma^{(\alpha)}_{2,3}
  \bm{U}^{(\alpha)}
  ,
  \label{eq:F-by-U-1b-slip-gen}
  \\
  \bm{T}^{(\alpha)}
  &=&
  8\pi\mu a_{\alpha}^3
  \Gamma^{(\alpha)}_{0,3}
  \bm{\Omega}^{(\alpha)}
  ,
  \label{eq:T-by-O-1b-slip-gen}
  \\
  \bm{S}^{(\alpha)}
  &=&
  \frac{20}{3}
  \pi\mu a_{\alpha}^3
  \Gamma^{(\alpha)}_{2,5}
  \bm{E}^{(\alpha)}
  .
  \label{eq:S-by-E-1b-slip-gen}
\end{eqnarray}
\end{subequations}
(See Eqs. (\ref{eq:F-by-U-1b-slip}), (\ref{eq:T-by-O-1b-slip}), 
and (\ref{eq:S-by-E-1b-slip}) in the previous section.) 
Reading Eq. (\ref{eq:dist-vel-FTS-1b-slip}) 
as multipole expansion of the velocity field, 
Fax\'{e}n's laws for slip sphere are derived as
\begin{eqnarray}
  \bm{F}^{(\alpha)}
  &=&
  6\pi\mu a_{\alpha}
  \Gamma^{(\alpha)}_{2,3}
  \left[
    \bm{U}^{(\alpha)}
    -
    \left(
      1
      +
      \Gamma^{(\alpha)}_{0,2}
      \frac{a_{\alpha}^2}{6}
      \nabla^2
    \right)
    \bm{u}'(\bm{x}_{\alpha})
  \right]
  ,
  \label{eq:faxen-F-slip}
  \\
  \bm{T}^{(\alpha)}
  &=&
  8\pi\mu a_{\alpha}^3
  \Gamma^{(\alpha)}_{0,3}
  \left[
    \bm{\Omega}^{(\alpha)}
    -
    \frac{1}{2}
    \left(
      \bm{\nabla}
      \times
      \bm{u}'
    \right)
    (\bm{x}_{\alpha})
  \right]
  ,
  \label{eq:faxen-T-slip}
  \\
  \bm{S}^{(\alpha)}
  &=&
  \frac{20}{3}
  \pi\mu a_{\alpha}^3
  \Gamma^{(\alpha)}_{2,5}
  \left[
    \bm{E}^{(\alpha)}
  \right.
  \nonumber\\
  &&
  \left.
    -
    \left(
      1
      +
      \Gamma^{(\alpha)}_{0,2}
      \frac{a_{\alpha}^2\nabla^2}{10}
    \right)
    \frac{1}{2}
    \left(
      \bm{\nabla}\bm{u}'
      +
      (\bm{\nabla}\bm{u})^{\dagger}
    \right)
    (\bm{x}_{\alpha})
  \right]
  ,
  \label{eq:faxen-S-slip}
\end{eqnarray}
where $\bm{u}'$ is the velocity field in absent of particle $\alpha$. 
For later use, 
we rewrite Eq. (\ref{eq:dist-vel-FTS-1b-slip}) 
in the resistance form by replacing 
$\bm{F}^{(\alpha)}$, $\bm{T}^{(\alpha)}$, and $\bm{S}^{(\alpha)}$ by 
$\bm{U}^{(\alpha)}$, $\bm{\Omega}^{(\alpha)}$, and $\bm{E}^{(\alpha)}$ 
from Eqs. (\ref{eq:F-by-U-1b-slip-gen}), 
(\ref{eq:T-by-O-1b-slip-gen}), and (\ref{eq:S-by-E-1b-slip-gen}) 
as 
\begin{eqnarray}
  \bm{u}(\bm{x})
  &=&
  \frac{3a_{\alpha}}{4}
  \Gamma^{(\alpha)}_{2,3}
  \left(
    1
    +
    \Gamma^{(\alpha)}_{0,2}
    \frac{a_{\alpha}^2}{6}
    \nabla^2
  \right)
  \bm{J}(\bm{x}-\bm{x}_{\alpha})
  \cdot
  \bm{U}^{(\alpha)}
  \nonumber\\
  &&
  +
  a_{\alpha}^3
  \Gamma^{(\alpha)}_{0,3}
  \bm{R}(\bm{x}-\bm{x}_{\alpha})
  \cdot
  \bm{\Omega}^{(\alpha)}
  \nonumber\\
  &&
  -
  \frac{5a_{\alpha}^3}{6}
  \Gamma^{(\alpha)}_{2,5}
  \left(
    1
    +
    \Gamma^{(\alpha)}_{0,2}
    \frac{a_{\alpha}^2\nabla^2}{10}
  \right)
  \bm{K}(\bm{x}-\bm{x}_{\alpha})
  :\bm{E}^{(\alpha)}
  .
  \label{eq:dist-vel-in-res-form}
\end{eqnarray}

\subsection{Translating Spheres in Axisymmetric Motion}
Here we set the relative vector between particle $1$ and $2$ 
in $z$ direction as 
\begin{equation}
  \bm{r}
  =
  \bm{x}_2
  -
  \bm{x}_1
  =
  (0,0,r)
  .
\end{equation}
For the function $X^{A}$, we set the velocity of the particle $1$ 
parallel to $\bm{r}$ as 
\begin{equation}
  \bm{U}^{(1)}
  =
  (0,0,U^{(1)})
  .
  \label{eq:mofr-XA-U}
\end{equation}

From Fax\'{e}n's law for the force (\ref{eq:faxen-F-slip}) 
with the disturbance field (\ref{eq:dist-vel-in-res-form}) 
with Eq. (\ref{eq:mofr-XA-U}), 
we have the force on the particle $2$ 
due to the translating particle $1$ as
\begin{eqnarray}
  F^{(2)}_i
  &=&
  6\pi\mu a_2
  \Gamma^{(2)}_{2,3}
  U^{(2)}_i
  \nonumber\\
  &&
  -
  6\pi\mu a_2
  \left[
    \frac{3}{2}
    \Gamma^{(2)}_{2,3}
    \Gamma^{(1)}_{2,3}
    \frac{a_1}{r}
    -
    \frac{1}{2}
    \Gamma^{(2)}_{2,3}
    \Gamma^{(1)}_{0,3}
    \frac{a_1^3}{r^3}
  \right.
  \nonumber\\
  &&
  \left.
    -
    \frac{1}{2}
    \Gamma^{(2)}_{0,3}
    \Gamma^{(1)}_{2,3}
    \frac{a_1a_2^2}{r^3}
  \right]
  U^{(1)}
  \delta_{iz}
  .
\end{eqnarray}
In terms of the scalar functions $X^{A}_{\alpha\beta}$, 
the force is expressed as 
\begin{eqnarray}
  F^{(2)}_i
  &=&
  6\pi\mu a_2
  X^{A}_{22}(s,\lambda)
  U^{(2)}\delta_{iz}
  \nonumber\\
  &&
  +
  3\pi\mu (a_2 + a_1)
  X^{A}_{21}(s,\lambda)
  U^{(1)}\delta_{iz}
  ,
\end{eqnarray}
where $s$ and $\lambda$ are defined in Eq. (\ref{eq:two-body-s-and-lambda}). 
Therefore,
\begin{subequations}
\begin{eqnarray}
  X^{A}_{22}(s,\lambda)
  &=&
  \Gamma^{(2)}_{2,3}
  ,\\
  X^{A}_{21}(s,\lambda)
  &=&
  \frac{-2\lambda}{1+\lambda}
  \left(
    \frac{
      3
      \Gamma^{(2)}_{2,3}
      \Gamma^{(1)}_{2,3}
    }{(1+\lambda)s}
    -
    \frac{
      4
      \Gamma^{(2)}_{2,3}
      \Gamma^{(1)}_{0,3}
      +
      4
      \lambda^2
      \Gamma^{(2)}_{0,3}
      \Gamma^{(1)}_{2,3}
    }{(1+\lambda)^3s^3}
  \right)
  .
  \nonumber\\
\end{eqnarray}
\end{subequations}
From the symmetry of $X^{A}_{\alpha\beta}$ 
in Eq. (\ref{eq:XA-particle-exchange}), we have 
\begin{eqnarray}
  X^{A}_{12}(s,\lambda)
  &=&
  \frac{-2}{1+\lambda}
  \left(
    \frac{
      3\lambda
      \Gamma^{(1)}_{2,3}
      \Gamma^{(2)}_{2,3}
    }{(1+\lambda)s}
    -
    \frac{
      4\lambda^3
      \Gamma^{(1)}_{2,3}
      \Gamma^{(2)}_{0,3}
      +
      4\lambda
      \Gamma^{(1)}_{0,3}
      \Gamma^{(2)}_{2,3}
    }{(1+\lambda)^3s^3}
  \right)
  .
  \nonumber\\
\end{eqnarray}
From the expression of $X^{A}_{12}$ in Eq. (\ref{eq:XA12-by-f}), 
we have $f^{XA}_k$ as 
\begin{subequations}
\begin{eqnarray}
  f^{XA}_{1}
  &=&
  3
  \Gamma^{(1)}_{2,3}
  \Gamma^{(2)}_{2,3}
  \lambda
  ,
  \label{eq:mofr-fXA-1}
  \\
  f^{XA}_{3}
  &=&
  -
  4\lambda
  \Gamma^{(1)}_{0,3}
  \Gamma^{(2)}_{2,3}
  -
  4\lambda^3
  \Gamma^{(1)}_{2,3}
  \Gamma^{(2)}_{0,3}
  .
  \label{eq:mofr-fXA-3}
\end{eqnarray}
For the self part $X^{A}_{11}$, we have 
\begin{equation}
  f^{XA}_{0}
  =
  \Gamma^{(1)}_{2,3}
  .
  \label{eq:mofr-fXA-0}
\end{equation}
\end{subequations}
These coefficients (and those for the rest of the functions below) 
will be compared with the results 
by twin multipole expansions in Sec. \ref{sec:twin-multipole-expansions}.

From Fax\'{e}n's law for the torque (\ref{eq:faxen-T-slip}), 
we have torque on the particle $2$ 
due to the translating particle $1$ as
\begin{equation}
  T^{(2)}_i
  =
  0
  ,
\end{equation}
because $\bm{\Omega}^{\alpha} = 0$ in the present problem 
and $\partial_{j}u^{(1)}_k$ is symmetric about the indices $j,k$. 
This fact reflects that there is no $X^{B}$ function in Eq. (\ref{eq:B-by-YB}).

From Fax\'{e}n's law for the stresslet (\ref{eq:faxen-S-slip}), 
\begin{eqnarray}
  S^{(2)}_{ij}
  &=&
  \frac{20}{3}\pi\mu a_2^3
  \Gamma^{(2)}_{2,5}
  E^{(2)}_{ij}
  \nonumber\\
  &&
  -
  \frac{20}{3}\pi\mu a_2^3
  \left(
    -
    \frac{9}{4}
    \frac{a_1}{r^2}
    \Gamma^{(2)}_{2,5}
    \Gamma^{(1)}_{2,3}
    +
    \frac{9}{4}
    \frac{a_1^3}{r^4}
    \Gamma^{(2)}_{2,5}
    \Gamma^{(1)}_{0,3}
  \right.
  \nonumber\\
  &&
  \left.
    +
    \frac{27a_1a_2^2}{20r^4}
    \Gamma^{(2)}_{0,5}
    \Gamma^{(1)}_{2,3}
  \right)
  U^{(1)}
  \left(
    \delta_{iz}
    \delta_{jz}
    -
    \frac{\delta_{ij}}{3}
  \right)
  .
\end{eqnarray}
In terms of the scalar functions $X^{G}_{\alpha\beta}$, 
the stresslet is expressed as 
\begin{eqnarray}
  S^{(2)}_{ij}
  &=&
  \mu\pi
  \left(a_2+a_1\right)^2
  X^{G}_{21}
  U^{(1)}
  \left(
    \delta_{iz}
    \delta_{jz}
    -
    \frac{1}{3}
    \delta_{ij}
  \right)
  ,
\end{eqnarray}
so that
\begin{eqnarray}
  X^{G}_{21}
  &=&
  \frac{-4\lambda^3}{(1+\lambda)^2}
  \left[
    -
    \frac{15}{(1+\lambda)^2s^2}
    \Gamma^{(2)}_{2,5}
    \Gamma^{(1)}_{2,3}
    +
    \frac{60}{(1+\lambda)^4s^4}
    \Gamma^{(2)}_{2,5}
    \Gamma^{(1)}_{0,3}
  \right.
  \nonumber\\
  &&\quad
  \left.
    +
    \frac{
      36
      \lambda^2
    }{(1+\lambda)^4s^4}
    \Gamma^{(2)}_{0,5}
    \Gamma^{(1)}_{2,3}
  \right]
  .
\end{eqnarray}
From the symmetry of $X^{G}_{\alpha\beta}$ 
in Eq. (\ref{eq:XG-particle-exchange}), we have 
\begin{equation}
  X^{G}_{12}
  =
  \frac{-4}{(1+\lambda)^2}
  \left[
    \frac{
      15\lambda
      \Gamma^{(1)}_{2,5}
      \Gamma^{(2)}_{2,3}
    }{(1+\lambda)^2s^2}
    -
    \frac{
      60\lambda^3
      \Gamma^{(1)}_{2,5}
      \Gamma^{(2)}_{0,3}
      +
      36
      \lambda
      \Gamma^{(1)}_{0,5}
      \Gamma^{(2)}_{2,3}
    }{(1+\lambda)^4s^4}
  \right]
  .
\end{equation}
From the expression of $X^{G}_{12}$ in Eq. (\ref{eq:XG12-by-f}), we have 
$f^{XG}_k$ as 
\begin{subequations}
\begin{eqnarray}
  f^{XG}_{0}
  &=&
  0
  ,
  \label{eq:mofr-fXG-0}
  \\
  f^{XG}_{2}
  &=&
  15
  \Gamma^{(1)}_{2,5}
  \Gamma^{(2)}_{2,3}
  \lambda
  ,
  \label{eq:mofr-fXG-2}
  \\
  f^{XG}_{4}
  &=&
  -
  36
  \lambda
  \Gamma^{(1)}_{0,5}
  \Gamma^{(2)}_{2,3}
  -
  60\lambda^3
  \Gamma^{(1)}_{2,5}
  \Gamma^{(2)}_{0,3}
  .
  \label{eq:mofr-fXG-4}
\end{eqnarray}
\end{subequations}

\subsection{Translating Spheres in Asymmetric Motion}
Next, we study the asymmetric motion of the spheres 
to their center-to-center vector, that is, the velocity 
$\bm{U}^{(1)}$ is in $y$-direction as
\begin{equation}
  \bm{U}^{(1)}
  =
  (0,U^{(1)},0)
  .
  \label{eq:mofr-YA-U}
\end{equation}
Note that, for $\bm{r}=(0,0,r)$, 
from Eq. (\ref{eq:A-by-XA-and-YA}), we have 
\begin{equation}
  \widehat{\mathsf{A}}_{\alpha\beta}\cdot\bm{U}^{(\beta)}
  =
  \left[
    \begin{array}{c}
      Y^{A}_{\alpha\beta} U^{(\beta)}_x \\
      Y^{A}_{\alpha\beta} U^{(\beta)}_y \\
      X^{A}_{\alpha\beta} U^{(\beta)}_z
    \end{array}
  \right]
  .
\end{equation}

From Fax\'{e}n's law for the force (\ref{eq:faxen-F-slip}) 
with the disturbance field (\ref{eq:dist-vel-in-res-form}) 
with Eq. (\ref{eq:mofr-YA-U}), 
we have the force on the particle $2$ 
due to the translating particle $1$ as
\begin{eqnarray}
  F^{(2)}_i
  &=&
  6\pi\mu a_2
  \Gamma^{(2)}_{2,3}
  U^{(2)}_i
  -
  6\pi\mu a_2
  \left(
    \frac{3a_1}{4r}
    \Gamma^{(2)}_{2,3}
    \Gamma^{(1)}_{2,3}
  \right.
  \nonumber\\
  &&
  \left.
    +
    \frac{1}{4}
    \frac{a_1^3}{r^3}
    \Gamma^{(2)}_{2,3}
    \Gamma^{(1)}_{0,3}
    +
    \frac{a_2^2}{4}
    \frac{a_1}{r^3}
    \Gamma^{(2)}_{0,3}
    \Gamma^{(1)}_{2,3}
  \right)
  U^{(1)}
  \delta_{iy}
  .
\end{eqnarray}
In terms of the scalar functions $Y^{A}_{\alpha\beta}$, 
the force is expressed as 
\begin{eqnarray}
  F^{(2)}_i
  &=&
  6\pi\mu a_2
  Y^{A}_{22}(s,\lambda)
  U^{(2)}\delta_{iy}
  \nonumber\\
  &&
  +
  3\pi\mu (a_2 + a_1)
  Y^{A}_{21}(s,\lambda)
  U^{(1)}\delta_{iy}
  .
\end{eqnarray}
Therefore,
\begin{subequations}
\begin{eqnarray}
  &&
  Y^{A}_{22}(s,\lambda)
  =
  \Gamma^{(2)}_{2,3}
  ,\\
  &&
  Y^{A}_{21}(s,\lambda)
  =
  -
  \frac{2\lambda}{1+\lambda}
  \left(
    \frac{3}{2}
    \frac{1}{(1+\lambda)s}
    \Gamma^{(2)}_{2,3}
    \Gamma^{(1)}_{2,3}
  \right.
  \nonumber\\
  &&
  \quad
  \left.
    +
    \frac{2}{(1+\lambda)^3s^3}
    \left(
      \Gamma^{(2)}_{2,3}
      \Gamma^{(1)}_{0,3}
      +
      \lambda^2
      \Gamma^{(2)}_{0,3}
      \Gamma^{(1)}_{2,3}
    \right)
  \right)
  .
\end{eqnarray}
\end{subequations}
From the symmetry of $Y^{A}_{\alpha\beta}$ 
in Eq. (\ref{eq:YA-particle-exchange}), we have
\begin{subequations}
\begin{eqnarray}
  &&
  Y^{A}_{11}(s,\lambda)
  =
  \Gamma^{(1)}_{2,3}
  ,\\
  &&
  Y^{A}_{12}(s,\lambda)
  =
  -
  \frac{2}{1+\lambda}
  \left(
    \frac{3}{2}
    \frac{\lambda}{(1+\lambda)s}
    \Gamma^{(1)}_{2,3}
    \Gamma^{(2)}_{2,3}
  \right.
  \nonumber\\
  &&
  \quad
  \left.
    +
    \frac{2}{(1+\lambda)^3s^3}
    \left(
      \lambda^3
      \Gamma^{(1)}_{2,3}
      \Gamma^{(2)}_{0,3}
      +
      \lambda
      \Gamma^{(1)}_{0,3}
      \Gamma^{(2)}_{2,3}
    \right)
  \right)
  .
\end{eqnarray}
\end{subequations}
From the expression of $Y^{A}_{12}$ in Eq. (\ref{eq:YA12-by-f}), we have 
$f^{YA}_k$ as 
\begin{subequations}
\begin{eqnarray}
  f^{YA}_{0}
  &=&
  \Gamma^{(1)}_{2,3}
  ,
  \label{eq:mofr-fYA-0}
  \\
  f^{YA}_{1}
  &=&
  \frac{3}{2}
  \Gamma^{(1)}_{2,3}
  \Gamma^{(2)}_{2,3}
  \lambda
  ,
  \label{eq:mofr-fYA-1}
  \\
  f^{YA}_{3}
  &=&
  2
  \Gamma^{(1)}_{2,3}
  \Gamma^{(2)}_{0,3}
  \lambda^3
  +
  2
  \Gamma^{(1)}_{0,3}
  \Gamma^{(2)}_{2,3}
  \lambda
  .
  \label{eq:mofr-fYA-3}
\end{eqnarray}
\end{subequations}

From Fax\'{e}n's law for the torque (\ref{eq:faxen-T-slip}), 
we have the torque on the particle $2$ 
due to the translating particle $1$ as
\begin{equation}
  T^{(2)}_i
  =
  -
  6\pi\mu a_2^3
  \frac{a_1}{r^2}
  \Gamma^{(2)}_{0,3}
  \Gamma^{(1)}_{2,3}
  U^{(1)}
  \delta_{ix}
  .
\end{equation}
In terms of the scalar functions $Y^{B}_{\alpha\beta}$, 
the torque is expressed as 
\begin{eqnarray}
  T^{(2)}_i
  &=&
  4\pi\mu a_2^2
  Y^{B}_{22}
  \delta_{ix}
  U^{(2)}
  +
  \pi\mu
  (a_2+a_1)^2
  Y^{B}_{21}
  \delta_{ix}
  U^{(1)}
  .
\end{eqnarray}
Therefore,
\begin{subequations}
\begin{eqnarray}
  Y^{B}_{22}
  &=&
  0
  ,\\
  Y^{B}_{21}
  &=&
  \frac{-4}{(1+\lambda)^2}
  \frac{6\lambda^3}{(1+\lambda)^2s^2}
  \Gamma^{(2)}_{0,3}
  \Gamma^{(1)}_{2,3}
  .
\end{eqnarray}
\end{subequations}
From the symmetry of $Y^{B}_{\alpha\beta}$ 
in Eq. (\ref{eq:YB-particle-exchange}), we have
\begin{subequations}
\begin{eqnarray}
  Y^{B}_{11}
  &=&
  0
  ,\\
  Y^{B}_{12}
  &=&
  \frac{-4}{(1+\lambda)^2}
  \frac{-6\lambda}{(1+\lambda)^2s^2}
  \Gamma^{(1)}_{0,3}
  \Gamma^{(2)}_{2,3}
  .
\end{eqnarray}
\end{subequations}
From the expression of $Y^{B}_{11}$ and $Y^{B}_{12}$ 
in Eqs. (\ref{eq:YB11-by-f}) and (\ref{eq:YB12-by-f}), we have 
$f^{YB}_k$ as 
\begin{subequations}
\begin{eqnarray}
  f^{YB}_{0}
  &=&
  0,
  \label{eq:mofr-fYB-0}
  \\
  f^{YB}_{1}
  &=&
  0,
  \label{eq:mofr-fYB-1}
  \\
  f^{YB}_{2}
  &=&
  -
  6\lambda
  \Gamma^{(1)}_{0,3}
  \Gamma^{(2)}_{2,3}
  .
  \label{eq:mofr-fYB-2}
\end{eqnarray}
\end{subequations}

From Fax\'{e}n's law for the stresslet (\ref{eq:faxen-S-slip}), 
the stresslet is given by
\begin{eqnarray}
  S^{(2)}_{ij}
  &=&
  \frac{20}{3}\pi\mu a_2^3
  \Gamma^{(2)}_{2,5}
  \left(
    \frac{3a_1^3}{4r^4}
    \Gamma^{(1)}_{0,3}
    +
    \Gamma^{(2)}_{0,2}
    \frac{a_2}{10}
    \frac{9a_1}{2r^4}
    \Gamma^{(1)}_{2,3}
  \right)
  \nonumber\\
  &&
  \times
  U^{(1)}
  \left(
    \delta_{iy}
    \delta_{jz}
    +
    \delta_{jy}
    \delta_{iz}
  \right)
  .
\end{eqnarray}
Note that
\begin{eqnarray}
  S^{(2)}_{ij}
  &=&
  4\pi\mu a_2^2
  G^{22}_{ijk}
  U^{(2)}_{k}
  +
  \pi\mu (a_2+a_1)^2
  G^{21}_{ijk}
  U^{(1)}_{k}
  ,
\end{eqnarray}
where 
\begin{equation}
  G^{(\alpha\beta)}_{ijk}
  U_{k}
  =
  Y^{G}_{\alpha\beta}
  \left(
    \delta_{iz}
    \delta_{jy}
    +
    \delta_{jz}
    \delta_{iy}
  \right)
  U
  ,
\end{equation}
for $\bm{e} = (0,0,1)$ and $\bm{U} = (0, U, 0)$. 
Therefore, we have 
\begin{subequations}
\begin{eqnarray}
  Y^{G}_{22}
  &=&
  0
  ,\\
  Y^{G}_{21}
  &=&
  \frac{20\lambda^2}{(1+\lambda)^2}
  \left(
    \frac{4\lambda}{(1+\lambda)^4s^4}
    \Gamma^{(2)}_{2,5}
    \Gamma^{(1)}_{0,3}
  \right.
  \nonumber\\
  &&
  \left.
    +
    \frac{12}{5}
    \frac{\lambda^3}{(1+\lambda)^4s^4}
    \Gamma^{(2)}_{0,5}
    \Gamma^{(1)}_{2,3}
  \right)
  .
\end{eqnarray}
\end{subequations}
From the symmetry of $Y^{G}_{\alpha\beta}$ 
in Eq. (\ref{eq:YG-particle-exchange}), we have 
\begin{subequations}
\begin{eqnarray}
  Y^{G}_{11}
  &=&
  0
  ,\\
  Y^{G}_{12}
  &=&
  \frac{-4}{(1+\lambda)^2}
  \left(
    \frac{20\lambda^{3}}{(1+\lambda)^4s^4}
    \Gamma^{(1)}_{2,5}
    \Gamma^{(2)}_{0,3}
  \right.
  \nonumber\\
  &&
  \left.
    +
    \frac{12\lambda}{(1+\lambda)^4s^4}
    \Gamma^{(1)}_{0,5}
    \Gamma^{(2)}_{2,3}
  \right)
  .
\end{eqnarray}
\end{subequations}
From the expression of $Y^{G}_{12}$ in Eq. (\ref{eq:YG12-by-f}), we have 
$f^{YG}_k$ as 
\begin{subequations}
\begin{eqnarray}
  f^{YG}_{0} &=& 0,
  \label{eq:mofr-fYG-0}
  \\
  f^{YG}_{2} &=& 0,
  \label{eq:mofr-fYG-2}
  \\
  f^{YG}_{4} &=&
  20\lambda^{3}
  \Gamma^{(1)}_{2,5}
  \Gamma^{(2)}_{0,3}
  +
  12\lambda
  \Gamma^{(1)}_{0,5}
  \Gamma^{(2)}_{2,3}
  .
  \label{eq:mofr-fYG-4}
\end{eqnarray}
\end{subequations}

\subsection{Rotating Spheres}
Next, we consider rotating spheres. 
In the two-body problem with $\bm{r} = (0,0,r)$, 
we set the angular velocity $\bm{\Omega}^{(1)}$ 
for the axisymmetric case by 
\begin{subequations}
\begin{equation}
  \Omega^{(1)}_i
  =
  \Omega^{(1)}
  \delta_{iz}
  ,
  \label{eq:mofr-XC-O}
\end{equation}
and for the asymmetric case to the axis $\bm{r}$ by 
\begin{equation}
  \Omega^{(1)}_i
  =
  \Omega^{(1)}
  \delta_{iy}
  .
  \label{eq:mofr-YC-O}
\end{equation}
\end{subequations}

\paragraph{Torque in Axisymmetric Motion}
From Fax\'{e}n's law for the torque (\ref{eq:faxen-T-slip}) 
with the disturbance field (\ref{eq:dist-vel-in-res-form}) 
with Eq. (\ref{eq:mofr-XC-O}), 
we have the torque on the particle $2$ 
due to the translating particle $1$ as
\begin{equation}
  T^{(2)}_i
  =
  8\pi\mu a_2^3
  \Gamma^{(2)}_{0,3}
  \Omega^{(2)}_i
  -
  8\pi\mu a_2^3
  \Gamma^{(2)}_{0,3}
  \frac{a_1^3}{r^3}
  \Gamma^{(1)}_{0,3}
  \delta_{iz}
  \Omega^{(1)}
  .
\end{equation}
In terms of the scalar functions $X^{C}_{\alpha\beta}$, 
the torque is expressed as 
\begin{eqnarray}
  T^{(2)}_{i}
  &=&
  8\pi\mu a_2^3
  X^{C}_{22}
  \Omega^{(2)}
  \delta_{iz}
  +
  \pi\mu (a_2+a_1)^3
  X^{C}_{21}
  \Omega^{(1)}
  \delta_{iz}
  .
\end{eqnarray}
Therefore,
\begin{subequations}
\begin{eqnarray}
  X^{C}_{22}
  &=&
  \Gamma^{(2)}_{0,3}
  ,\\
  X^{C}_{21}
  &=&
  -
  \frac{8\lambda^3}{(1+\lambda)^3}
  \frac{8}{(1+\lambda)^3s^3}
  \Gamma^{(2)}_{0,3}
  \Gamma^{(1)}_{0,3}
  .
\end{eqnarray}
\end{subequations}
From the symmetry of $X^{C}_{\alpha\beta}$ 
in Eq. (\ref{eq:XC-particle-exchange}), we have 
\begin{eqnarray}
  X^{C}_{12}(\lambda)
  &=&
  -
  \frac{8}{(1+\lambda)^3}
  \frac{8\lambda^3}{(1+\lambda)^3s^3}
  \Gamma^{(1)}_{0,3}
  \Gamma^{(2)}_{0,3}
  .
\end{eqnarray}
From the expression of $X^{C}_{12}$ in Eq. (\ref{eq:XC12-by-f}), we have 
$f^{XC}_k$ as 
\begin{subequations}
\begin{eqnarray}
  f^{XC}_{0}
  &=&
  \Gamma^{(1)}_{0,3}
  ,
  \label{eq:mofr-fXC-0}
  \\
  f^{XC}_{1} &=& 0,
  \label{eq:mofr-fXC-1}
  \\
  f^{XC}_{3} &=&
  8\lambda^3
  \Gamma^{(1)}_{0,3}
  \Gamma^{(2)}_{0,3}
  .
  \label{eq:mofr-fXC-3}
\end{eqnarray}
\end{subequations}

\paragraph{Torque in Asymmetric Motion}
For the asymmetric motion to the center-to-center vector, 
from Fax\'{e}n's law for the torque (\ref{eq:faxen-T-slip}) 
with the disturbance field (\ref{eq:dist-vel-in-res-form}) 
with Eq. (\ref{eq:mofr-YC-O}), 
we have the torque on the particle $2$ 
due to the translating particle $1$ as
\begin{equation}
  T^{(2)}_i
  =
  8\pi\mu a_2^3
  \Gamma^{(2)}_{0,3}
  \Omega^{(2)}_i
  +
  4\pi\mu a_2^3
  \Gamma^{(2)}_{0,3}
  \Gamma^{(1)}_{0,3}
  \frac{a_1^3}{r^3}
  \delta_{iy}
  \Omega^{(1)}
  .
\end{equation}
In terms of the scalar functions $Y^{C}_{\alpha\beta}$, 
the torque is expressed as 
\begin{eqnarray}
  T^{(2)}_{i}
  &=&
  8\pi\mu a_2^3
  Y^{C}_{22}
  \Omega^{(2)}
  \delta_{iy}
  +
  \pi\mu (a_2+a_1)^3
  Y^{C}_{21}
  \Omega^{(1)}
  \delta_{iy}
  .
\end{eqnarray}
Therefore,
\begin{subequations}
\begin{eqnarray}
  Y^{C}_{22}
  &=&
  \Gamma^{(2)}_{0,3}
  ,\\
  Y^{C}_{21}
  &=&
  \frac{4\lambda^3}{(1+\lambda)^3}
  \Gamma^{(2)}_{0,3}
  \Gamma^{(1)}_{0,3}
  \frac{8}{(1+\lambda)^3s^3}
  .
\end{eqnarray}
\end{subequations}
From the symmetry of $Y^{C}_{\alpha\beta}$ 
in Eq. (\ref{eq:YC-particle-exchange}), we have 
\begin{eqnarray}
  Y^{C}_{12}(\lambda)
  &=&
  \frac{4}{(1+\lambda)^3}
  \Gamma^{(1)}_{0,3}
  \Gamma^{(2)}_{0,3}
  \frac{8\lambda^3}{(1+\lambda)^3s^3}
  .
\end{eqnarray}
From the expression of $Y^{C}_{12}$ in Eq. (\ref{eq:YC12-by-f}), we have 
$f^{YC}_k$ as 
\begin{subequations}
\begin{eqnarray}
  f^{YC}_{1} &=& 0,
  \label{eq:mofr-fYC-1}
  \\
  f^{YC}_{3} &=&
  4\lambda^3
  \Gamma^{(1)}_{0,3}
  \Gamma^{(2)}_{0,3}
  .
  \label{eq:mofr-fYC-3}
\end{eqnarray}
\end{subequations}

\paragraph{Stresslet}
Because $\partial_ju^{(1)}_i$ for the axisymmetric motion 
is anti-symmetric for $i$ and $j$, 
there is no contribution to the stresslet. 
For the asymmetric motion to the axis, 
from Fax\'{e}n's law for the stresslet (\ref{eq:faxen-S-slip}) 
with the disturbance field (\ref{eq:dist-vel-in-res-form}), 
\begin{equation}
  S^{(2)}_{ij}
  =
  10\pi\mu a_2^3
  \frac{a_1^3}{r^3}
  \Gamma^{(2)}_{2,5}
  \Gamma^{(1)}_{0,3}
  \left(
    \delta_{iz}
    \delta_{jx}
    +
    \delta_{ix}
    \delta_{jz}
  \right)
  \Omega^{(1)}
  .
\end{equation}
The stresslet on particle $2$ caused by particle $1$ is given by 
\begin{equation}
  \pi\mu (a_2+a_1)^3
  H^{(21)}_{ijk}
  \Omega^{(1)}_{k}
  ,
\end{equation}
where, for $\bm{r}=(0,0,r)$ and $\Omega^{(1)}_{k} = \Omega^{(1)} \delta_{ky}$, 
\begin{equation}
  H^{(21)}_{ijk}
  \Omega^{(1)}_{k}
  =
  Y^{H}_{21}
  \left(
    \delta_{iz}
    \delta_{jx}
    +
    \delta_{jz}
    \delta_{ix}
  \right)
  \Omega^{(1)}
  .
\end{equation}
Therefore, 
\begin{equation}
  Y^{H}_{21}
  =
  \frac{10}{(1+\lambda)^3}
  \frac{8}{(1+\lambda)^3s^3}
  \Gamma^{(2)}_{2,5}
  \Gamma^{(1)}_{0,3}
  .
\end{equation}
From the symmetry of $Y^{H}_{\alpha\beta}$ 
in Eq. (\ref{eq:YH-particle-exchange}), we have
\begin{equation}
  Y^{H}_{12}
  =
  \frac{10}{(1+\lambda)^3}
  \frac{8\lambda^3}{(1+\lambda)^3s^3}
  \Gamma^{(1)}_{2,5}
  \Gamma^{(2)}_{0,3}
  .
\end{equation}
From the expression of $Y^{H}_{12}$ in Eq. (\ref{eq:YH12-by-f}), we have 
$f^{YH}_k$ as 
\begin{subequations}
\begin{eqnarray}
  f^{YH}_{1} &=& 0,
  \label{eq:mofr-fYH-1}
  \\
  f^{YH}_{3} &=&
  10\lambda^3
  \Gamma^{(1)}_{2,5}
  \Gamma^{(2)}_{0,3}
  .
  \label{eq:mofr-fYH-3}
\end{eqnarray}
\end{subequations}

\subsection{Spheres in Strain Flow}
Next, we consider the problem under the strain flow. 
Let us define three types of strain by 
\begin{subequations}
\begin{eqnarray}
  E^{X}_{kl}
  &=&
  E^{X}
  \left(
    \delta_{kz}
    \delta_{lz}
    -
    \frac{\delta_{kl}}{3}
  \right)
  ,
  \label{eq:mofr-EX-for-XM}
  \\
  E^{Y}_{kl}
  &=&
  E^{Y}
  \left(
    \delta_{kz}
    \delta_{lx}
    +
    \delta_{kx}
    \delta_{lz}
  \right)
  ,
  \label{eq:mofr-EY-for-YM}
  \\
  E^{Z}_{kl}
  &=&
  E^{Z}
  \left(
    \delta_{kx}
    \delta_{lx}
    -
    \delta_{ky}
    \delta_{ly}
  \right)
  ,
  \label{eq:mofr-EZ-for-ZM}
\end{eqnarray}
\end{subequations}
which correspond to the scalar functions 
$X^{M}_{\alpha\beta}$, $Y^{M}_{\alpha\beta}$, and $Z^{M}_{\alpha\beta}$, 
respectively. 

In the following, 
we will see $\bm{S}^{(2;1)}$, 
the stresslet on particle $2$ caused by particle $1$, 
which is related to the resistance functions 
$X^{M}_{12}$, $Y^{M}_{12}$, and $Z^{M}_{12}$. 
From Fax\'{e}n's law for the stresslet (\ref{eq:faxen-S-slip}), 
it is given by 
\begin{equation}
  S^{(2;1)}_{ij}
  =
  \frac{20}{3}\pi\mu a_2^3
  \Gamma^{(2)}_{2,5}
  \left[
    -
    \left(
      1
      +
      \Gamma^{(2)}_{0,2}
      \frac{a_2^2\nabla^2}{10}
    \right)
    \frac{1}{2}
    \left[
      \partial_i
      u^{(1)}_j
      +
      \partial_j
      u^{(1)}_i
    \right](\bm{x}_2)
  \right]
  .
  \label{eq:mofr-S-for-M}
\end{equation}

\paragraph{Function $X^{M}$}
Substituting 
the disturbance field (\ref{eq:dist-vel-in-res-form}) 
with $E^{X}_{kl}$ (\ref{eq:mofr-EX-for-XM}) 
into Eq. (\ref{eq:mofr-S-for-M}), 
we have 
\begin{eqnarray}
  S^{(2;1)}_{ij}
  &=&
  \frac{20}{3}\pi\mu a_2^3
  \left[
    \frac{5a_1^3}{r^3}
    \Gamma^{(2)}_{2,5}
    \Gamma^{(1)}_{2,5}
  \right.
  \nonumber\\
  &&
  \left.
    -
    \frac{6}{r^5}
    \left(
      a_1^5
      \Gamma^{(2)}_{2,5}
      \Gamma^{(1)}_{0,5}
      +
      a_2^2
      a_1^3
      \Gamma^{(2)}_{0,5}
      \Gamma^{(1)}_{2,5}
    \right)
  \right]
  \nonumber\\
  &&
  \times
  \left(
    \delta_{iz}\delta_{jz}
    -
    \frac{\delta_{ij}}{3}
  \right)
  E^{X}
  .
\end{eqnarray}
In terms of the scalar function $X^{M}_{21}$, 
it is written as 
\begin{equation}
  S^{(2;1)}_{ij}
  =
  \frac{5}{6}
  \pi\mu
  (a_2+a_1)^3
  X^{M}_{21}
  \left(
    \delta_{iz}\delta_{jz}
    -
    \frac{\delta_{ij}}{3}
  \right)
  E^{X}
  .
\end{equation}
Therefore, 
\begin{eqnarray}
  X^{M}_{21}
  &=&
  8
  \frac{\lambda^3}{(1+\lambda)^3}
  \left[
    \frac{40}{(1+\lambda)^3s^3}
    \Gamma^{(2)}_{2,5}
    \Gamma^{(1)}_{2,5}
  \right.
  \nonumber\\
  &&
  \left.
    -\frac{192}{(1+\lambda)^5s^5}
    \left(
      \Gamma^{(2)}_{2,5}
      \Gamma^{(1)}_{0,5}
      +
      \lambda^2
      \Gamma^{(2)}_{0,5}
      \Gamma^{(1)}_{2,5}
    \right)
  \right]
  .
\end{eqnarray}
From the symmetry of $X^{M}_{\alpha\beta}$ 
in Eq. (\ref{eq:XM-particle-exchange}), we have 
\begin{eqnarray}
  X^{M}_{12}
  &=&
  \frac{8}{(1+\lambda)^3}
  \left[
    \frac{40\lambda^3}{(1+\lambda)^3s^3}
    \Gamma^{(1)}_{2,5}
    \Gamma^{(2)}_{2,5}
  \right.
  \nonumber\\
  &&
  \left.
    -
    \frac{192}{(1+\lambda)^5s^5}
    \left(
      \lambda^5
      \Gamma^{(1)}_{2,5}
      \Gamma^{(2)}_{0,5}
      +
      \lambda^3
      \Gamma^{(1)}_{0,5}
      \Gamma^{(2)}_{2,5}
    \right)
  \right]
  .
\end{eqnarray}
From the expression of $X^{M}_{12}$ in Eq. (\ref{eq:XM12-by-f}), we have 
$f^{XM}_k$ as 
\begin{subequations}
\begin{eqnarray}
  f^{XM}_{1} &=& 0,
  \label{eq:mofr-fXM-1}
  \\
  f^{XM}_{3} &=&
  40\lambda^3
  \Gamma^{(1)}_{2,5}
  \Gamma^{(2)}_{2,5}
  ,
  \label{eq:mofr-fXM-3}
  \\
  f^{XM}_{5} &=&
  -
  192
  \left(
    \lambda^5
    \Gamma^{(1)}_{2,5}
    \Gamma^{(2)}_{0,5}
    +
    \lambda^3
    \Gamma^{(1)}_{0,5}
    \Gamma^{(2)}_{2,5}
  \right)
  .
  \label{eq:mofr-fXM-5}
\end{eqnarray}
\end{subequations}

\paragraph{Function $Y^{M}$}
Substituting 
the disturbance field (\ref{eq:dist-vel-in-res-form}) 
with $E^{Y}_{kl}$ (\ref{eq:mofr-EY-for-YM}) 
into Eq. (\ref{eq:mofr-S-for-M}), 
we have 
\begin{eqnarray}
  S^{(2;1)}_{ij}
  &=&
  \frac{20}{3}\pi\mu a_2^3
  \left[
    -
    \frac{5}{2}
    \frac{a_1^3}{r^3}
    \Gamma^{(2)}_{2,5}
    \Gamma^{(1)}_{2,5}
  \right.
  \nonumber\\
  &&
  \left.
    +
    \frac{4}{r^5}
    \left(
      a_1^5
      \Gamma^{(2)}_{2,5}
      \Gamma^{(1)}_{0,5}
      +
      a_2^2
      a_1^3
      \Gamma^{(2)}_{0,5}
      \Gamma^{(1)}_{2,5}
    \right)
  \right]
  \nonumber\\
  &&
  \times
  \left(
    \delta_{ix}\delta_{jz}
    +
    \delta_{iz}\delta_{jx}
  \right)
  E^{Y}
  .
\end{eqnarray}
In terms of the scalar function $Y^{M}_{21}$, 
it is written as 
\begin{equation}
  S^{(2;1)}_{ij}
  =
  \frac{5}{6}
  \pi\mu
  (a_2+a_1)^3
  Y^{M}_{21}
  \left(
    \delta_{iz}\delta_{jx}
    +
    \delta_{ix}\delta_{jz}
  \right)
  E^{Y}
  .
\end{equation}
Therefore, 
\begin{eqnarray}
  Y^{M}_{21}
  &=&
  8
  \frac{\lambda^3}{(1+\lambda)^3}
  \left[
    -
    \frac{20}{(1+\lambda)^3s^3}
    \Gamma^{(2)}_{2,5}
    \Gamma^{(1)}_{2,5}
  \right.
  \nonumber\\
  &&
  \left.
    +
    \frac{128}{(1+\lambda)^5s^5}
    \left(
      \Gamma^{(2)}_{2,5}
      \Gamma^{(1)}_{0,5}
      +
      \lambda^2
      \Gamma^{(2)}_{0,5}
      \Gamma^{(1)}_{2,5}
    \right)
  \right]
  .
\end{eqnarray}
From the symmetry of $Y^{M}_{\alpha\beta}$ 
in Eq. (\ref{eq:YM-particle-exchange}), we have 
\begin{eqnarray}
  Y^{M}_{12}
  &=&
  \frac{8}{(1+\lambda)^3}
  \left[
    -
    \frac{20\lambda^3}{(1+\lambda)^3s^3}
    \Gamma^{(1)}_{2,5}
    \Gamma^{(2)}_{2,5}
  \right.
  \nonumber\\
  &&
  \left.
    +
    \frac{128}{(1+\lambda)^5s^5}
    \left(
      \lambda^{5}
      \Gamma^{(1)}_{2,5}
      \Gamma^{(2)}_{0,5}
      +
      \lambda^{3}
      \Gamma^{(1)}_{0,5}
      \Gamma^{(2)}_{2,5}
    \right)
  \right]
  .
\end{eqnarray}
From the expression of $Y^{M}_{12}$ in Eq. (\ref{eq:YM12-by-f}), we have 
$f^{YM}_k$ as 
\begin{subequations}
\begin{eqnarray}
  f^{YM}_{1} &=& 0,
  \label{eq:mofr-fYM-1}
  \\
  f^{YM}_{3} &=&
  -
  20\lambda^3
  \Gamma^{(1)}_{2,5}
  \Gamma^{(2)}_{2,5}
  ,
  \label{eq:mofr-fYM-3}
  \\
  f^{YM}_{5} &=&
  128
  \left(
    \lambda^{5}
    \Gamma^{(1)}_{2,5}
    \Gamma^{(2)}_{0,5}
    +
    \lambda^{3}
    \Gamma^{(1)}_{0,5}
    \Gamma^{(2)}_{2,5}
  \right)
  .
  \label{eq:mofr-fYM-5}
\end{eqnarray}
\end{subequations}

\paragraph{Function $Z^{M}$}
Substituting 
the disturbance field (\ref{eq:dist-vel-in-res-form}) 
with $E^{Z}_{kl}$ (\ref{eq:mofr-EZ-for-ZM}) 
into Eq. (\ref{eq:mofr-S-for-M}), 
we have 
\begin{eqnarray}
  S^{(2;1)}_{ij}
  &=&
  -
  \frac{20}{3}\pi\mu a_2^3
  \frac{1}{r^5}
  \left(
    a_1^5
    \Gamma^{(2)}_{2,5}
    \Gamma^{(1)}_{0,5}
    +
    a_2^2
    a_1^3
    \Gamma^{(2)}_{0,5}
    \Gamma^{(1)}_{2,5}
  \right)
  \nonumber\\
  &&
  \times
  \left(
    \delta_{ix}\delta_{jx}
    -
    \delta_{iy}\delta_{jy}
  \right)
  E^{Z}
  .
\end{eqnarray}
In terms of the scalar function $Z^{M}_{21}$, 
it is written as 
\begin{equation}
  S^{(2;1)}_{ij}
  =
  \frac{5}{6}
  \pi\mu
  (a_2+a_1)^3
  Z^{M}_{21}
  \left(
    \delta_{ix}\delta_{jx}
    -
    \delta_{iy}\delta_{jy}
  \right)
  E^{Z}
  .
\end{equation}
Therefore, 
\begin{equation}
  Z^{M}_{21}
  =
  -
  8
  \frac{\lambda^3}{(1+\lambda)^3}
  \frac{32}{(1+\lambda)^5s^5}
  \left(
    \Gamma^{(2)}_{2,5}
    \Gamma^{(1)}_{0,5}
    +
    \lambda^2
    \Gamma^{(2)}_{0,5}
    \Gamma^{(1)}_{2,5}
  \right)
  .
\end{equation}
From the symmetry of $Z^{M}_{\alpha\beta}$ 
in Eq. (\ref{eq:ZM-particle-exchange}), we have 
\begin{equation}
  Z^{M}_{12}
  =
  \frac{-8}{(1+\lambda)^3}
  \frac{32}{(1+\lambda)^5s^5}
  \left(
    \lambda^5
    \Gamma^{(1)}_{2,5}
    \Gamma^{(2)}_{0,5}
    +
    \lambda^3
    \Gamma^{(1)}_{0,5}
    \Gamma^{(2)}_{2,5}
  \right)
  .
\end{equation}
From the expression of $Z^{M}_{12}$ in Eq. (\ref{eq:ZM12-by-f}), we have 
$f^{ZM}_k$ as 
\begin{subequations}
\begin{eqnarray}
  f^{ZM}_{1} &=& 0,
  \label{eq:mofr-fZM-1}
  \\
  f^{ZM}_{3} &=& 0,
  \label{eq:mofr-fZM-3}
  \\
  f^{ZM}_{5} &=&
  32
  \left(
    \lambda^5
    \Gamma^{(1)}_{2,5}
    \Gamma^{(2)}_{0,5}
    +
    \lambda^3
    \Gamma^{(1)}_{0,5}
    \Gamma^{(2)}_{2,5}
  \right)
  .
  \label{eq:mofr-fZM-5}
\end{eqnarray}
\end{subequations}

\newpage


\end{document}